\documentclass[twocolumn,floatfix,prl,aps,showpacs]{revtex4-2}
\usepackage{graphicx,amsmath,amssymb,color,ulem}
\usepackage[utf8]{inputenc}
\usepackage{nicefrac}
\usepackage{multirow, makecell, ragged2e}
\usepackage[titletoc,title]{appendix}
\usepackage[colorlinks,bookmarks=true,citecolor=blue,linkcolor=red,urlcolor=blue]{hyperref}

\newcommand{\be}{\begin{equation}}
\newcommand{\ee}{\end{equation}}

\newcommand{\ba}{\begin{eqnarray}}
\newcommand{\ea}{\end{eqnarray}}

\renewcommand{\vec}[1]{\mbox{\boldmath$#1$}}

\begin{document}

\title{
  Composite Fermion Pairing Induced by Landau Level Mixing
}
\author{Tongzhou Zhao$^1$,  Ajit C. Balram$^{2,3}$, and J. K. Jain$^4$}

\affiliation{$^1$Institute of Physics, Chinese Academy of Sciences, Beijing 100190, China}
\affiliation{$^{2}$Institute of Mathematical Sciences, CIT Campus, Chennai 600113, India}
\affiliation{$^{3}$Homi Bhabha National Institute, Training School Complex, Anushaktinagar, Mumbai 400094, India}   
 \affiliation{$^4$Department of Physics, 104 Davey Lab, Pennsylvania State University, University Park, Pennsylvania 16802, USA}

\date{\today}

\begin{abstract} 
	Pairing of composite fermions provides a possible mechanism for fractional quantum Hall effect at even denominator fractions and is believed to serve as a platform for realizing quasiparticles with non-Abelian braiding statistics. We present results from fixed-phase diffusion Monte Carlo calculations which predict that substantial Landau level mixing can induce a pairing of composite fermions at filling factors $\nu=1/2$ and $\nu=1/4$ in the $l=-3$ relative angular momentum channel, thereby destabilizing the composite-fermion Fermi seas to produce non-Abelian fractional quantum Hall states. 

\end{abstract}
\maketitle

The recent observation by Wang {\it et al.}~\cite{Wang22a,Berkowitz22} of fractional quantum Hall effect (FQHE) at filling factor $\nu=3/4$ has come as a surprise, because a priori one would have expected a composite-fermion (CF) Fermi sea here~\cite{Halperin93,Halperin20b,Shayegan20}, where composite fermions are bound states of electrons and an even number of quantized vortices~\cite{Jain89,Jain07,Jain20}. The half-filled Landau level (LL) state at $\nu=1/2$ is known to be a Fermi sea of composite fermions with two quantized vortices bound to them. A Fermi sea of composite fermions carrying four vortices has also been confirmed unambiguously at $\nu=1/4$ through commensurability oscillations~\cite{Hossain19}. This implies, by particle-hole (PH) symmetry, a CF Fermi sea (CFFS) also at $\nu=1-1/4=3/4$. Further support to a CFFS at these fractions comes from the observation of FQHE at several fractions belonging to the sequences $\nu=s/(4s\pm 1)$ and $\nu=1-s/(4s\pm 1)$~\cite{Pan00,Pan02,Chung21}, which are integer quantum Hall states of composite fermions carrying four vortices; these terminate into CFFSs at $\nu=1/4$ and $\nu=3/4$ in the limit $s\rightarrow \infty$. 

FQHE at an even denominator fraction was first observed at $\nu=5/2$~\cite{Willett87,Pan99}, which corresponds to half filling in the second LL. It has been proposed that FQHE here arises from a pairing of composite fermions~\cite{Moore91, Greiter91, Greiter92a, Read00}, which is modeled in terms of the Moore-Read Pfaffian (MR-Pf) wave function~\cite{Moore91} representing a chiral $p$-wave pairing of composite fermions. (Even denominator FQHE in the $\mathcal{N}=1$ LL of bilayer graphene is analogous to the $5/2$ state in GaAs quantum wells (QWs)~\cite{Papic11, Zibrov16, Huang22, Balram21b}.) Why is there a difference between the physics at half filling in the lowest and the second LLs? For this purpose one must consider the CF-CF interaction, which derives from the electron-electron interaction. Extensive comparisons with exact diagonalization studies as well as experiments have shown that the model of non-interacting composite fermions is qualitatively valid when the short-range part of the interelectron interaction is dominant, which is the case in the lowest LL (LLL)~\cite{Jain20}. The short range part of the electron-electron interaction is weaker in the second LL (as measured by the Haldane pseudopotentials~\cite{Haldane83}), rendering the interaction between composite fermions slightly attractive, and thereby causing a pairing instability of the CFFS~\cite{Scarola00b}. The excitations of this state are predicted, akin to the Abrikosov vortices in a two-dimensional chiral $p$-wave superconductor, to be realizations of particles obeying non-Abelian braid statistics~\cite{Moore91,Read00,Nayak96,Nayak08,Ma22}. 

What can weaken the short range part of the inter-electron interaction in the LLL? One possibility is finite QW width. There is indeed evidence for FQHE at $\nu=1/4$ in very wide QWs~\cite{Luhman08,Shabani09a,Shabani09b,Shabani13}. Reference~\cite{Faugno19} has proposed that the modification of the interaction due to QW width makes the CFFS unstable to an $f$-wave pairing. However, the $3/4$ FQHE has been observed in rather narrow QWs (width of only 20 nm~\cite{Wang22a}), which sit comfortably in the CFFS region of the phase diagram evaluated in Ref.~\cite{Faugno19}. 

With the QW width ruled out as a relevant factor, one is left with LL mixing (LLM) as the possible cause for FQHE at $\nu=3/4$. The FQHE at $\nu=3/4$ has been observed in hole-type samples~\cite{Wang22a}, which, because of the larger hole mass, and hence smaller cyclotron energy, have much stronger LLM than electron-type samples. Indeed, the LLM parameter is $\kappa\simeq 10$ and $14$ for the two samples of Ref.~\cite{Wang22a}, where $\kappa=(e^2/\epsilon \ell)/(\hbar\omega_c)$ is the ratio of the Coulomb energy to the cyclotron energy (here $\epsilon$ is the dielectric constant of the semiconductor, $\ell=\sqrt{\hbar c/eB}$ is the magnetic length at magnetic field $B$, and $ \hbar\omega_c=\hbar eB/m_{b}c$ is the cyclotron energy of particles with band mass $m_{b}$).

It is clear that LLM will screen the short range part of the interelectron interaction. Can it induce pairing of composite fermions? The answer to this question relies on the ability to calculate accurately small energy differences in the presence of significant LLM. Below we consider quantitatively the possibility of CF pairing driven by LLM at filling factors $\nu=1/2$ and $\nu=1/4$. (For technical reasons stated below, the state at $\nu=3/4$ is not amenable to our calculation.)  For $\kappa\rightarrow 0$, one can employ a perturbative approach ~\cite{MacDonald84,Melik-Alaverdian95,Murthy02,Bishara09,Wojs10, Sodemann13, Simon13,Peterson13,Peterson14}, wherein LLM enters through a renormalization of the two-body interaction, while also introducing three and higher body interactions. Given that large $\kappa$ values are of interest, we instead use a fixed-phase diffusion Monte Carlo method (FPDMC)~\cite{Ortiz93, Melik-Alaverdian97,Melik-Alaverdian01}, which provides a non-perturbative treatment of LLM. While this method has its own approximations (mentioned below), it provides strict variational upper bounds for the energies of various states, and has given a fairly reasonable account of experiments on spin transitions ~\cite{Zhang16} and the competition between the FQHE and the crystal phase ~\cite{Zhao18,Ma20,Rosales21}.  Our calculations suggest that the 1/2 and 1/4 CFFSs are unstable to pairing in the presence of substantial LLM. At both of these filling factors, we find that the most favored pairing channel for composite fermions  is $l=-3$, which belongs in the same phase as the anti-Pfaffian (APf) state.

We note here that LLM has been considered previously in the context of the 5/2 state. Here, the MR-Pf is energetically equivalent to, although topologically distinct from, its hole partner called the APf~\cite{Levin07, Lee07} in the absence of LLM. Much theoretical work has investigated how LLM will break the tie between the two~\cite{Bishara09, Wojs10, Rezayi11, Peterson13, Pakrouski15, Rezayi17, Sreejith17, Herviou22}. In contrast to the situation at $\nu=5/2$, where both the MR-Pf and the APf states are present even in the absence of LLM, the present work asks if LLM can induce pairing, and hence FQHE, where none was present in the absence of LLM.

We begin by enumerating all of the candidate states that we study in this Letter. For the incompressible states, we consider the MR-Pf state, its generalizations~\cite{Read00}, and several Jain parton states~\cite{Jain89b}, all of which are non-Abelian. For all the calculations we will be employing the spherical geometry~\cite{Haldane83} which has $N$ electrons on the surface of a sphere moving under the influence of a radial magnetic field produced by a magnetic monopole of strength $Q$ at the center, which emanates a total flux of $2Q\phi_0$, with $\phi_0=hc/e$ defining the flux quantum. The state with $n$ filled LLs is denoted by $\Phi_n$ (with $\Phi_{-n}=[\Phi_{n}]^{*}\equiv\Phi_{\bar{n}}$) in what follows, with $\Phi_1=\prod_{j<k}(u_jv_k-v_ju_k)$, where $u_i=\cos(\theta_{i}/2)e^{i\phi_{i}/2}$ and $v_i=\sin(\theta_{i}/2)e^{-i\phi_{i}/2}$ are the spinor coordinates of the $i^{\rm th}$ electron with $\theta_{i}$ and $\phi_{i}$ being its polar and azimuthal angles on the sphere. The states occur at $2Q={\nu}^{-1}N-\mathcal{S}$ where the shift $\mathcal{S}$ is one of the topological properties of a FQHE state~\cite{Wen92}. Here are the explicit states:

(i) The CFFS of composite fermions  carrying $2p$ vortices is given by $\Psi^{\rm CFFS}= {\cal P}_{\rm LLL}\Phi^{{\rm eFS}} \Phi_1^{2p}$, where $\Phi^{{\rm eFS}}$ is the electron-Fermi sea wave function in the spherical geometry at zero flux, and ${\cal P}_{\rm LLL}$ refers to projection into the LLL, for which we use the Jain Kamilla (JK) method described in Refs.~\cite{Jain97,Jain97b}. We also consider the unprojected CFFS, labeled as unp-CFFS. 
We shall consider only filled shell states~\cite{Rezayi94, Balram15b} that occur for particle numbers $N=4, 9, 16, 25$.

(ii)  For wave functions for the CF pairs in relative angular momentum $l$ we follow Ref.~\cite{Read00}. For positive $l$, we write
\begin{equation}
	\Psi^{{\rm Pf}_l}_{\nu=1/2p} = \text{Pf}\left[\frac{(u^{*}_i v^{*}_j-v^{*}_iu^{*}_j)^{(l-1)}}{(u_iv_j-v_iu_j)} \right]\Phi_1^{2p},
	\label{eq: pairing_Read_Green_lg0}
\end{equation}
whereas for negative $l$, we have 
\begin{equation}
	\Psi^{{\rm Pf}_{-|l|}}_{\nu=1/2p}= \text{Pf}\left[\frac{(u_iv_j-v_iu_j)^{(|l|-1)}}{(u^{*}_iv^{*}_j-v^{*}_iu^{*}_j)} \right]\Phi_1^{2p}.
	\label{eq: pairing_Read_Green_ll0}
\end{equation}
Here ${\rm Pf}[M_{i,j}]\sim \mathcal{A}(M_{1,2}M_{3,4}\cdots M_{N-1,N})$, where $N$ is even, $M_{ij}$ is an antisymmetric matrix and $\mathcal{A}$ represents antisymmetrization.  $\Psi^{\rm Pf_1}$ represents the MR-Pf wave function which occurs at shift $\mathcal{S}=2p+1$. $\Psi^{\rm Pf_{-1}}$ represents the PH-symmetric (PHS)-Pf wave function~\cite{Jolicoeur07,Zucker16,Balram18,Mishmash18}, and occurs at the same shift $\mathcal{S}=2p-1$ as the PHS-Pf state proposed by Son~\cite{Son15}. 
$\Psi^{\rm Pf_{-3}}$ occurs at the same shift $\mathcal{S}=2p-3$ as the APf. Another paired state with shift $\mathcal{S}=2p-3$, lying in the APf phase, is given by\cite{Yutushui20}:
\begin{equation}
\Psi^{{\rm Pf'}_{-3}}_{\nu=1/2p}= \text{Pf}\left[\frac{(u_iv_j-v_iu_j)}{(u^{*}_iv^{*}_j-v^{*}_iu^{*}_j)^2} \right]\Phi_1^{2p}.
\label{eq: pairing_Yutushui_Mross}
\end{equation}
We will use the unprojected wave functions to fix the phase, except for $\Psi^{{\rm Pf}_1}$ which already resides in the LLL. 

(iii) The unprojected Jain $221$ parton wave function~\cite{Jain89b} at $\nu=1/2$ is given by $\Psi_{1/2}^{{\rm unp}-221} = \Phi^2_2\Phi_1$. This is a non-Abelian state~\cite{Wen91} representing an $f$-wave pairing of composite fermions ~\cite{Balram18, Faugno19}. It is the exact ground state for a short-range Hamiltonian~\cite{Wu17, Bandyopadhyay18} and is possibly relevant for 1/2 FQHE in $\mathcal{N}=3$ LL of monolayer graphene~\cite{Kim19,Sharma22}. We do not consider the LLL-projected 221 state as that requires the construction of this state in the Fock space, which can be accomplished only for very small systems. For $\nu=1/4$, the closely related 22111 parton state can be conveniently projected into the LLL as $\Psi_{1/4}^{22111} \equiv [{\cal P}_{\rm LLL}\Phi_2\Phi_1^2]^2/\Phi_1= [\Psi_{2/5}]^2/\Phi_1$, which can be evaluated for fairly large systems by the standard projection methods~\cite{Jain97,Jain97b}. 

(iv) The unprojected $\bar{2}\bar{2}111$ parton state~\cite{Balram18} at $\nu=1/2$ is given by $\Psi_{1/2}^{{\rm unp}-\bar{2}\bar{2}111} = \Phi^2_{\bar{2}}\Phi_1^3$.  This state has the same shift $\mathcal{S}$ as the APf, and its low-lying entanglement spectrum is identical to that of the APf for small systems~\cite{Balram18}. These features suggest that the $\Psi_{1/2}^{{\rm unp}-\bar{2}\bar{2}111}$ lies in the same phase as the APf. The state can be projected into the LLL as $\Psi_{1/2}^{\bar{2}\bar{2}111}\equiv [{\cal P}_{\rm LLL}\Phi_{\bar{2}}\Phi_1^2]^2/\Phi_1$=$[\Psi_{2/3}]^2/\Phi_1$, which can be explicitly performed for fairly large systems by the JK projection with the reverse-vortex attachment~\cite{Davenport12, Balram15a}. The unprojected and projected $\bar{2}\bar{2}11111$ parton states for $\nu=1/4$, which lie in the same universality class as the APf, can be constructed similarly: $\Psi_{1/4}^{{\rm unp}-\bar{2}\bar{2}11111} = \Phi^2_{\bar{2}}\Phi_1^5$ and $\Psi_{1/4}^{\bar{2}\bar{2}11111}\equiv [{\cal P}_{\rm LLL}\Phi_{\bar{2}}\Phi_1^2]^2 \Phi_1$=$[\Psi_{2/3}]^2 \Phi_1$.

The APf wave function at $\nu=1/2$ is obtained from the MR-Pf wave function by performing PH transformation, and the APf wave function at $\nu=1/4$ can be accessed by multiplying it by $\Phi_1^2$. We do not consider the APf state because no convenient wave function is known for it (it must be constructed by an explicit PH transformation in the Fock space representation) and also because previous work has shown that at $\nu=1/2$ the energies of the MR-Pf and the APf wave functions remain very close even with LLM~\cite{Sreejith17}. 

The above ``trial" wave functions are used to fix the ``phase" in the FPDMC calculation that incorporates LLM.  The basic outline is as follows (see Ref.~\cite{Ortiz93} and the supplemental material (Supplemental Material)~\cite{SM_LLM_Zhao22} for more details). Given a trial wave function $\Psi(\mathcal R)$ where $\{\mathcal R\}$ collectively denotes the positions of all particles, we first write $\Psi(\mathcal R)=\Phi(\mathcal R)e^{i\varphi(\mathcal R)}$ where $\Phi(\mathcal R)=|\Psi(\mathcal R)|$ is non-negative, and $\varphi(\mathcal R)$ is the phase of the wave function. The variational energy of the system of interacting electrons in a magnetic field described by the vector potential $\vec{A}$ is given by $\langle \Psi(\mathcal R)|H|\Psi(\mathcal R) \rangle=\langle \Phi(\mathcal R)|H_R|\Phi(\mathcal R) \rangle$ with $H_R= \sum_{j=1}^N \left[ \vec{p}_j^2 +[\hbar\vec{\nabla}_j\varphi(\mathcal R)+(e/c)\vec{A}(\vec{r}_j)]^2\right]/2m_{b}+V_{\rm Coulomb}(\mathcal R)$. We now assume that the phase $\varphi(\mathcal R)$ remains fixed. The lowest energy in this phase sector is obtained by varying $\Phi(\mathcal R)$. The energy minimization is accomplished by applying the standard DMC method~\cite{Reynolds82,Foulkes01} to the imaginary time Schr\"odinger equation $-\hbar{\partial\over\partial \tau}\Phi(\mathcal R,\tau)=\left[ H_R(\mathcal R)-E_T)\right]\Phi(\mathcal R,\tau)$, where the fixed phase appears effectively through a vector potential.  Details of the FPDMC method, as well as its application to the spherical geometry, are given in Refs.~\cite{Ortiz93, Melik-Alaverdian97,Melik-Alaverdian01,Zhang16,Zhao18}. 
The principal shortcoming of this method is that the accuracy of the energy depends on the choice of the phase. Previous studies~\cite{Guclu05, Zhang16, Zhao18} have indicated that the phase of an accurate LLL wave function remains a reasonably good approximation even in the presence of LLM. Here we also use ``unprojected" wave functions (which are not confined to the LLL) to fix the phase. Even though the wave function is modified in the FPDMC process, we will continue to label it by the initial trial wave function.

We will assume that the state is fully spin polarized, as expected at high magnetic fields. We will not include corrections due to finite QW thickness and consider a purely two-dimensional system. Our results are thus applicable to narrow QWs. All energies below are quoted in units of $e^2/\epsilon \ell$.

\begin{figure}[t]
	\includegraphics[width = 3.0in]{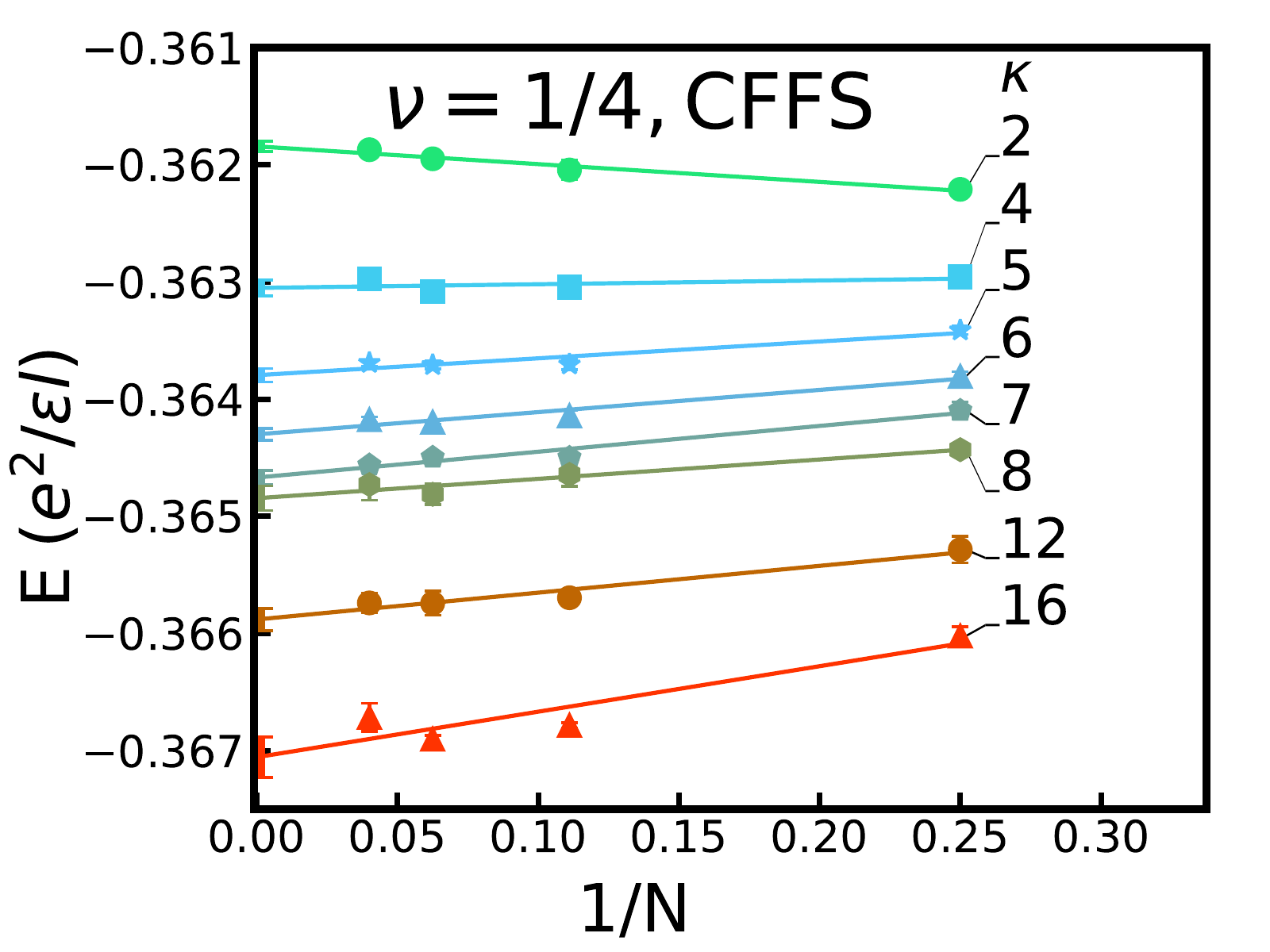}
	\includegraphics[width = 3.0in]{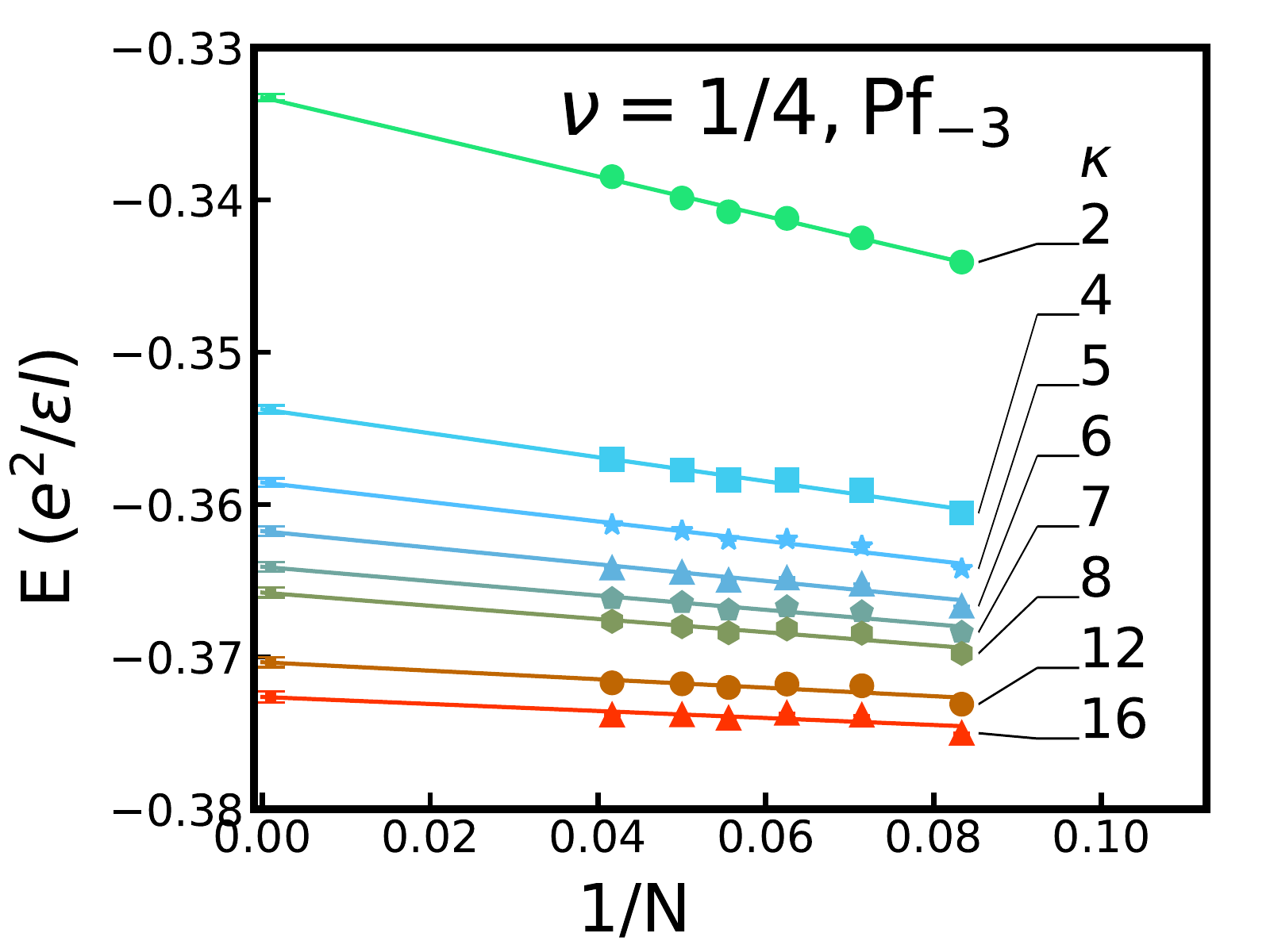}
	\caption{\label{fig: extrapolation} The energies of the CFFS (top) and the $\mathrm{Pf_{-3}}$ (bottom) states for $\nu=1/4$ as a function of $1/N$ for several different values of the LL mixing parameter $\kappa$ (shown on plots). The thermodynamic values of energies, whose uncertainties are labeled near the vertical axes, are obtained by linear regression. Extrapolations for all candidate states are shown in the Supplemental Material~\cite{SM_LLM_Zhao22}.
	}
\end{figure}

\begin{figure}[t]
	\includegraphics[width = 3.0in]{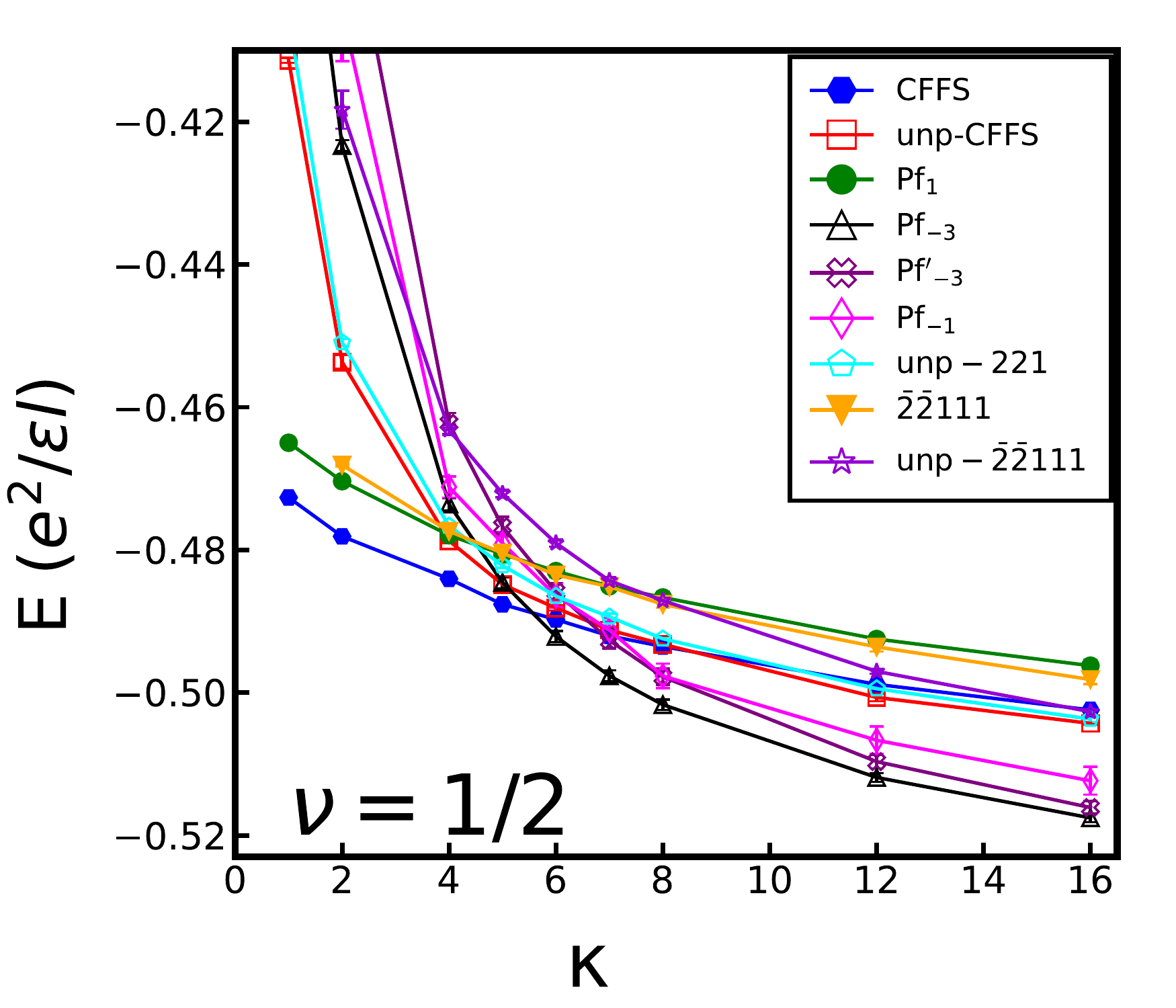}
	\includegraphics[width = 3.0in]{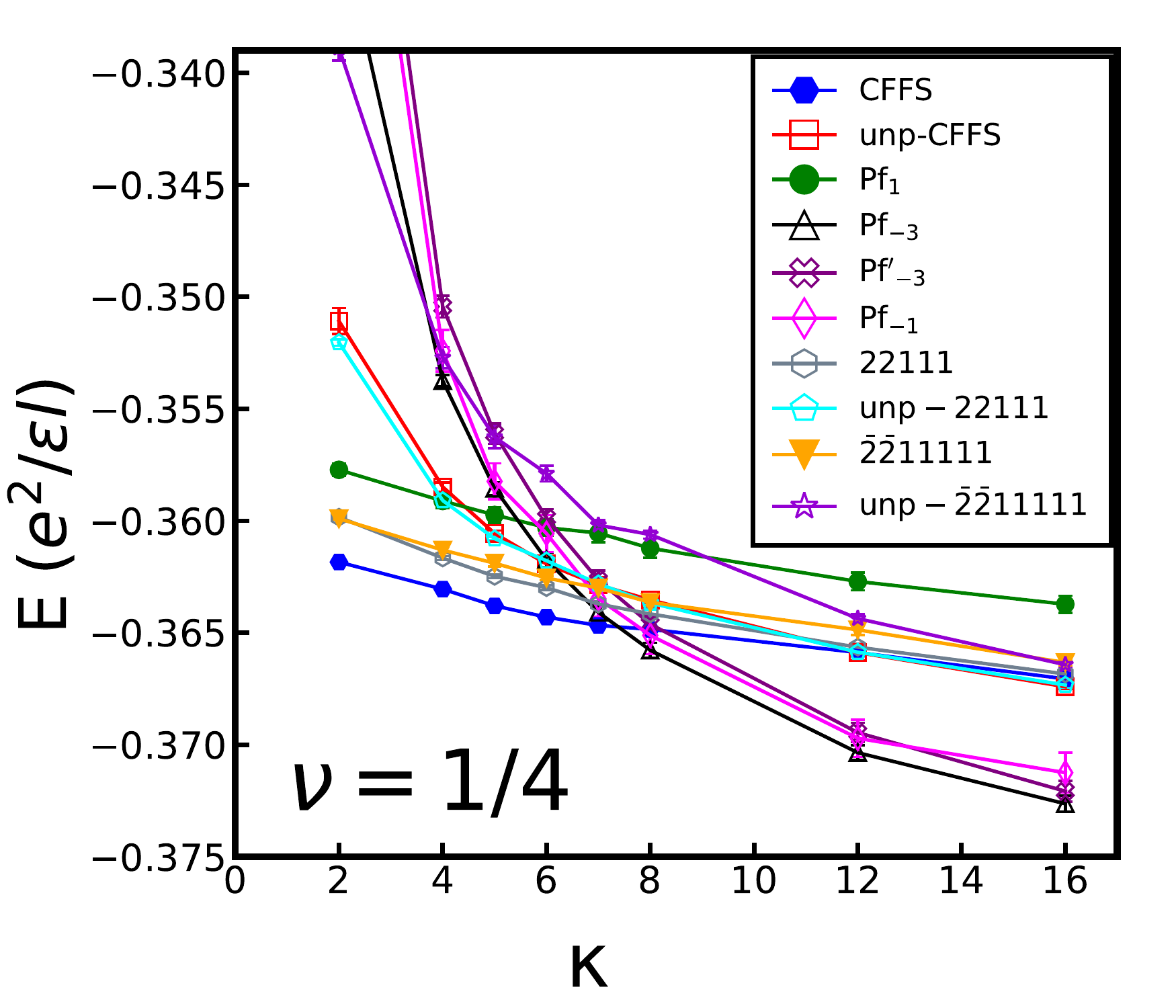}
	\caption{\label{fig: energy}This figure shows the thermodynamic energies as a function of the LL mixing parameter $\kappa$ when the phase sector is fixed using various trial states shown on the figures. For small $\kappa$ the lowest energy is obtained in the CFFS phase sector, but for large $\kappa$ the state derived from the $l=-3$ paired state wins at both $\nu=1/2$ and $\nu=1/4$. 
	}
\end{figure}

Figure~\ref{fig: extrapolation} presents the energies of the CFFS and the $\mathrm{Pf_{-3}}$ states for $\nu=1/4$ as a function of $1/N$ for several values of $\kappa$ (for  extrapolations of other states see the Supplemental Material~\cite{SM_LLM_Zhao22}). The energies include the contribution $-N^2/(2\sqrt{Q})~e^2/(\varepsilon \ell)$ from electron-background and background-background interaction, assuming a uniform neutralizing background charge. Because of the shift, the electron density has an $N$ dependence, which causes a correction to the energy; we multiply the energy of the system by $\sqrt{2Q\nu/N}$ to compensate for the finite-size effect~\cite{Morf87}.  This leads to better fits by linear regression in $1/N$ and reduces the error in the thermodynamic limit. The uncertainty in the thermodynamic limit originates primarily from the deviation of the finite-size energy from the fitted line.  The thermodynamic energies  of various states as a function of $\kappa$ are plotted in Figure~\ref{fig: energy}. In the absence of LLM ($\kappa=0$), the CFFS has significantly lower energy than the paired states, consistent with experiments that have confirmed the CFFS here~\cite{Halperin93, Halperin20, Shayegan20} and earlier numerical studies~\cite{Balram15c}. Our most important finding is that at both $\nu=1/2$ and $\nu=1/4$ the CFFS is unstable to pairing as $\kappa$ is increased. Further, at both of these filling factors, there is a level crossing into the $l=-3$ paired state at $\kappa \approx 6-7$. This state has a very high energy at small $\kappa$ (as is the case for all states that are not LLL projected), but its energy comes down rapidly with LLM. As shown in the Supplemental Material~\cite{SM_LLM_Zhao22}, even at large LLM, the pair-correlation function of the $l=-3$ paired states show oscillations that decay with distance and converge to the density, as anticipated for gapped liquid states~\cite{Kamilla97, Balram15b}.

As mentioned earlier, LLM weakens the short-distance repulsion between the electrons and may thus induce a weak residual attractive interaction between CFs leading to their pairing. We do not have a simple qualitative argument for why pairing in the $l=-3$ channel is preferred over other pairing channels. Only detailed calculations, like the ones presented here, can help identify the optimal pairing channel, as is also the case for the extensively studied CF pairing at $\nu=5/2$ (see Supplemental Material for further discussion~\cite{SM_LLM_Zhao22}); of course, the decisive verification will come only from experiments.

The results are sensitive to the trial wave function used to fix the phase even within the same topological sector. For example, the energies starting from the projected and unprojected 22111 or CFFS states are significantly different for small $\kappa$, although they tend to be similar for large $\kappa$. That implies that the precise value of $\kappa$ where the phase transition takes place from the CFFS to the paired state is only approximate. Finite width corrections are also likely to affect the transition. These points notwithstanding, our calculations make what we believe to be a plausible case that a transition will take place as a function of $\kappa$ into a paired state. We note here that FQHE at $\nu=1/2$ has been observed in wide QWs~\cite{Suen92, Suen92b, Shabani13, Liu14b, Liu14d, Mueed15}; some calculations have suggested a two-component Abelian Halperin-331 state~\cite{Halperin84, He93, Peterson10} while others the MR-Pf or the APf~\cite{Zhu16, Zhao21}. In contrast, for our current problem where we are considering the role of LLM at zero width, the MR-Pf (Pf$_1$) is not competitive for any $\kappa$.

The topological properties of $\Psi^{{\rm Pf}_{-3}}$, which is in the same phase as the APf, have been enumerated in earlier articles~\cite{Levin07, Lee07}. All candidate states support quasiparticles with fractional charge $e/4p$. The APf state supports an upstream neutral mode, which is experimentally measurable~\cite{Dutta22}; this can distinguish it from the MR-Pf and 22$1^{2p+1}$ states (with the caveat that edge reconstruction can produce upstream neutral modes in these states as well). A decisive measurement would be the thermal Hall conductance~\cite{Banerjee18}, which is given by $c[\pi^2k_B^2/(3h)]T$, where the chiral central charge is $c=1+l/2$ for the state with CF pairing in the relative angular momentum $l$ channel. 

Unfortunately, the above calculation cannot be performed directly at $\nu=3/4$, because the hole conjugates of the unprojected wave functions are not defined, and even for the LLL projected states the hole conjugates can be constructed only for very small systems, as this requires working with their explicit Fock space representations. Nonetheless, our results support the idea that LLM is responsible for a paired FQHE here. Ref.~\cite{Sreejith17} found that even though the energies of the MR-Pf and APf wave functions vary substantially with $\kappa$, they remain surprisingly close, and the same is true of the gaps of the 1/3 and 2/3 FQHE states. It is therefore a plausible first guess that the $3/4$ FQHE state stabilized in Ref.~\cite{Wang22a} may be in the same universality class as the hole partner of the $l=-3$ paired state. 

We have not considered the possibility of the crystal state in our calculations. Previous theoretical (see \cite{Zhao18} and references therein), as well as experimental studies (see \cite{Rosales21b} and references therein), have indicated that sufficient LLM can also stabilize the crystal phase. At what $\kappa$ the crystal phase appears at $\nu=1/4$ and $\nu=1/2$ is left for a future study. 

Before ending, we note that values of $\kappa>7$ at $\nu=1/2$ have been achieved in hole-type GaAs QWs as well as AlAs QWs~\cite{Sajoto90,Chung17,Ma20,Rosales21b,Ma21}. No evidence has yet been seen for FQHE at $\nu=1/2$ or $\nu=1/4$ in narrow QWs. It may be that our calculation underestimates the critical $\kappa$ for pairing instability. It is also possible that better quality samples would be needed for the observation of these states; after all, FQHE at $\nu=3/4$ has also revealed itself only in the highest quality samples that have become available recently~\cite{Chung21}. 

In summary, we have found theoretically that LLM can cause a pairing of composite fermions to produce non-Abelian FQHE states. Specifically, we predict that the CFFSs at $\nu=1/2$ and $\nu=1/4$ will transition, with increasing LL mixing, into $l=-3$ paired states of composite fermions carrying two and four vortices, respectively. We further speculate that the observed FQHE at $\nu=3/4$ is the hole partner of the latter. We hope that our work will motivate further study of the even denominator states in the LLL in the presence of high LLM.

\begin{acknowledgments}
	We acknowledge many fruitful discussions with Mansour Shayegan, which motivated this work. 
	T.Z.Z. was supported by  National Key Research and Development Program of China (Grant No. 2017YFA0302900) and National Natural Science Foundation of China (No.11888101). A.C.B. acknowledges the Science and Engineering Research Board (SERB) of the Department of Science and Technology (DST) for funding support via the Start-up Grant SRG/2020/000154. J.K.J. was supported by the U. S. Department of Energy, Office of Basic Energy Sciences, under Grant no. DE-SC-0005042. 
\end{acknowledgments}


\begin{thebibliography}{115}%
	\makeatletter
	\providecommand \@ifxundefined [1]{%
		\@ifx{#1\undefined}
	}%
	\providecommand \@ifnum [1]{%
		\ifnum #1\expandafter \@firstoftwo
		\else \expandafter \@secondoftwo
		\fi
	}%
	\providecommand \@ifx [1]{%
		\ifx #1\expandafter \@firstoftwo
		\else \expandafter \@secondoftwo
		\fi
	}%
	\providecommand \natexlab [1]{#1}%
	\providecommand \enquote  [1]{``#1''}%
	\providecommand \bibnamefont  [1]{#1}%
	\providecommand \bibfnamefont [1]{#1}%
	\providecommand \citenamefont [1]{#1}%
	\providecommand \href@noop [0]{\@secondoftwo}%
	\providecommand \href [0]{\begingroup \@sanitize@url \@href}%
	\providecommand \@href[1]{\@@startlink{#1}\@@href}%
	\providecommand \@@href[1]{\endgroup#1\@@endlink}%
	\providecommand \@sanitize@url [0]{\catcode `\\12\catcode `\$12\catcode
		`\&12\catcode `\#12\catcode `\^12\catcode `\_12\catcode `\%12\relax}%
	\providecommand \@@startlink[1]{}%
	\providecommand \@@endlink[0]{}%
	\providecommand \url  [0]{\begingroup\@sanitize@url \@url }%
	\providecommand \@url [1]{\endgroup\@href {#1}{\urlprefix }}%
	\providecommand \urlprefix  [0]{URL }%
	\providecommand \Eprint [0]{\href }%
	\providecommand \doibase [0]{https://doi.org/}%
	\providecommand \selectlanguage [0]{\@gobble}%
	\providecommand \bibinfo  [0]{\@secondoftwo}%
	\providecommand \bibfield  [0]{\@secondoftwo}%
	\providecommand \translation [1]{[#1]}%
	\providecommand \BibitemOpen [0]{}%
	\providecommand \bibitemStop [0]{}%
	\providecommand \bibitemNoStop [0]{.\EOS\space}%
	\providecommand \EOS [0]{\spacefactor3000\relax}%
	\providecommand \BibitemShut  [1]{\csname bibitem#1\endcsname}%
	\let\auto@bib@innerbib\@empty
	\bibitem [{\citenamefont {Wang}\ \emph {et~al.}(2022)\citenamefont {Wang},
		\citenamefont {Gupta}, \citenamefont {Singh}, \citenamefont {Chung},
		\citenamefont {Pfeiffer}, \citenamefont {West}, \citenamefont {Baldwin},
		\citenamefont {Winkler},\ and\ \citenamefont {Shayegan}}]{Wang22a}%
	\BibitemOpen
	\bibfield  {author} {\bibinfo {author} {\bibfnamefont {C.}~\bibnamefont
			{Wang}}, \bibinfo {author} {\bibfnamefont {A.}~\bibnamefont {Gupta}},
		\bibinfo {author} {\bibfnamefont {S.~K.}\ \bibnamefont {Singh}}, \bibinfo
		{author} {\bibfnamefont {Y.~J.}\ \bibnamefont {Chung}}, \bibinfo {author}
		{\bibfnamefont {L.~N.}\ \bibnamefont {Pfeiffer}}, \bibinfo {author}
		{\bibfnamefont {K.~W.}\ \bibnamefont {West}}, \bibinfo {author}
		{\bibfnamefont {K.~W.}\ \bibnamefont {Baldwin}}, \bibinfo {author}
		{\bibfnamefont {R.}~\bibnamefont {Winkler}},\ and\ \bibinfo {author}
		{\bibfnamefont {M.}~\bibnamefont {Shayegan}},\ }\bibfield  {title} {\bibinfo
		{title} {Even-denominator fractional quantum {Hall} state at filling factor
			$\ensuremath{\nu}=3/4$},\ }\href
	{https://doi.org/10.1103/PhysRevLett.129.156801} {\bibfield  {journal}
		{\bibinfo  {journal} {Phys. Rev. Lett.}\ }\textbf {\bibinfo {volume} {129}},\
		\bibinfo {pages} {156801} (\bibinfo {year} {2022})}\BibitemShut {NoStop}%
	\bibitem [{\citenamefont {Berkowitz}(2022)}]{Berkowitz22}%
	\BibitemOpen
	\bibfield  {author} {\bibinfo {author} {\bibfnamefont {R.}~\bibnamefont
			{Berkowitz}},\ }\bibfield  {title} {\bibinfo {title} {An exotic fractional
			quantum {Hall} state},\ }\href {https://doi.org/10.1103/Physics.15.s134}
	{\bibfield  {journal} {\bibinfo  {journal} {Physics}\ }\textbf {\bibinfo
			{volume} {15}},\ \bibinfo {pages} {s134} (\bibinfo {year}
		{2022})}\BibitemShut {NoStop}%
	\bibitem [{\citenamefont {Halperin}\ \emph {et~al.}(1993)\citenamefont
		{Halperin}, \citenamefont {Lee},\ and\ \citenamefont {Read}}]{Halperin93}%
	\BibitemOpen
	\bibfield  {author} {\bibinfo {author} {\bibfnamefont {B.~I.}\ \bibnamefont
			{Halperin}}, \bibinfo {author} {\bibfnamefont {P.~A.}\ \bibnamefont {Lee}},\
		and\ \bibinfo {author} {\bibfnamefont {N.}~\bibnamefont {Read}},\ }\bibfield
	{title} {\bibinfo {title} {Theory of the half-filled {Landau} level},\ }\href
	{https://doi.org/10.1103/PhysRevB.47.7312} {\bibfield  {journal} {\bibinfo
			{journal} {Phys. Rev. B}\ }\textbf {\bibinfo {volume} {47}},\ \bibinfo
		{pages} {7312} (\bibinfo {year} {1993})}\BibitemShut {NoStop}%
	\bibitem [{\citenamefont {Halperin}(2020)}]{Halperin20b}%
	\BibitemOpen
	\bibfield  {author} {\bibinfo {author} {\bibfnamefont {B.~I.}\ \bibnamefont
			{Halperin}},\ }\bibfield  {title} {\bibinfo {title} {The {Half}-{Full}
			{Landau} {Level}},\ }in\ \href {https://doi.org/10.1142/9789811217494_0003}
	{\emph {\bibinfo {booktitle} {Fractional Quantum Hall Effects: New
				Developments}}},\ \bibinfo {editor} {edited by\ \bibinfo {editor}
		{\bibfnamefont {B.~I.}\ \bibnamefont {Halperin}}\ and\ \bibinfo {editor}
		{\bibfnamefont {J.~K.}\ \bibnamefont {Jain}}}\ (\bibinfo  {publisher} {World
		Scientific Pub Co Inc, Singapore},\ \bibinfo {year} {2020})\ Chap.~\bibinfo
	{chapter} {2}, pp.\ \bibinfo {pages} {79--132}\BibitemShut {NoStop}%
	\bibitem [{\citenamefont {Shayegan}(2020)}]{Shayegan20}%
	\BibitemOpen
	\bibfield  {author} {\bibinfo {author} {\bibfnamefont {M.}~\bibnamefont
			{Shayegan}},\ }\bibfield  {title} {\bibinfo {title} {Probing {Composite}
			{Fermions} {Near} {Half}-{Filled} {Landau} {Levels}},\ }in\ \href
	{https://doi.org/10.1142/9789811217494_0003} {\emph {\bibinfo {booktitle}
			{Fractional Quantum Hall Effects: New Developments}}},\ \bibinfo {editor}
	{edited by\ \bibinfo {editor} {\bibfnamefont {B.~I.}\ \bibnamefont
			{Halperin}}\ and\ \bibinfo {editor} {\bibfnamefont {J.~K.}\ \bibnamefont
			{Jain}}}\ (\bibinfo  {publisher} {World Scientific Pub Co Inc, Singapore},\
	\bibinfo {year} {2020})\ Chap.~\bibinfo {chapter} {3}, pp.\ \bibinfo {pages}
	{133--181}\BibitemShut {NoStop}%
	\bibitem [{\citenamefont {Jain}(1989{\natexlab{a}})}]{Jain89}%
	\BibitemOpen
	\bibfield  {author} {\bibinfo {author} {\bibfnamefont {J.~K.}\ \bibnamefont
			{Jain}},\ }\bibfield  {title} {\bibinfo {title} {Composite-fermion approach
			for the fractional quantum {Hall} effect},\ }\href
	{https://doi.org/10.1103/PhysRevLett.63.199} {\bibfield  {journal} {\bibinfo
			{journal} {Phys. Rev. Lett.}\ }\textbf {\bibinfo {volume} {63}},\ \bibinfo
		{pages} {199} (\bibinfo {year} {1989}{\natexlab{a}})}\BibitemShut {NoStop}%
	\bibitem [{\citenamefont {Jain}(2007)}]{Jain07}%
	\BibitemOpen
	\bibfield  {author} {\bibinfo {author} {\bibfnamefont {J.~K.}\ \bibnamefont
			{Jain}},\ }\href@noop {} {\emph {\bibinfo {title} {Composite Fermions}}}\
	(\bibinfo  {publisher} {Cambridge University Press, New York, US},\ \bibinfo
	{year} {2007})\BibitemShut {NoStop}%
	\bibitem [{\citenamefont {Jain}(2020)}]{Jain20}%
	\BibitemOpen
	\bibfield  {author} {\bibinfo {author} {\bibfnamefont {J.~K.}\ \bibnamefont
			{Jain}},\ }\bibfield  {title} {\bibinfo {title} {Thirty {Years} of
			{Composite} {Fermions} and {Beyond}},\ }in\ \href
	{https://doi.org/10.1142/9789811217494_0001} {\emph {\bibinfo {booktitle}
			{Fractional Quantum Hall Effects: New Developments}}},\ \bibinfo {editor}
	{edited by\ \bibinfo {editor} {\bibfnamefont {B.~I.}\ \bibnamefont
			{Halperin}}\ and\ \bibinfo {editor} {\bibfnamefont {J.~K.}\ \bibnamefont
			{Jain}}}\ (\bibinfo  {publisher} {World Scientific Pub Co Inc, Singapore},\
	\bibinfo {year} {2020})\ Chap.~\bibinfo {chapter} {3}, pp.\ \bibinfo {pages}
	{1--78}\BibitemShut {NoStop}%
	\bibitem [{\citenamefont {Hossain}\ \emph {et~al.}(2019)\citenamefont
		{Hossain}, \citenamefont {Ma}, \citenamefont {Mueed}, \citenamefont
		{Kamburov}, \citenamefont {Pfeiffer}, \citenamefont {West}, \citenamefont
		{Baldwin}, \citenamefont {Winkler},\ and\ \citenamefont
		{Shayegan}}]{Hossain19}%
	\BibitemOpen
	\bibfield  {author} {\bibinfo {author} {\bibfnamefont {M.~S.}\ \bibnamefont
			{Hossain}}, \bibinfo {author} {\bibfnamefont {M.~K.}\ \bibnamefont {Ma}},
		\bibinfo {author} {\bibfnamefont {M.~A.}\ \bibnamefont {Mueed}}, \bibinfo
		{author} {\bibfnamefont {D.}~\bibnamefont {Kamburov}}, \bibinfo {author}
		{\bibfnamefont {L.~N.}\ \bibnamefont {Pfeiffer}}, \bibinfo {author}
		{\bibfnamefont {K.~W.}\ \bibnamefont {West}}, \bibinfo {author}
		{\bibfnamefont {K.~W.}\ \bibnamefont {Baldwin}}, \bibinfo {author}
		{\bibfnamefont {R.}~\bibnamefont {Winkler}},\ and\ \bibinfo {author}
		{\bibfnamefont {M.}~\bibnamefont {Shayegan}},\ }\bibfield  {title} {\bibinfo
		{title} {Geometric resonance of four-flux composite fermions},\ }\href
	{https://doi.org/10.1103/PhysRevB.100.041112} {\bibfield  {journal} {\bibinfo
			{journal} {Phys. Rev. B}\ }\textbf {\bibinfo {volume} {100}},\ \bibinfo
		{pages} {041112} (\bibinfo {year} {2019})}\BibitemShut {NoStop}%
	\bibitem [{\citenamefont {Pan}\ \emph {et~al.}(2000)\citenamefont {Pan},
		\citenamefont {Stormer}, \citenamefont {Tsui}, \citenamefont {Pfeiffer},
		\citenamefont {Baldwin},\ and\ \citenamefont {West}}]{Pan00}%
	\BibitemOpen
	\bibfield  {author} {\bibinfo {author} {\bibfnamefont {W.}~\bibnamefont
			{Pan}}, \bibinfo {author} {\bibfnamefont {H.~L.}\ \bibnamefont {Stormer}},
		\bibinfo {author} {\bibfnamefont {D.~C.}\ \bibnamefont {Tsui}}, \bibinfo
		{author} {\bibfnamefont {L.~N.}\ \bibnamefont {Pfeiffer}}, \bibinfo {author}
		{\bibfnamefont {K.~W.}\ \bibnamefont {Baldwin}},\ and\ \bibinfo {author}
		{\bibfnamefont {K.~W.}\ \bibnamefont {West}},\ }\bibfield  {title} {\bibinfo
		{title} {Effective mass of the four-flux composite fermion at
			$\ensuremath{\nu}=1/4$},\ }\href {https://doi.org/10.1103/PhysRevB.61.R5101}
	{\bibfield  {journal} {\bibinfo  {journal} {Phys. Rev. B}\ }\textbf {\bibinfo
			{volume} {61}},\ \bibinfo {pages} {R5101} (\bibinfo {year}
		{2000})}\BibitemShut {NoStop}%
	\bibitem [{\citenamefont {Pan}\ \emph {et~al.}(2002)\citenamefont {Pan},
		\citenamefont {Stormer}, \citenamefont {Tsui}, \citenamefont {Pfeiffer},
		\citenamefont {Baldwin},\ and\ \citenamefont {West}}]{Pan02}%
	\BibitemOpen
	\bibfield  {author} {\bibinfo {author} {\bibfnamefont {W.}~\bibnamefont
			{Pan}}, \bibinfo {author} {\bibfnamefont {H.~L.}\ \bibnamefont {Stormer}},
		\bibinfo {author} {\bibfnamefont {D.~C.}\ \bibnamefont {Tsui}}, \bibinfo
		{author} {\bibfnamefont {L.~N.}\ \bibnamefont {Pfeiffer}}, \bibinfo {author}
		{\bibfnamefont {K.~W.}\ \bibnamefont {Baldwin}},\ and\ \bibinfo {author}
		{\bibfnamefont {K.~W.}\ \bibnamefont {West}},\ }\bibfield  {title} {\bibinfo
		{title} {Transition from an electron solid to the sequence of fractional
			quantum {Hall} states at very low {Landau} level filling factor},\ }\href
	{https://doi.org/10.1103/PhysRevLett.88.176802} {\bibfield  {journal}
		{\bibinfo  {journal} {Phys. Rev. Lett.}\ }\textbf {\bibinfo {volume} {88}},\
		\bibinfo {pages} {176802} (\bibinfo {year} {2002})}\BibitemShut {NoStop}%
	\bibitem [{\citenamefont {Chung}\ \emph {et~al.}(2021)\citenamefont {Chung},
		\citenamefont {Villegas~Rosales}, \citenamefont {Baldwin}, \citenamefont
		{Madathil}, \citenamefont {West}, \citenamefont {Shayegan},\ and\
		\citenamefont {Pfeiffer}}]{Chung21}%
	\BibitemOpen
	\bibfield  {author} {\bibinfo {author} {\bibfnamefont {Y.~J.}\ \bibnamefont
			{Chung}}, \bibinfo {author} {\bibfnamefont {K.~A.}\ \bibnamefont
			{Villegas~Rosales}}, \bibinfo {author} {\bibfnamefont {K.~W.}\ \bibnamefont
			{Baldwin}}, \bibinfo {author} {\bibfnamefont {P.~T.}\ \bibnamefont
			{Madathil}}, \bibinfo {author} {\bibfnamefont {K.~W.}\ \bibnamefont {West}},
		\bibinfo {author} {\bibfnamefont {M.}~\bibnamefont {Shayegan}},\ and\
		\bibinfo {author} {\bibfnamefont {L.~N.}\ \bibnamefont {Pfeiffer}},\
	}\bibfield  {title} {\bibinfo {title} {Ultra-high-quality two-dimensional
			electron systems},\ }\bibfield  {journal} {\bibinfo  {journal} {Nature
			Materials}\ }\href {https://doi.org/10.1038/s41563-021-00942-3}
	{10.1038/s41563-021-00942-3} (\bibinfo {year} {2021})\BibitemShut {NoStop}%
	\bibitem [{\citenamefont {Willett}\ \emph {et~al.}(1987)\citenamefont
		{Willett}, \citenamefont {Eisenstein}, \citenamefont {St\"ormer},
		\citenamefont {Tsui}, \citenamefont {Gossard},\ and\ \citenamefont
		{English}}]{Willett87}%
	\BibitemOpen
	\bibfield  {author} {\bibinfo {author} {\bibfnamefont {R.}~\bibnamefont
			{Willett}}, \bibinfo {author} {\bibfnamefont {J.~P.}\ \bibnamefont
			{Eisenstein}}, \bibinfo {author} {\bibfnamefont {H.~L.}\ \bibnamefont
			{St\"ormer}}, \bibinfo {author} {\bibfnamefont {D.~C.}\ \bibnamefont {Tsui}},
		\bibinfo {author} {\bibfnamefont {A.~C.}\ \bibnamefont {Gossard}},\ and\
		\bibinfo {author} {\bibfnamefont {J.~H.}\ \bibnamefont {English}},\
	}\bibfield  {title} {\bibinfo {title} {Observation of an even-denominator
			quantum number in the fractional quantum {Hall} effect},\ }\href
	{https://doi.org/10.1103/PhysRevLett.59.1776} {\bibfield  {journal} {\bibinfo
			{journal} {Phys. Rev. Lett.}\ }\textbf {\bibinfo {volume} {59}},\ \bibinfo
		{pages} {1776} (\bibinfo {year} {1987})}\BibitemShut {NoStop}%
	\bibitem [{\citenamefont {Pan}\ \emph {et~al.}(1999)\citenamefont {Pan},
		\citenamefont {Xia}, \citenamefont {Shvarts}, \citenamefont {Adams},
		\citenamefont {Stormer}, \citenamefont {Tsui}, \citenamefont {Pfeiffer},
		\citenamefont {Baldwin},\ and\ \citenamefont {West}}]{Pan99}%
	\BibitemOpen
	\bibfield  {author} {\bibinfo {author} {\bibfnamefont {W.}~\bibnamefont
			{Pan}}, \bibinfo {author} {\bibfnamefont {J.-S.}\ \bibnamefont {Xia}},
		\bibinfo {author} {\bibfnamefont {V.}~\bibnamefont {Shvarts}}, \bibinfo
		{author} {\bibfnamefont {D.~E.}\ \bibnamefont {Adams}}, \bibinfo {author}
		{\bibfnamefont {H.~L.}\ \bibnamefont {Stormer}}, \bibinfo {author}
		{\bibfnamefont {D.~C.}\ \bibnamefont {Tsui}}, \bibinfo {author}
		{\bibfnamefont {L.~N.}\ \bibnamefont {Pfeiffer}}, \bibinfo {author}
		{\bibfnamefont {K.~W.}\ \bibnamefont {Baldwin}},\ and\ \bibinfo {author}
		{\bibfnamefont {K.~W.}\ \bibnamefont {West}},\ }\bibfield  {title} {\bibinfo
		{title} {Exact quantization of the even-denominator fractional quantum {Hall}
			state at
			$\mathit{\ensuremath{\nu}}\phantom{\rule{0ex}{0ex}}=\phantom{\rule{0ex}{0ex}}5/2$
			{Landau} level filling factor},\ }\href
	{https://doi.org/10.1103/PhysRevLett.83.3530} {\bibfield  {journal} {\bibinfo
			{journal} {Phys. Rev. Lett.}\ }\textbf {\bibinfo {volume} {83}},\ \bibinfo
		{pages} {3530} (\bibinfo {year} {1999})}\BibitemShut {NoStop}%
	\bibitem [{\citenamefont {Moore}\ and\ \citenamefont {Read}(1991)}]{Moore91}%
	\BibitemOpen
	\bibfield  {author} {\bibinfo {author} {\bibfnamefont {G.}~\bibnamefont
			{Moore}}\ and\ \bibinfo {author} {\bibfnamefont {N.}~\bibnamefont {Read}},\
	}\bibfield  {title} {\bibinfo {title} {Nonabelions in the fractional quantum
			{Hall} effect},\ }\href {https://doi.org/10.1016/0550-3213(91)90407-O}
	{\bibfield  {journal} {\bibinfo  {journal} {Nucl. Phys. B}\ }\textbf
		{\bibinfo {volume} {360}},\ \bibinfo {pages} {362 } (\bibinfo {year}
		{1991})}\BibitemShut {NoStop}%
	\bibitem [{\citenamefont {Greiter}\ \emph {et~al.}(1991)\citenamefont
		{Greiter}, \citenamefont {Wen},\ and\ \citenamefont {Wilczek}}]{Greiter91}%
	\BibitemOpen
	\bibfield  {author} {\bibinfo {author} {\bibfnamefont {M.}~\bibnamefont
			{Greiter}}, \bibinfo {author} {\bibfnamefont {X.-G.}\ \bibnamefont {Wen}},\
		and\ \bibinfo {author} {\bibfnamefont {F.}~\bibnamefont {Wilczek}},\
	}\bibfield  {title} {\bibinfo {title} {Paired {Hall} state at half filling},\
	}\href {https://doi.org/10.1103/PhysRevLett.66.3205} {\bibfield  {journal}
		{\bibinfo  {journal} {Phys. Rev. Lett.}\ }\textbf {\bibinfo {volume} {66}},\
		\bibinfo {pages} {3205} (\bibinfo {year} {1991})}\BibitemShut {NoStop}%
	\bibitem [{\citenamefont {Greiter}\ \emph {et~al.}(1992)\citenamefont
		{Greiter}, \citenamefont {Wen},\ and\ \citenamefont {Wilczek}}]{Greiter92a}%
	\BibitemOpen
	\bibfield  {author} {\bibinfo {author} {\bibfnamefont {M.}~\bibnamefont
			{Greiter}}, \bibinfo {author} {\bibfnamefont {X.}~\bibnamefont {Wen}},\ and\
		\bibinfo {author} {\bibfnamefont {F.}~\bibnamefont {Wilczek}},\ }\bibfield
	{title} {\bibinfo {title} {Paired {Hall} states},\ }\href
	{https://doi.org/http://dx.doi.org/10.1016/0550-3213(92)90401-V} {\bibfield
		{journal} {\bibinfo  {journal} {Nucl. Phys. B}\ }\textbf {\bibinfo {volume}
			{374}},\ \bibinfo {pages} {567 } (\bibinfo {year} {1992})}\BibitemShut
	{NoStop}%
	\bibitem [{\citenamefont {Read}\ and\ \citenamefont {Green}(2000)}]{Read00}%
	\BibitemOpen
	\bibfield  {author} {\bibinfo {author} {\bibfnamefont {N.}~\bibnamefont
			{Read}}\ and\ \bibinfo {author} {\bibfnamefont {D.}~\bibnamefont {Green}},\
	}\bibfield  {title} {\bibinfo {title} {Paired states of fermions in two
			dimensions with breaking of parity and time-reversal symmetries and the
			fractional quantum {Hall} effect},\ }\href
	{https://doi.org/10.1103/PhysRevB.61.10267} {\bibfield  {journal} {\bibinfo
			{journal} {Phys. Rev. B}\ }\textbf {\bibinfo {volume} {61}},\ \bibinfo
		{pages} {10267} (\bibinfo {year} {2000})}\BibitemShut {NoStop}%
	\bibitem [{\citenamefont {Papi\ifmmode~\acute{c}\else \'{c}\fi{}}\ \emph
		{et~al.}(2011)\citenamefont {Papi\ifmmode~\acute{c}\else \'{c}\fi{}},
		\citenamefont {Abanin}, \citenamefont {Barlas},\ and\ \citenamefont
		{Bhatt}}]{Papic11}%
	\BibitemOpen
	\bibfield  {author} {\bibinfo {author} {\bibfnamefont {Z.}~\bibnamefont
			{Papi\ifmmode~\acute{c}\else \'{c}\fi{}}}, \bibinfo {author} {\bibfnamefont
			{D.~A.}\ \bibnamefont {Abanin}}, \bibinfo {author} {\bibfnamefont
			{Y.}~\bibnamefont {Barlas}},\ and\ \bibinfo {author} {\bibfnamefont {R.~N.}\
			\bibnamefont {Bhatt}},\ }\bibfield  {title} {\bibinfo {title} {Tunable
			interactions and phase transitions in {Dirac} materials in a magnetic
			field},\ }\href {https://doi.org/10.1103/PhysRevB.84.241306} {\bibfield
		{journal} {\bibinfo  {journal} {Phys. Rev. B}\ }\textbf {\bibinfo {volume}
			{84}},\ \bibinfo {pages} {241306} (\bibinfo {year} {2011})}\BibitemShut
	{NoStop}%
	\bibitem [{\citenamefont {{Zibrov}}\ \emph {et~al.}(2017)\citenamefont
		{{Zibrov}}, \citenamefont {{Kometter}}, \citenamefont {{Zhou}}, \citenamefont
		{{Spanton}}, \citenamefont {{Taniguchi}}, \citenamefont {{Watanabe}},
		\citenamefont {{Zaletel}},\ and\ \citenamefont {{Young}}}]{Zibrov16}%
	\BibitemOpen
	\bibfield  {author} {\bibinfo {author} {\bibfnamefont {A.~A.}\ \bibnamefont
			{{Zibrov}}}, \bibinfo {author} {\bibfnamefont {C.~R.}\ \bibnamefont
			{{Kometter}}}, \bibinfo {author} {\bibfnamefont {H.}~\bibnamefont {{Zhou}}},
		\bibinfo {author} {\bibfnamefont {E.~M.}\ \bibnamefont {{Spanton}}}, \bibinfo
		{author} {\bibfnamefont {T.}~\bibnamefont {{Taniguchi}}}, \bibinfo {author}
		{\bibfnamefont {K.}~\bibnamefont {{Watanabe}}}, \bibinfo {author}
		{\bibfnamefont {M.~P.}\ \bibnamefont {{Zaletel}}},\ and\ \bibinfo {author}
		{\bibfnamefont {A.~F.}\ \bibnamefont {{Young}}},\ }\bibfield  {title}
	{\bibinfo {title} {Tunable interacting composite fermion phases in a
			half-filled bilayer-graphene {Landau} level},\ }\href
	{https://doi.org/10.1038/nature23893} {\bibfield  {journal} {\bibinfo
			{journal} {Nature}\ }\textbf {\bibinfo {volume} {549}},\ \bibinfo {pages}
		{360} (\bibinfo {year} {2017})}\BibitemShut {NoStop}%
	\bibitem [{\citenamefont {Huang}\ \emph {et~al.}(2022)\citenamefont {Huang},
		\citenamefont {Fu}, \citenamefont {Hickey}, \citenamefont {Alem},
		\citenamefont {Lin}, \citenamefont {Watanabe}, \citenamefont {Taniguchi},\
		and\ \citenamefont {Zhu}}]{Huang22}%
	\BibitemOpen
	\bibfield  {author} {\bibinfo {author} {\bibfnamefont {K.}~\bibnamefont
			{Huang}}, \bibinfo {author} {\bibfnamefont {H.}~\bibnamefont {Fu}}, \bibinfo
		{author} {\bibfnamefont {D.~R.}\ \bibnamefont {Hickey}}, \bibinfo {author}
		{\bibfnamefont {N.}~\bibnamefont {Alem}}, \bibinfo {author} {\bibfnamefont
			{X.}~\bibnamefont {Lin}}, \bibinfo {author} {\bibfnamefont {K.}~\bibnamefont
			{Watanabe}}, \bibinfo {author} {\bibfnamefont {T.}~\bibnamefont
			{Taniguchi}},\ and\ \bibinfo {author} {\bibfnamefont {J.}~\bibnamefont
			{Zhu}},\ }\bibfield  {title} {\bibinfo {title} {Valley isospin controlled
			fractional quantum hall states in bilayer graphene},\ }\href
	{https://doi.org/10.1103/PhysRevX.12.031019} {\bibfield  {journal} {\bibinfo
			{journal} {Phys. Rev. X}\ }\textbf {\bibinfo {volume} {12}},\ \bibinfo
		{pages} {031019} (\bibinfo {year} {2022})}\BibitemShut {NoStop}%
	\bibitem [{\citenamefont {Balram}(2022)}]{Balram21b}%
	\BibitemOpen
	\bibfield  {author} {\bibinfo {author} {\bibfnamefont {A.~C.}\ \bibnamefont
			{Balram}},\ }\bibfield  {title} {\bibinfo {title} {Transitions from {Abelian}
			composite fermion to non-{Abelian} parton fractional quantum {Hall} states in
			the zeroth {Landau} level of bilayer graphene},\ }\href
	{https://doi.org/10.1103/PhysRevB.105.L121406} {\bibfield  {journal}
		{\bibinfo  {journal} {Phys. Rev. B}\ }\textbf {\bibinfo {volume} {105}},\
		\bibinfo {pages} {L121406} (\bibinfo {year} {2022})}\BibitemShut {NoStop}%
	\bibitem [{\citenamefont {Haldane}(1983)}]{Haldane83}%
	\BibitemOpen
	\bibfield  {author} {\bibinfo {author} {\bibfnamefont {F.~D.~M.}\
			\bibnamefont {Haldane}},\ }\bibfield  {title} {\bibinfo {title} {Fractional
			quantization of the {Hall} effect: A hierarchy of incompressible quantum
			fluid states},\ }\href {https://doi.org/10.1103/PhysRevLett.51.605}
	{\bibfield  {journal} {\bibinfo  {journal} {Phys. Rev. Lett.}\ }\textbf
		{\bibinfo {volume} {51}},\ \bibinfo {pages} {605} (\bibinfo {year}
		{1983})}\BibitemShut {NoStop}%
	\bibitem [{\citenamefont {Scarola}\ \emph {et~al.}(2000)\citenamefont
		{Scarola}, \citenamefont {Park},\ and\ \citenamefont {Jain}}]{Scarola00b}%
	\BibitemOpen
	\bibfield  {author} {\bibinfo {author} {\bibfnamefont {V.~W.}\ \bibnamefont
			{Scarola}}, \bibinfo {author} {\bibfnamefont {K.}~\bibnamefont {Park}},\ and\
		\bibinfo {author} {\bibfnamefont {J.~K.}\ \bibnamefont {Jain}},\ }\bibfield
	{title} {\bibinfo {title} {Cooper instability of composite fermions},\
	}\href@noop {} {\bibfield  {journal} {\bibinfo  {journal} {Nature}\ }\textbf
		{\bibinfo {volume} {406}},\ \bibinfo {pages} {863} (\bibinfo {year}
		{2000})}\BibitemShut {NoStop}%
	\bibitem [{\citenamefont {Nayak}\ and\ \citenamefont
		{Wilczek}(1996)}]{Nayak96}%
	\BibitemOpen
	\bibfield  {author} {\bibinfo {author} {\bibfnamefont {C.}~\bibnamefont
			{Nayak}}\ and\ \bibinfo {author} {\bibfnamefont {F.}~\bibnamefont
			{Wilczek}},\ }\bibfield  {title} {\bibinfo {title} {2n-quasihole states
			realize 2(n-1)-dimensional spinor braiding statistics in paired quantum
			{Hall} states},\ }\href {https://doi.org/10.1016/0550-3213(96)00430-0}
	{\bibfield  {journal} {\bibinfo  {journal} {Nucl. Phys. B}\ }\textbf
		{\bibinfo {volume} {479}},\ \bibinfo {pages} {529} (\bibinfo {year}
		{1996})}\BibitemShut {NoStop}%
	\bibitem [{\citenamefont {Nayak}\ \emph {et~al.}(2008)\citenamefont {Nayak},
		\citenamefont {Simon}, \citenamefont {Stern}, \citenamefont {Freedman},\ and\
		\citenamefont {Das~Sarma}}]{Nayak08}%
	\BibitemOpen
	\bibfield  {author} {\bibinfo {author} {\bibfnamefont {C.}~\bibnamefont
			{Nayak}}, \bibinfo {author} {\bibfnamefont {S.~H.}\ \bibnamefont {Simon}},
		\bibinfo {author} {\bibfnamefont {A.}~\bibnamefont {Stern}}, \bibinfo
		{author} {\bibfnamefont {M.}~\bibnamefont {Freedman}},\ and\ \bibinfo
		{author} {\bibfnamefont {S.}~\bibnamefont {Das~Sarma}},\ }\bibfield  {title}
	{\bibinfo {title} {Non-{Abelian} anyons and topological quantum
			computation},\ }\href {https://doi.org/10.1103/RevModPhys.80.1083} {\bibfield
		{journal} {\bibinfo  {journal} {Rev. Mod. Phys.}\ }\textbf {\bibinfo
			{volume} {80}},\ \bibinfo {pages} {1083} (\bibinfo {year}
		{2008})}\BibitemShut {NoStop}%
	\bibitem [{\citenamefont {Ma}\ \emph {et~al.}(2022)\citenamefont {Ma},
		\citenamefont {Peterson}, \citenamefont {Scarola},\ and\ \citenamefont
		{Yang}}]{Ma22}%
	\BibitemOpen
	\bibfield  {author} {\bibinfo {author} {\bibfnamefont {K.~K.~W.}\
			\bibnamefont {Ma}}, \bibinfo {author} {\bibfnamefont {M.~R.}\ \bibnamefont
			{Peterson}}, \bibinfo {author} {\bibfnamefont {V.~W.}\ \bibnamefont
			{Scarola}},\ and\ \bibinfo {author} {\bibfnamefont {K.}~\bibnamefont
			{Yang}},\ }\href {https://doi.org/10.48550/ARXIV.2208.07908} {\bibinfo
		{title} {Fractional quantum {Hall} effect at the filling factor $\nu=5/2$}}
	(\bibinfo {year} {2022})\BibitemShut {NoStop}%
	\bibitem [{\citenamefont {Luhman}\ \emph {et~al.}(2008)\citenamefont {Luhman},
		\citenamefont {Pan}, \citenamefont {Tsui}, \citenamefont {Pfeiffer},
		\citenamefont {Baldwin},\ and\ \citenamefont {West}}]{Luhman08}%
	\BibitemOpen
	\bibfield  {author} {\bibinfo {author} {\bibfnamefont {D.~R.}\ \bibnamefont
			{Luhman}}, \bibinfo {author} {\bibfnamefont {W.}~\bibnamefont {Pan}},
		\bibinfo {author} {\bibfnamefont {D.~C.}\ \bibnamefont {Tsui}}, \bibinfo
		{author} {\bibfnamefont {L.~N.}\ \bibnamefont {Pfeiffer}}, \bibinfo {author}
		{\bibfnamefont {K.~W.}\ \bibnamefont {Baldwin}},\ and\ \bibinfo {author}
		{\bibfnamefont {K.~W.}\ \bibnamefont {West}},\ }\bibfield  {title} {\bibinfo
		{title} {Observation of a fractional quantum {Hall} state at
			$\ensuremath{\nu}=1/4$ in a wide {Ga}{As} quantum well},\ }\href
	{https://doi.org/10.1103/PhysRevLett.101.266804} {\bibfield  {journal}
		{\bibinfo  {journal} {Phys. Rev. Lett.}\ }\textbf {\bibinfo {volume} {101}},\
		\bibinfo {pages} {266804} (\bibinfo {year} {2008})}\BibitemShut {NoStop}%
	\bibitem [{\citenamefont {Shabani}\ \emph
		{et~al.}(2009{\natexlab{a}})\citenamefont {Shabani}, \citenamefont {Gokmen},\
		and\ \citenamefont {Shayegan}}]{Shabani09a}%
	\BibitemOpen
	\bibfield  {author} {\bibinfo {author} {\bibfnamefont {J.}~\bibnamefont
			{Shabani}}, \bibinfo {author} {\bibfnamefont {T.}~\bibnamefont {Gokmen}},\
		and\ \bibinfo {author} {\bibfnamefont {M.}~\bibnamefont {Shayegan}},\
	}\bibfield  {title} {\bibinfo {title} {Correlated states of electrons in wide
			quantum wells at low fillings: The role of charge distribution symmetry},\
	}\href {https://doi.org/10.1103/PhysRevLett.103.046805} {\bibfield  {journal}
		{\bibinfo  {journal} {Phys. Rev. Lett.}\ }\textbf {\bibinfo {volume} {103}},\
		\bibinfo {pages} {046805} (\bibinfo {year} {2009}{\natexlab{a}})}\BibitemShut
	{NoStop}%
	\bibitem [{\citenamefont {Shabani}\ \emph
		{et~al.}(2009{\natexlab{b}})\citenamefont {Shabani}, \citenamefont {Gokmen},
		\citenamefont {Chiu},\ and\ \citenamefont {Shayegan}}]{Shabani09b}%
	\BibitemOpen
	\bibfield  {author} {\bibinfo {author} {\bibfnamefont {J.}~\bibnamefont
			{Shabani}}, \bibinfo {author} {\bibfnamefont {T.}~\bibnamefont {Gokmen}},
		\bibinfo {author} {\bibfnamefont {Y.~T.}\ \bibnamefont {Chiu}},\ and\
		\bibinfo {author} {\bibfnamefont {M.}~\bibnamefont {Shayegan}},\ }\bibfield
	{title} {\bibinfo {title} {Evidence for developing fractional quantum {Hall}
			states at even denominator $1/2$ and $1/4$ fillings in asymmetric wide
			quantum wells},\ }\href {https://doi.org/10.1103/PhysRevLett.103.256802}
	{\bibfield  {journal} {\bibinfo  {journal} {Phys. Rev. Lett.}\ }\textbf
		{\bibinfo {volume} {103}},\ \bibinfo {pages} {256802} (\bibinfo {year}
		{2009}{\natexlab{b}})}\BibitemShut {NoStop}%
	\bibitem [{\citenamefont {Shabani}\ \emph {et~al.}(2013)\citenamefont
		{Shabani}, \citenamefont {Liu}, \citenamefont {Shayegan}, \citenamefont
		{Pfeiffer}, \citenamefont {West},\ and\ \citenamefont {Baldwin}}]{Shabani13}%
	\BibitemOpen
	\bibfield  {author} {\bibinfo {author} {\bibfnamefont {J.}~\bibnamefont
			{Shabani}}, \bibinfo {author} {\bibfnamefont {Y.}~\bibnamefont {Liu}},
		\bibinfo {author} {\bibfnamefont {M.}~\bibnamefont {Shayegan}}, \bibinfo
		{author} {\bibfnamefont {L.~N.}\ \bibnamefont {Pfeiffer}}, \bibinfo {author}
		{\bibfnamefont {K.~W.}\ \bibnamefont {West}},\ and\ \bibinfo {author}
		{\bibfnamefont {K.~W.}\ \bibnamefont {Baldwin}},\ }\bibfield  {title}
	{\bibinfo {title} {Phase diagrams for the stability of the
			$\ensuremath{\nu}=\frac{1}{2}$ fractional quantum {Hall} effect in electron
			systems confined to symmetric, wide {Ga}{As} quantum wells},\ }\href
	{https://doi.org/10.1103/PhysRevB.88.245413} {\bibfield  {journal} {\bibinfo
			{journal} {Phys. Rev. B}\ }\textbf {\bibinfo {volume} {88}},\ \bibinfo
		{pages} {245413} (\bibinfo {year} {2013})}\BibitemShut {NoStop}%
	\bibitem [{\citenamefont {Faugno}\ \emph {et~al.}(2019)\citenamefont {Faugno},
		\citenamefont {Balram}, \citenamefont {Barkeshli},\ and\ \citenamefont
		{Jain}}]{Faugno19}%
	\BibitemOpen
	\bibfield  {author} {\bibinfo {author} {\bibfnamefont {W.~N.}\ \bibnamefont
			{Faugno}}, \bibinfo {author} {\bibfnamefont {A.~C.}\ \bibnamefont {Balram}},
		\bibinfo {author} {\bibfnamefont {M.}~\bibnamefont {Barkeshli}},\ and\
		\bibinfo {author} {\bibfnamefont {J.~K.}\ \bibnamefont {Jain}},\ }\bibfield
	{title} {\bibinfo {title} {Prediction of a non-{Abelian} fractional quantum
			{Hall} state with $f$-wave pairing of composite fermions in wide quantum
			wells},\ }\href {https://doi.org/10.1103/PhysRevLett.123.016802} {\bibfield
		{journal} {\bibinfo  {journal} {Phys. Rev. Lett.}\ }\textbf {\bibinfo
			{volume} {123}},\ \bibinfo {pages} {016802} (\bibinfo {year}
		{2019})}\BibitemShut {NoStop}%
	\bibitem [{\citenamefont {MacDonald}(1984)}]{MacDonald84}%
	\BibitemOpen
	\bibfield  {author} {\bibinfo {author} {\bibfnamefont {A.~H.}\ \bibnamefont
			{MacDonald}},\ }\bibfield  {title} {\bibinfo {title} {Influence of
			{Landau}-level mixing on the charge-density-wave state of a two-dimensional
			electron gas in a strong magnetic field},\ }\href
	{https://doi.org/10.1103/PhysRevB.30.4392} {\bibfield  {journal} {\bibinfo
			{journal} {Phys. Rev. B}\ }\textbf {\bibinfo {volume} {30}},\ \bibinfo
		{pages} {4392} (\bibinfo {year} {1984})}\BibitemShut {NoStop}%
	\bibitem [{\citenamefont {Melik-Alaverdian}\ and\ \citenamefont
		{Bonesteel}(1995)}]{Melik-Alaverdian95}%
	\BibitemOpen
	\bibfield  {author} {\bibinfo {author} {\bibfnamefont {V.}~\bibnamefont
			{Melik-Alaverdian}}\ and\ \bibinfo {author} {\bibfnamefont {N.~E.}\
			\bibnamefont {Bonesteel}},\ }\bibfield  {title} {\bibinfo {title} {Composite
			fermions and {Landau}-level mixing in the fractional quantum {Hall} effect},\
	}\href {https://doi.org/10.1103/PhysRevB.52.R17032} {\bibfield  {journal}
		{\bibinfo  {journal} {Phys. Rev. B}\ }\textbf {\bibinfo {volume} {52}},\
		\bibinfo {pages} {R17032} (\bibinfo {year} {1995})}\BibitemShut {NoStop}%
	\bibitem [{\citenamefont {Murthy}\ and\ \citenamefont
		{Shankar}(2002)}]{Murthy02}%
	\BibitemOpen
	\bibfield  {author} {\bibinfo {author} {\bibfnamefont {G.}~\bibnamefont
			{Murthy}}\ and\ \bibinfo {author} {\bibfnamefont {R.}~\bibnamefont
			{Shankar}},\ }\bibfield  {title} {\bibinfo {title} {Hamiltonian theory of the
			fractional quantum {Hall} effect: Effect of {Landau} level mixing},\ }\href
	{https://doi.org/10.1103/PhysRevB.65.245309} {\bibfield  {journal} {\bibinfo
			{journal} {Phys. Rev. B}\ }\textbf {\bibinfo {volume} {65}},\ \bibinfo
		{pages} {245309} (\bibinfo {year} {2002})}\BibitemShut {NoStop}%
	\bibitem [{\citenamefont {Bishara}\ and\ \citenamefont
		{Nayak}(2009)}]{Bishara09}%
	\BibitemOpen
	\bibfield  {author} {\bibinfo {author} {\bibfnamefont {W.}~\bibnamefont
			{Bishara}}\ and\ \bibinfo {author} {\bibfnamefont {C.}~\bibnamefont
			{Nayak}},\ }\bibfield  {title} {\bibinfo {title} {Effect of {Landau} level
			mixing on the effective interaction between electrons in the fractional
			quantum {Hall} regime},\ }\href {https://doi.org/10.1103/PhysRevB.80.121302}
	{\bibfield  {journal} {\bibinfo  {journal} {Phys. Rev. B}\ }\textbf {\bibinfo
			{volume} {80}},\ \bibinfo {pages} {121302} (\bibinfo {year}
		{2009})}\BibitemShut {NoStop}%
	\bibitem [{\citenamefont {W\'ojs}\ \emph {et~al.}(2010)\citenamefont {W\'ojs},
		\citenamefont {T\ifmmode~\mbox{\H{o}}\else \H{o}\fi{}ke},\ and\ \citenamefont
		{Jain}}]{Wojs10}%
	\BibitemOpen
	\bibfield  {author} {\bibinfo {author} {\bibfnamefont {A.}~\bibnamefont
			{W\'ojs}}, \bibinfo {author} {\bibfnamefont {C.}~\bibnamefont
			{T\ifmmode~\mbox{\H{o}}\else \H{o}\fi{}ke}},\ and\ \bibinfo {author}
		{\bibfnamefont {J.~K.}\ \bibnamefont {Jain}},\ }\bibfield  {title} {\bibinfo
		{title} {Landau-level mixing and the emergence of {Pfaffian} excitations for
			the $\nu=5/2$ fractional quantum {Hall} effect},\ }\href
	{https://doi.org/10.1103/PhysRevLett.105.096802} {\bibfield  {journal}
		{\bibinfo  {journal} {Phys. Rev. Lett.}\ }\textbf {\bibinfo {volume} {105}},\
		\bibinfo {pages} {096802} (\bibinfo {year} {2010})}\BibitemShut {NoStop}%
	\bibitem [{\citenamefont {Sodemann}\ and\ \citenamefont
		{MacDonald}(2013)}]{Sodemann13}%
	\BibitemOpen
	\bibfield  {author} {\bibinfo {author} {\bibfnamefont {I.}~\bibnamefont
			{Sodemann}}\ and\ \bibinfo {author} {\bibfnamefont {A.~H.}\ \bibnamefont
			{MacDonald}},\ }\bibfield  {title} {\bibinfo {title} {{Landau} level mixing
			and the fractional quantum {Hall} effect},\ }\href
	{https://doi.org/10.1103/PhysRevB.87.245425} {\bibfield  {journal} {\bibinfo
			{journal} {Phys. Rev. B}\ }\textbf {\bibinfo {volume} {87}},\ \bibinfo
		{pages} {245425} (\bibinfo {year} {2013})}\BibitemShut {NoStop}%
	\bibitem [{\citenamefont {Simon}\ and\ \citenamefont {Rezayi}(2013)}]{Simon13}%
	\BibitemOpen
	\bibfield  {author} {\bibinfo {author} {\bibfnamefont {S.~H.}\ \bibnamefont
			{Simon}}\ and\ \bibinfo {author} {\bibfnamefont {E.~H.}\ \bibnamefont
			{Rezayi}},\ }\bibfield  {title} {\bibinfo {title} {{Landau} level mixing in
			the perturbative limit},\ }\href {https://doi.org/10.1103/PhysRevB.87.155426}
	{\bibfield  {journal} {\bibinfo  {journal} {Phys. Rev. B}\ }\textbf {\bibinfo
			{volume} {87}},\ \bibinfo {pages} {155426} (\bibinfo {year}
		{2013})}\BibitemShut {NoStop}%
	\bibitem [{\citenamefont {Peterson}\ and\ \citenamefont
		{Nayak}(2013)}]{Peterson13}%
	\BibitemOpen
	\bibfield  {author} {\bibinfo {author} {\bibfnamefont {M.~R.}\ \bibnamefont
			{Peterson}}\ and\ \bibinfo {author} {\bibfnamefont {C.}~\bibnamefont
			{Nayak}},\ }\bibfield  {title} {\bibinfo {title} {More realistic
			{Hamiltonians} for the fractional quantum {Hall} regime in {Ga}{As} and
			graphene},\ }\href {https://doi.org/10.1103/PhysRevB.87.245129} {\bibfield
		{journal} {\bibinfo  {journal} {Phys. Rev. B}\ }\textbf {\bibinfo {volume}
			{87}},\ \bibinfo {pages} {245129} (\bibinfo {year} {2013})}\BibitemShut
	{NoStop}%
	\bibitem [{\citenamefont {Peterson}\ and\ \citenamefont
		{Nayak}(2014)}]{Peterson14}%
	\BibitemOpen
	\bibfield  {author} {\bibinfo {author} {\bibfnamefont {M.~R.}\ \bibnamefont
			{Peterson}}\ and\ \bibinfo {author} {\bibfnamefont {C.}~\bibnamefont
			{Nayak}},\ }\bibfield  {title} {\bibinfo {title} {Effects of {Landau} level
			mixing on the fractional quantum {Hall} effect in monolayer graphene},\
	}\href {https://doi.org/10.1103/PhysRevLett.113.086401} {\bibfield  {journal}
		{\bibinfo  {journal} {Phys. Rev. Lett.}\ }\textbf {\bibinfo {volume} {113}},\
		\bibinfo {pages} {086401} (\bibinfo {year} {2014})}\BibitemShut {NoStop}%
	\bibitem [{\citenamefont {Ortiz}\ \emph {et~al.}(1993)\citenamefont {Ortiz},
		\citenamefont {Ceperley},\ and\ \citenamefont {Martin}}]{Ortiz93}%
	\BibitemOpen
	\bibfield  {author} {\bibinfo {author} {\bibfnamefont {G.}~\bibnamefont
			{Ortiz}}, \bibinfo {author} {\bibfnamefont {D.~M.}\ \bibnamefont
			{Ceperley}},\ and\ \bibinfo {author} {\bibfnamefont {R.~M.}\ \bibnamefont
			{Martin}},\ }\bibfield  {title} {\bibinfo {title} {New stochastic method for
			systems with broken time-reversal symmetry: 2d fermions in a magnetic
			field},\ }\href {https://doi.org/10.1103/PhysRevLett.71.2777} {\bibfield
		{journal} {\bibinfo  {journal} {Phys. Rev. Lett.}\ }\textbf {\bibinfo
			{volume} {71}},\ \bibinfo {pages} {2777} (\bibinfo {year}
		{1993})}\BibitemShut {NoStop}%
	\bibitem [{\citenamefont {Melik-Alaverdian}\ \emph {et~al.}(1997)\citenamefont
		{Melik-Alaverdian}, \citenamefont {Bonesteel},\ and\ \citenamefont
		{Ortiz}}]{Melik-Alaverdian97}%
	\BibitemOpen
	\bibfield  {author} {\bibinfo {author} {\bibfnamefont {V.}~\bibnamefont
			{Melik-Alaverdian}}, \bibinfo {author} {\bibfnamefont {N.~E.}\ \bibnamefont
			{Bonesteel}},\ and\ \bibinfo {author} {\bibfnamefont {G.}~\bibnamefont
			{Ortiz}},\ }\bibfield  {title} {\bibinfo {title} {Quantum {Hall} fluids on
			the {Haldane} sphere: A diffusion {Monte} {Carlo} study},\ }\href
	{https://doi.org/10.1103/PhysRevLett.79.5286} {\bibfield  {journal} {\bibinfo
			{journal} {Phys. Rev. Lett.}\ }\textbf {\bibinfo {volume} {79}},\ \bibinfo
		{pages} {5286} (\bibinfo {year} {1997})}\BibitemShut {NoStop}%
	\bibitem [{\citenamefont {Melik-Alaverdian}\ \emph {et~al.}(2001)\citenamefont
		{Melik-Alaverdian}, \citenamefont {Ortiz},\ and\ \citenamefont
		{Bonesteel}}]{Melik-Alaverdian01}%
	\BibitemOpen
	\bibfield  {author} {\bibinfo {author} {\bibfnamefont {V.}~\bibnamefont
			{Melik-Alaverdian}}, \bibinfo {author} {\bibfnamefont {G.}~\bibnamefont
			{Ortiz}},\ and\ \bibinfo {author} {\bibfnamefont {N.}~\bibnamefont
			{Bonesteel}},\ }\bibfield  {title} {\bibinfo {title} {Quantum projector
			method on curved manifolds},\ }\href
	{https://doi.org/10.1023/A:1010326231389} {\bibfield  {journal} {\bibinfo
			{journal} {Journal of Statistical Physics}\ }\textbf {\bibinfo {volume}
			{104}},\ \bibinfo {pages} {449} (\bibinfo {year} {2001})}\BibitemShut
	{NoStop}%
	\bibitem [{\citenamefont {Zhang}\ \emph {et~al.}(2016)\citenamefont {Zhang},
		\citenamefont {W\'ojs},\ and\ \citenamefont {Jain}}]{Zhang16}%
	\BibitemOpen
	\bibfield  {author} {\bibinfo {author} {\bibfnamefont {Y.}~\bibnamefont
			{Zhang}}, \bibinfo {author} {\bibfnamefont {A.}~\bibnamefont {W\'ojs}},\ and\
		\bibinfo {author} {\bibfnamefont {J.~K.}\ \bibnamefont {Jain}},\ }\bibfield
	{title} {\bibinfo {title} {Landau-level mixing and particle-hole symmetry
			breaking for spin transitions in the fractional quantum {Hall} effect},\
	}\href {https://doi.org/10.1103/PhysRevLett.117.116803} {\bibfield  {journal}
		{\bibinfo  {journal} {Phys. Rev. Lett.}\ }\textbf {\bibinfo {volume} {117}},\
		\bibinfo {pages} {116803} (\bibinfo {year} {2016})}\BibitemShut {NoStop}%
	\bibitem [{\citenamefont {Zhao}\ \emph {et~al.}(2018)\citenamefont {Zhao},
		\citenamefont {Zhang},\ and\ \citenamefont {Jain}}]{Zhao18}%
	\BibitemOpen
	\bibfield  {author} {\bibinfo {author} {\bibfnamefont {J.}~\bibnamefont
			{Zhao}}, \bibinfo {author} {\bibfnamefont {Y.}~\bibnamefont {Zhang}},\ and\
		\bibinfo {author} {\bibfnamefont {J.~K.}\ \bibnamefont {Jain}},\ }\bibfield
	{title} {\bibinfo {title} {Crystallization in the fractional quantum {Hall}
			regime induced by {Landau}-level mixing},\ }\href
	{https://doi.org/10.1103/PhysRevLett.121.116802} {\bibfield  {journal}
		{\bibinfo  {journal} {Phys. Rev. Lett.}\ }\textbf {\bibinfo {volume} {121}},\
		\bibinfo {pages} {116802} (\bibinfo {year} {2018})}\BibitemShut {NoStop}%
	\bibitem [{\citenamefont {Ma}\ \emph {et~al.}(2020)\citenamefont {Ma},
		\citenamefont {Villegas~Rosales}, \citenamefont {Deng}, \citenamefont
		{Chung}, \citenamefont {Pfeiffer}, \citenamefont {West}, \citenamefont
		{Baldwin}, \citenamefont {Winkler},\ and\ \citenamefont {Shayegan}}]{Ma20}%
	\BibitemOpen
	\bibfield  {author} {\bibinfo {author} {\bibfnamefont {M.~K.}\ \bibnamefont
			{Ma}}, \bibinfo {author} {\bibfnamefont {K.~A.}\ \bibnamefont
			{Villegas~Rosales}}, \bibinfo {author} {\bibfnamefont {H.}~\bibnamefont
			{Deng}}, \bibinfo {author} {\bibfnamefont {Y.~J.}\ \bibnamefont {Chung}},
		\bibinfo {author} {\bibfnamefont {L.~N.}\ \bibnamefont {Pfeiffer}}, \bibinfo
		{author} {\bibfnamefont {K.~W.}\ \bibnamefont {West}}, \bibinfo {author}
		{\bibfnamefont {K.~W.}\ \bibnamefont {Baldwin}}, \bibinfo {author}
		{\bibfnamefont {R.}~\bibnamefont {Winkler}},\ and\ \bibinfo {author}
		{\bibfnamefont {M.}~\bibnamefont {Shayegan}},\ }\bibfield  {title} {\bibinfo
		{title} {Thermal and quantum melting phase diagrams for a
			magnetic-field-induced wigner solid},\ }\href
	{https://doi.org/10.1103/PhysRevLett.125.036601} {\bibfield  {journal}
		{\bibinfo  {journal} {Phys. Rev. Lett.}\ }\textbf {\bibinfo {volume} {125}},\
		\bibinfo {pages} {036601} (\bibinfo {year} {2020})}\BibitemShut {NoStop}%
	\bibitem [{\citenamefont {Villegas~Rosales}\ \emph
		{et~al.}(2021{\natexlab{a}})\citenamefont {Villegas~Rosales}, \citenamefont
		{Madathil}, \citenamefont {Chung}, \citenamefont {Pfeiffer}, \citenamefont
		{West}, \citenamefont {Baldwin},\ and\ \citenamefont {Shayegan}}]{Rosales21}%
	\BibitemOpen
	\bibfield  {author} {\bibinfo {author} {\bibfnamefont {K.~A.}\ \bibnamefont
			{Villegas~Rosales}}, \bibinfo {author} {\bibfnamefont {P.~T.}\ \bibnamefont
			{Madathil}}, \bibinfo {author} {\bibfnamefont {Y.~J.}\ \bibnamefont {Chung}},
		\bibinfo {author} {\bibfnamefont {L.~N.}\ \bibnamefont {Pfeiffer}}, \bibinfo
		{author} {\bibfnamefont {K.~W.}\ \bibnamefont {West}}, \bibinfo {author}
		{\bibfnamefont {K.~W.}\ \bibnamefont {Baldwin}},\ and\ \bibinfo {author}
		{\bibfnamefont {M.}~\bibnamefont {Shayegan}},\ }\bibfield  {title} {\bibinfo
		{title} {Fractional quantum hall effect energy gaps: Role of electron layer
			thickness},\ }\href {https://doi.org/10.1103/PhysRevLett.127.056801}
	{\bibfield  {journal} {\bibinfo  {journal} {Phys. Rev. Lett.}\ }\textbf
		{\bibinfo {volume} {127}},\ \bibinfo {pages} {056801} (\bibinfo {year}
		{2021}{\natexlab{a}})}\BibitemShut {NoStop}%
	\bibitem [{\citenamefont {Levin}\ \emph {et~al.}(2007)\citenamefont {Levin},
		\citenamefont {Halperin},\ and\ \citenamefont {Rosenow}}]{Levin07}%
	\BibitemOpen
	\bibfield  {author} {\bibinfo {author} {\bibfnamefont {M.}~\bibnamefont
			{Levin}}, \bibinfo {author} {\bibfnamefont {B.~I.}\ \bibnamefont
			{Halperin}},\ and\ \bibinfo {author} {\bibfnamefont {B.}~\bibnamefont
			{Rosenow}},\ }\bibfield  {title} {\bibinfo {title} {Particle-hole symmetry
			and the {Pfaffian} state},\ }\href
	{https://doi.org/10.1103/PhysRevLett.99.236806} {\bibfield  {journal}
		{\bibinfo  {journal} {Phys. Rev. Lett.}\ }\textbf {\bibinfo {volume} {99}},\
		\bibinfo {pages} {236806} (\bibinfo {year} {2007})}\BibitemShut {NoStop}%
	\bibitem [{\citenamefont {Lee}\ \emph {et~al.}(2007)\citenamefont {Lee},
		\citenamefont {Ryu}, \citenamefont {Nayak},\ and\ \citenamefont
		{Fisher}}]{Lee07}%
	\BibitemOpen
	\bibfield  {author} {\bibinfo {author} {\bibfnamefont {S.-S.}\ \bibnamefont
			{Lee}}, \bibinfo {author} {\bibfnamefont {S.}~\bibnamefont {Ryu}}, \bibinfo
		{author} {\bibfnamefont {C.}~\bibnamefont {Nayak}},\ and\ \bibinfo {author}
		{\bibfnamefont {M.~P.~A.}\ \bibnamefont {Fisher}},\ }\bibfield  {title}
	{\bibinfo {title} {Particle-hole symmetry and the $\nu=5/2$ quantum {Hall}
			state},\ }\href {https://doi.org/10.1103/PhysRevLett.99.236807} {\bibfield
		{journal} {\bibinfo  {journal} {Phys. Rev. Lett.}\ }\textbf {\bibinfo
			{volume} {99}},\ \bibinfo {pages} {236807} (\bibinfo {year}
		{2007})}\BibitemShut {NoStop}%
	\bibitem [{\citenamefont {Rezayi}\ and\ \citenamefont
		{Simon}(2011)}]{Rezayi11}%
	\BibitemOpen
	\bibfield  {author} {\bibinfo {author} {\bibfnamefont {E.~H.}\ \bibnamefont
			{Rezayi}}\ and\ \bibinfo {author} {\bibfnamefont {S.~H.}\ \bibnamefont
			{Simon}},\ }\bibfield  {title} {\bibinfo {title} {Breaking of particle-hole
			symmetry by {Landau} level mixing in the $\nu=5/2$ quantized {Hall} state},\
	}\href {https://doi.org/10.1103/PhysRevLett.106.116801} {\bibfield  {journal}
		{\bibinfo  {journal} {Phys. Rev. Lett.}\ }\textbf {\bibinfo {volume} {106}},\
		\bibinfo {pages} {116801} (\bibinfo {year} {2011})}\BibitemShut {NoStop}%
	\bibitem [{\citenamefont {Pakrouski}\ \emph {et~al.}(2015)\citenamefont
		{Pakrouski}, \citenamefont {Peterson}, \citenamefont {Jolicoeur},
		\citenamefont {Scarola}, \citenamefont {Nayak},\ and\ \citenamefont
		{Troyer}}]{Pakrouski15}%
	\BibitemOpen
	\bibfield  {author} {\bibinfo {author} {\bibfnamefont {K.}~\bibnamefont
			{Pakrouski}}, \bibinfo {author} {\bibfnamefont {M.~R.}\ \bibnamefont
			{Peterson}}, \bibinfo {author} {\bibfnamefont {T.}~\bibnamefont {Jolicoeur}},
		\bibinfo {author} {\bibfnamefont {V.~W.}\ \bibnamefont {Scarola}}, \bibinfo
		{author} {\bibfnamefont {C.}~\bibnamefont {Nayak}},\ and\ \bibinfo {author}
		{\bibfnamefont {M.}~\bibnamefont {Troyer}},\ }\bibfield  {title} {\bibinfo
		{title} {Phase diagram of the $\nu=5/2$ fractional quantum {Hall} effect:
			Effects of {Landau}-level mixing and nonzero width},\ }\href
	{https://doi.org/10.1103/PhysRevX.5.021004} {\bibfield  {journal} {\bibinfo
			{journal} {Phys. Rev. X}\ }\textbf {\bibinfo {volume} {5}},\ \bibinfo {pages}
		{021004} (\bibinfo {year} {2015})}\BibitemShut {NoStop}%
	\bibitem [{\citenamefont {Rezayi}(2017)}]{Rezayi17}%
	\BibitemOpen
	\bibfield  {author} {\bibinfo {author} {\bibfnamefont {E.~H.}\ \bibnamefont
			{Rezayi}},\ }\bibfield  {title} {\bibinfo {title} {Landau level mixing and
			the ground state of the $\ensuremath{\nu}=5/2$ quantum {Hall} effect},\
	}\href {https://doi.org/10.1103/PhysRevLett.119.026801} {\bibfield  {journal}
		{\bibinfo  {journal} {Phys. Rev. Lett.}\ }\textbf {\bibinfo {volume} {119}},\
		\bibinfo {pages} {026801} (\bibinfo {year} {2017})}\BibitemShut {NoStop}%
	\bibitem [{\citenamefont {Sreejith}\ \emph {et~al.}(2017)\citenamefont
		{Sreejith}, \citenamefont {Zhang},\ and\ \citenamefont {Jain}}]{Sreejith17}%
	\BibitemOpen
	\bibfield  {author} {\bibinfo {author} {\bibfnamefont {G.~J.}\ \bibnamefont
			{Sreejith}}, \bibinfo {author} {\bibfnamefont {Y.}~\bibnamefont {Zhang}},\
		and\ \bibinfo {author} {\bibfnamefont {J.~K.}\ \bibnamefont {Jain}},\
	}\bibfield  {title} {\bibinfo {title} {Surprising robustness of particle-hole
			symmetry for composite-fermion liquids},\ }\href
	{https://doi.org/10.1103/PhysRevB.96.125149} {\bibfield  {journal} {\bibinfo
			{journal} {Phys. Rev. B}\ }\textbf {\bibinfo {volume} {96}},\ \bibinfo
		{pages} {125149} (\bibinfo {year} {2017})}\BibitemShut {NoStop}%
	\bibitem [{\citenamefont {Herviou}\ and\ \citenamefont
		{Mila}(2022)}]{Herviou22}%
	\BibitemOpen
	\bibfield  {author} {\bibinfo {author} {\bibfnamefont {L.}~\bibnamefont
			{Herviou}}\ and\ \bibinfo {author} {\bibfnamefont {F.}~\bibnamefont {Mila}},\
	}\href {https://doi.org/10.48550/ARXIV.2210.04925} {\bibinfo {title}
		{Possible restoration of particle-hole symmetry in the 5/2 quantized {Hall}
			state at small magnetic field}} (\bibinfo {year} {2022})\BibitemShut
	{NoStop}%
	\bibitem [{\citenamefont {Jain}(1989{\natexlab{b}})}]{Jain89b}%
	\BibitemOpen
	\bibfield  {author} {\bibinfo {author} {\bibfnamefont {J.~K.}\ \bibnamefont
			{Jain}},\ }\bibfield  {title} {\bibinfo {title} {Incompressible quantum
			{Hall} states},\ }\href {https://doi.org/10.1103/PhysRevB.40.8079} {\bibfield
		{journal} {\bibinfo  {journal} {Phys. Rev. B}\ }\textbf {\bibinfo {volume}
			{40}},\ \bibinfo {pages} {8079} (\bibinfo {year}
		{1989}{\natexlab{b}})}\BibitemShut {NoStop}%
	\bibitem [{\citenamefont {Wen}\ and\ \citenamefont {Zee}(1992)}]{Wen92}%
	\BibitemOpen
	\bibfield  {author} {\bibinfo {author} {\bibfnamefont {X.~G.}\ \bibnamefont
			{Wen}}\ and\ \bibinfo {author} {\bibfnamefont {A.}~\bibnamefont {Zee}},\
	}\bibfield  {title} {\bibinfo {title} {Shift and spin vector: New topological
			quantum numbers for the {Hall} fluids},\ }\href
	{https://doi.org/10.1103/PhysRevLett.69.953} {\bibfield  {journal} {\bibinfo
			{journal} {Phys. Rev. Lett.}\ }\textbf {\bibinfo {volume} {69}},\ \bibinfo
		{pages} {953} (\bibinfo {year} {1992})}\BibitemShut {NoStop}%
	\bibitem [{\citenamefont {Jain}\ and\ \citenamefont
		{Kamilla}(1997{\natexlab{a}})}]{Jain97}%
	\BibitemOpen
	\bibfield  {author} {\bibinfo {author} {\bibfnamefont {J.~K.}\ \bibnamefont
			{Jain}}\ and\ \bibinfo {author} {\bibfnamefont {R.~K.}\ \bibnamefont
			{Kamilla}},\ }\bibfield  {title} {\bibinfo {title} {Composite fermions in the
			{Hilbert} space of the lowest electronic {Landau} level},\ }\href
	{https://doi.org/10.1142/S0217979297001301} {\bibfield  {journal} {\bibinfo
			{journal} {Int. J. Mod. Phys. B}\ }\textbf {\bibinfo {volume} {11}},\
		\bibinfo {pages} {2621} (\bibinfo {year} {1997}{\natexlab{a}})}\BibitemShut
	{NoStop}%
	\bibitem [{\citenamefont {Jain}\ and\ \citenamefont
		{Kamilla}(1997{\natexlab{b}})}]{Jain97b}%
	\BibitemOpen
	\bibfield  {author} {\bibinfo {author} {\bibfnamefont {J.~K.}\ \bibnamefont
			{Jain}}\ and\ \bibinfo {author} {\bibfnamefont {R.~K.}\ \bibnamefont
			{Kamilla}},\ }\bibfield  {title} {\bibinfo {title} {Quantitative study of
			large composite-fermion systems},\ }\href
	{https://doi.org/10.1103/PhysRevB.55.R4895} {\bibfield  {journal} {\bibinfo
			{journal} {Phys. Rev. B}\ }\textbf {\bibinfo {volume} {55}},\ \bibinfo
		{pages} {R4895} (\bibinfo {year} {1997}{\natexlab{b}})}\BibitemShut {NoStop}%
	\bibitem [{\citenamefont {Rezayi}\ and\ \citenamefont {Read}(1994)}]{Rezayi94}%
	\BibitemOpen
	\bibfield  {author} {\bibinfo {author} {\bibfnamefont {E.}~\bibnamefont
			{Rezayi}}\ and\ \bibinfo {author} {\bibfnamefont {N.}~\bibnamefont {Read}},\
	}\bibfield  {title} {\bibinfo {title} {Fermi-liquid-like state in a
			half-filled {Landau} level},\ }\href
	{https://doi.org/10.1103/PhysRevLett.72.900} {\bibfield  {journal} {\bibinfo
			{journal} {Phys. Rev. Lett.}\ }\textbf {\bibinfo {volume} {72}},\ \bibinfo
		{pages} {900} (\bibinfo {year} {1994})}\BibitemShut {NoStop}%
	\bibitem [{\citenamefont {Balram}\ \emph
		{et~al.}(2015{\natexlab{a}})\citenamefont {Balram}, \citenamefont
		{T\ifmmode~\mbox{\H{o}}\else \H{o}\fi{}ke},\ and\ \citenamefont
		{Jain}}]{Balram15b}%
	\BibitemOpen
	\bibfield  {author} {\bibinfo {author} {\bibfnamefont {A.~C.}\ \bibnamefont
			{Balram}}, \bibinfo {author} {\bibfnamefont {C.}~\bibnamefont
			{T\ifmmode~\mbox{\H{o}}\else \H{o}\fi{}ke}},\ and\ \bibinfo {author}
		{\bibfnamefont {J.~K.}\ \bibnamefont {Jain}},\ }\bibfield  {title} {\bibinfo
		{title} {Luttinger theorem for the strongly correlated {Fermi} liquid of
			composite fermions},\ }\href {https://doi.org/10.1103/PhysRevLett.115.186805}
	{\bibfield  {journal} {\bibinfo  {journal} {Phys. Rev. Lett.}\ }\textbf
		{\bibinfo {volume} {115}},\ \bibinfo {pages} {186805} (\bibinfo {year}
		{2015}{\natexlab{a}})}\BibitemShut {NoStop}%
	\bibitem [{\citenamefont {Jolicoeur}(2007)}]{Jolicoeur07}%
	\BibitemOpen
	\bibfield  {author} {\bibinfo {author} {\bibfnamefont {T.}~\bibnamefont
			{Jolicoeur}},\ }\bibfield  {title} {\bibinfo {title} {Non-abelian states with
			negative flux: A new series of quantum {Hall} states},\ }\href
	{https://doi.org/10.1103/PhysRevLett.99.036805} {\bibfield  {journal}
		{\bibinfo  {journal} {Phys. Rev. Lett.}\ }\textbf {\bibinfo {volume} {99}},\
		\bibinfo {pages} {036805} (\bibinfo {year} {2007})}\BibitemShut {NoStop}%
	\bibitem [{\citenamefont {Zucker}\ and\ \citenamefont
		{Feldman}(2016)}]{Zucker16}%
	\BibitemOpen
	\bibfield  {author} {\bibinfo {author} {\bibfnamefont {P.~T.}\ \bibnamefont
			{Zucker}}\ and\ \bibinfo {author} {\bibfnamefont {D.~E.}\ \bibnamefont
			{Feldman}},\ }\bibfield  {title} {\bibinfo {title} {Stabilization of the
			particle-hole {Pfaffian} order by {Landau}-level mixing and impurities that
			break particle-hole symmetry},\ }\href
	{https://doi.org/10.1103/PhysRevLett.117.096802} {\bibfield  {journal}
		{\bibinfo  {journal} {Phys. Rev. Lett.}\ }\textbf {\bibinfo {volume} {117}},\
		\bibinfo {pages} {096802} (\bibinfo {year} {2016})}\BibitemShut {NoStop}%
	\bibitem [{\citenamefont {Balram}\ \emph {et~al.}(2018)\citenamefont {Balram},
		\citenamefont {Barkeshli},\ and\ \citenamefont {Rudner}}]{Balram18}%
	\BibitemOpen
	\bibfield  {author} {\bibinfo {author} {\bibfnamefont {A.~C.}\ \bibnamefont
			{Balram}}, \bibinfo {author} {\bibfnamefont {M.}~\bibnamefont {Barkeshli}},\
		and\ \bibinfo {author} {\bibfnamefont {M.~S.}\ \bibnamefont {Rudner}},\
	}\bibfield  {title} {\bibinfo {title} {Parton construction of a wave function
			in the anti-{Pfaffian} phase},\ }\href
	{https://doi.org/10.1103/PhysRevB.98.035127} {\bibfield  {journal} {\bibinfo
			{journal} {Phys. Rev. B}\ }\textbf {\bibinfo {volume} {98}},\ \bibinfo
		{pages} {035127} (\bibinfo {year} {2018})}\BibitemShut {NoStop}%
	\bibitem [{\citenamefont {Mishmash}\ \emph {et~al.}(2018)\citenamefont
		{Mishmash}, \citenamefont {Mross}, \citenamefont {Alicea},\ and\
		\citenamefont {Motrunich}}]{Mishmash18}%
	\BibitemOpen
	\bibfield  {author} {\bibinfo {author} {\bibfnamefont {R.~V.}\ \bibnamefont
			{Mishmash}}, \bibinfo {author} {\bibfnamefont {D.~F.}\ \bibnamefont {Mross}},
		\bibinfo {author} {\bibfnamefont {J.}~\bibnamefont {Alicea}},\ and\ \bibinfo
		{author} {\bibfnamefont {O.~I.}\ \bibnamefont {Motrunich}},\ }\bibfield
	{title} {\bibinfo {title} {Numerical exploration of trial wave functions for
			the particle-hole-symmetric {Pfaffian}},\ }\href
	{https://doi.org/10.1103/PhysRevB.98.081107} {\bibfield  {journal} {\bibinfo
			{journal} {Phys. Rev. B}\ }\textbf {\bibinfo {volume} {98}},\ \bibinfo
		{pages} {081107} (\bibinfo {year} {2018})}\BibitemShut {NoStop}%
	\bibitem [{\citenamefont {Son}(2015)}]{Son15}%
	\BibitemOpen
	\bibfield  {author} {\bibinfo {author} {\bibfnamefont {D.~T.}\ \bibnamefont
			{Son}},\ }\bibfield  {title} {\bibinfo {title} {Is the composite fermion a
			{Dirac} particle?},\ }\href {https://doi.org/10.1103/PhysRevX.5.031027}
	{\bibfield  {journal} {\bibinfo  {journal} {Phys. Rev. X}\ }\textbf {\bibinfo
			{volume} {5}},\ \bibinfo {pages} {031027} (\bibinfo {year}
		{2015})}\BibitemShut {NoStop}%
	\bibitem [{\citenamefont {Yutushui}\ and\ \citenamefont
		{Mross}(2020)}]{Yutushui20}%
	\BibitemOpen
	\bibfield  {author} {\bibinfo {author} {\bibfnamefont {M.}~\bibnamefont
			{Yutushui}}\ and\ \bibinfo {author} {\bibfnamefont {D.~F.}\ \bibnamefont
			{Mross}},\ }\bibfield  {title} {\bibinfo {title} {Large-scale simulations of
			particle-hole-symmetric {Pfaffian} trial wave functions},\ }\href
	{https://doi.org/10.1103/PhysRevB.102.195153} {\bibfield  {journal} {\bibinfo
			{journal} {Phys. Rev. B}\ }\textbf {\bibinfo {volume} {102}},\ \bibinfo
		{pages} {195153} (\bibinfo {year} {2020})}\BibitemShut {NoStop}%
	\bibitem [{\citenamefont {Wen}(1991)}]{Wen91}%
	\BibitemOpen
	\bibfield  {author} {\bibinfo {author} {\bibfnamefont {X.~G.}\ \bibnamefont
			{Wen}},\ }\bibfield  {title} {\bibinfo {title} {Non-abelian statistics in the
			fractional quantum {Hall} states},\ }\href
	{https://doi.org/10.1103/PhysRevLett.66.802} {\bibfield  {journal} {\bibinfo
			{journal} {Phys. Rev. Lett.}\ }\textbf {\bibinfo {volume} {66}},\ \bibinfo
		{pages} {802} (\bibinfo {year} {1991})}\BibitemShut {NoStop}%
	\bibitem [{\citenamefont {Wu}\ \emph {et~al.}(2017)\citenamefont {Wu},
		\citenamefont {Shi},\ and\ \citenamefont {Jain}}]{Wu17}%
	\BibitemOpen
	\bibfield  {author} {\bibinfo {author} {\bibfnamefont {Y.}~\bibnamefont
			{Wu}}, \bibinfo {author} {\bibfnamefont {T.}~\bibnamefont {Shi}},\ and\
		\bibinfo {author} {\bibfnamefont {J.~K.}\ \bibnamefont {Jain}},\ }\bibfield
	{title} {\bibinfo {title} {Non-abelian parton fractional quantum {Hall}
			effect in multilayer graphene},\ }\href
	{https://doi.org/10.1021/acs.nanolett.7b01080} {\bibfield  {journal}
		{\bibinfo  {journal} {Nano Letters}\ }\textbf {\bibinfo {volume} {17}},\
		\bibinfo {pages} {4643} (\bibinfo {year} {2017})},\ \bibinfo {note} {pMID:
		28649831},\ \Eprint
	{https://arxiv.org/abs/http://dx.doi.org/10.1021/acs.nanolett.7b01080}
	{http://dx.doi.org/10.1021/acs.nanolett.7b01080} \BibitemShut {NoStop}%
	\bibitem [{\citenamefont {Bandyopadhyay}\ \emph {et~al.}(2018)\citenamefont
		{Bandyopadhyay}, \citenamefont {Chen}, \citenamefont {Ahari}, \citenamefont
		{Ortiz}, \citenamefont {Nussinov},\ and\ \citenamefont
		{Seidel}}]{Bandyopadhyay18}%
	\BibitemOpen
	\bibfield  {author} {\bibinfo {author} {\bibfnamefont {S.}~\bibnamefont
			{Bandyopadhyay}}, \bibinfo {author} {\bibfnamefont {L.}~\bibnamefont {Chen}},
		\bibinfo {author} {\bibfnamefont {M.~T.}\ \bibnamefont {Ahari}}, \bibinfo
		{author} {\bibfnamefont {G.}~\bibnamefont {Ortiz}}, \bibinfo {author}
		{\bibfnamefont {Z.}~\bibnamefont {Nussinov}},\ and\ \bibinfo {author}
		{\bibfnamefont {A.}~\bibnamefont {Seidel}},\ }\bibfield  {title} {\bibinfo
		{title} {Entangled {Pauli} principles: The {DNA} of quantum {Hall} fluids},\
	}\href {https://doi.org/10.1103/PhysRevB.98.161118} {\bibfield  {journal}
		{\bibinfo  {journal} {Phys. Rev. B}\ }\textbf {\bibinfo {volume} {98}},\
		\bibinfo {pages} {161118} (\bibinfo {year} {2018})}\BibitemShut {NoStop}%
	\bibitem [{\citenamefont {Kim}\ \emph {et~al.}(2019)\citenamefont {Kim},
		\citenamefont {Balram}, \citenamefont {Taniguchi}, \citenamefont {Watanabe},
		\citenamefont {Jain},\ and\ \citenamefont {Smet}}]{Kim19}%
	\BibitemOpen
	\bibfield  {author} {\bibinfo {author} {\bibfnamefont {Y.}~\bibnamefont
			{Kim}}, \bibinfo {author} {\bibfnamefont {A.~C.}\ \bibnamefont {Balram}},
		\bibinfo {author} {\bibfnamefont {T.}~\bibnamefont {Taniguchi}}, \bibinfo
		{author} {\bibfnamefont {K.}~\bibnamefont {Watanabe}}, \bibinfo {author}
		{\bibfnamefont {J.~K.}\ \bibnamefont {Jain}},\ and\ \bibinfo {author}
		{\bibfnamefont {J.~H.}\ \bibnamefont {Smet}},\ }\bibfield  {title} {\bibinfo
		{title} {Even denominator fractional quantum {Hall} states in higher {Landau}
			levels of graphene},\ }\href {https://doi.org/10.1038/s41567-018-0355-x}
	{\bibfield  {journal} {\bibinfo  {journal} {Nature Physics}\ }\textbf
		{\bibinfo {volume} {15}},\ \bibinfo {pages} {154} (\bibinfo {year}
		{2019})}\BibitemShut {NoStop}%
	\bibitem [{\citenamefont {Sharma}\ \emph {et~al.}(2023)\citenamefont {Sharma},
		\citenamefont {Pu}, \citenamefont {Balram},\ and\ \citenamefont
		{Jain}}]{Sharma22}%
	\BibitemOpen
	\bibfield  {author} {\bibinfo {author} {\bibfnamefont {A.}~\bibnamefont
			{Sharma}}, \bibinfo {author} {\bibfnamefont {S.}~\bibnamefont {Pu}}, \bibinfo
		{author} {\bibfnamefont {A.~C.}\ \bibnamefont {Balram}},\ and\ \bibinfo
		{author} {\bibfnamefont {J.~K.}\ \bibnamefont {Jain}},\ }\bibfield  {title}
	{\bibinfo {title} {Fractional quantum {Hall} effect with unconventional
			pairing in monolayer graphene},\ }\href
	{https://doi.org/10.1103/PhysRevLett.130.126201} {\bibfield  {journal}
		{\bibinfo  {journal} {Phys. Rev. Lett.}\ }\textbf {\bibinfo {volume} {130}},\
		\bibinfo {pages} {126201} (\bibinfo {year} {2023})}\BibitemShut {NoStop}%
	\bibitem [{\citenamefont {Davenport}\ and\ \citenamefont
		{Simon}(2012)}]{Davenport12}%
	\BibitemOpen
	\bibfield  {author} {\bibinfo {author} {\bibfnamefont {S.~C.}\ \bibnamefont
			{Davenport}}\ and\ \bibinfo {author} {\bibfnamefont {S.~H.}\ \bibnamefont
			{Simon}},\ }\bibfield  {title} {\bibinfo {title} {Spinful composite fermions
			in a negative effective field},\ }\href
	{https://doi.org/10.1103/PhysRevB.85.245303} {\bibfield  {journal} {\bibinfo
			{journal} {Phys. Rev. B}\ }\textbf {\bibinfo {volume} {85}},\ \bibinfo
		{pages} {245303} (\bibinfo {year} {2012})}\BibitemShut {NoStop}%
	\bibitem [{\citenamefont {Balram}\ \emph
		{et~al.}(2015{\natexlab{b}})\citenamefont {Balram}, \citenamefont {T\"oke},
		\citenamefont {W\'ojs},\ and\ \citenamefont {Jain}}]{Balram15a}%
	\BibitemOpen
	\bibfield  {author} {\bibinfo {author} {\bibfnamefont {A.~C.}\ \bibnamefont
			{Balram}}, \bibinfo {author} {\bibfnamefont {C.}~\bibnamefont {T\"oke}},
		\bibinfo {author} {\bibfnamefont {A.}~\bibnamefont {W\'ojs}},\ and\ \bibinfo
		{author} {\bibfnamefont {J.~K.}\ \bibnamefont {Jain}},\ }\bibfield  {title}
	{\bibinfo {title} {Fractional quantum {Hall} effect in graphene: Quantitative
			comparison between theory and experiment},\ }\href
	{https://doi.org/10.1103/PhysRevB.92.075410} {\bibfield  {journal} {\bibinfo
			{journal} {Phys. Rev. B}\ }\textbf {\bibinfo {volume} {92}},\ \bibinfo
		{pages} {075410} (\bibinfo {year} {2015}{\natexlab{b}})}\BibitemShut
	{NoStop}%
	\bibitem [{SM_()}]{SM_LLM_Zhao22}%
	\BibitemOpen
	\href@noop {} {}\bibinfo {note} {See Supplemental Material which contains i)
		a review of the fixed-phase diffusion Monte Carlo method and ii) the
		extrapolation to the thermodynamic limit of the energies and the
		pair-correlation functions of various candidate states as a function of the
		Landau level mixing, a discussion on the mechanism and nature of pairing and
		includes Refs.~\cite{Ambrumenil88, Sitko96, Lee01, Lee02, Balram16c,
			Balram17, Balram17b, Balram20b, Zhao21b, Hossain21, Zhao22,
			Melik-Alaverdian99,Guclu05a,Wu76,Laughlin83,Grimm71,Ceperley79}.}\BibitemShut
	{Stop}%
	\bibitem [{\citenamefont {Reynolds}\ \emph {et~al.}(1982)\citenamefont
		{Reynolds}, \citenamefont {Ceperley}, \citenamefont {Alder},\ and\
		\citenamefont {Lester~Jr.}}]{Reynolds82}%
	\BibitemOpen
	\bibfield  {author} {\bibinfo {author} {\bibfnamefont {P.~J.}\ \bibnamefont
			{Reynolds}}, \bibinfo {author} {\bibfnamefont {D.~M.}\ \bibnamefont
			{Ceperley}}, \bibinfo {author} {\bibfnamefont {B.~J.}\ \bibnamefont
			{Alder}},\ and\ \bibinfo {author} {\bibfnamefont {W.~A.}\ \bibnamefont
			{Lester~Jr.}},\ }\bibfield  {title} {\bibinfo {title} {Fixed‐node quantum
			{Monte} {Carlo} for molecules},\ }\href
	{https://doi.org/http://dx.doi.org/10.1063/1.443766} {\bibfield  {journal}
		{\bibinfo  {journal} {J. Chem. Phys.}\ }\textbf {\bibinfo {volume} {77}},\
		\bibinfo {pages} {5593} (\bibinfo {year} {1982})}\BibitemShut {NoStop}%
	\bibitem [{\citenamefont {Foulkes}\ \emph {et~al.}(2001)\citenamefont
		{Foulkes}, \citenamefont {Mitas}, \citenamefont {Needs},\ and\ \citenamefont
		{Rajagopal}}]{Foulkes01}%
	\BibitemOpen
	\bibfield  {author} {\bibinfo {author} {\bibfnamefont {W.~M.~C.}\
			\bibnamefont {Foulkes}}, \bibinfo {author} {\bibfnamefont {L.}~\bibnamefont
			{Mitas}}, \bibinfo {author} {\bibfnamefont {R.~J.}\ \bibnamefont {Needs}},\
		and\ \bibinfo {author} {\bibfnamefont {G.}~\bibnamefont {Rajagopal}},\
	}\bibfield  {title} {\bibinfo {title} {Quantum {Monte} {Carlo} simulations of
			solids},\ }\href {https://doi.org/10.1103/RevModPhys.73.33} {\bibfield
		{journal} {\bibinfo  {journal} {Rev. Mod. Phys.}\ }\textbf {\bibinfo {volume}
			{73}},\ \bibinfo {pages} {33} (\bibinfo {year} {2001})}\BibitemShut {NoStop}%
	\bibitem [{\citenamefont {G\"u\c{c}l\"u}\ and\ \citenamefont
		{Umrigar}(2005)}]{Guclu05}%
	\BibitemOpen
	\bibfield  {author} {\bibinfo {author} {\bibfnamefont {A.~D.}\ \bibnamefont
			{G\"u\c{c}l\"u}}\ and\ \bibinfo {author} {\bibfnamefont {C.~J.}\ \bibnamefont
			{Umrigar}},\ }\bibfield  {title} {\bibinfo {title} {Maximum-density droplet
			to lower-density droplet transition in quantum dots},\ }\href
	{https://doi.org/10.1103/PhysRevB.72.045309} {\bibfield  {journal} {\bibinfo
			{journal} {Phys. Rev. B}\ }\textbf {\bibinfo {volume} {72}},\ \bibinfo
		{pages} {045309} (\bibinfo {year} {2005})}\BibitemShut {NoStop}%
	\bibitem [{\citenamefont {Morf}\ and\ \citenamefont {Halperin}(1987)}]{Morf87}%
	\BibitemOpen
	\bibfield  {author} {\bibinfo {author} {\bibfnamefont {R.}~\bibnamefont
			{Morf}}\ and\ \bibinfo {author} {\bibfnamefont {B.~I.}\ \bibnamefont
			{Halperin}},\ }\bibfield  {title} {\bibinfo {title} {{Monte} {Carlo}
			evaluation of trial wavefunctions for the fractional quantized {Hall} effect:
			Spherical geometry},\ }\href {https://doi.org/10.1007/BF01304256} {\bibfield
		{journal} {\bibinfo  {journal} {Zeitschrift f{\"u}r Physik B Condensed
				Matter}\ }\textbf {\bibinfo {volume} {68}},\ \bibinfo {pages} {391} (\bibinfo
		{year} {1987})}\BibitemShut {NoStop}%
	\bibitem [{\citenamefont {Halperin}\ and\ \citenamefont
		{Jain}(2020)}]{Halperin20}%
	\BibitemOpen
	\bibinfo {editor} {\bibfnamefont {B.~I.}\ \bibnamefont {Halperin}}\ and\
	\bibinfo {editor} {\bibfnamefont {J.~K.}\ \bibnamefont {Jain}},\ eds.,\ \href
	{https://doi.org/10.1142/11751} {\emph {\bibinfo {title} {{Fractional}
				{Quantum} {Hall} {Effects} {New} {Developments}}}}\ (\bibinfo  {publisher}
	{World Scientific},\ \bibinfo {year} {2020})\ \Eprint
	{https://arxiv.org/abs/https://worldscientific.com/doi/pdf/10.1142/11751}
	{https://worldscientific.com/doi/pdf/10.1142/11751} \BibitemShut {NoStop}%
	\bibitem [{\citenamefont {Balram}\ \emph
		{et~al.}(2015{\natexlab{c}})\citenamefont {Balram}, \citenamefont
		{T\ifmmode~\mbox{\H{o}}\else \H{o}\fi{}ke}, \citenamefont {W\'ojs},\ and\
		\citenamefont {Jain}}]{Balram15c}%
	\BibitemOpen
	\bibfield  {author} {\bibinfo {author} {\bibfnamefont {A.~C.}\ \bibnamefont
			{Balram}}, \bibinfo {author} {\bibfnamefont {C.}~\bibnamefont
			{T\ifmmode~\mbox{\H{o}}\else \H{o}\fi{}ke}}, \bibinfo {author} {\bibfnamefont
			{A.}~\bibnamefont {W\'ojs}},\ and\ \bibinfo {author} {\bibfnamefont {J.~K.}\
			\bibnamefont {Jain}},\ }\bibfield  {title} {\bibinfo {title} {Spontaneous
			polarization of composite fermions in the $n=1$ {Landau} level of graphene},\
	}\href {https://doi.org/10.1103/PhysRevB.92.205120} {\bibfield  {journal}
		{\bibinfo  {journal} {Phys. Rev. B}\ }\textbf {\bibinfo {volume} {92}},\
		\bibinfo {pages} {205120} (\bibinfo {year} {2015}{\natexlab{c}})}\BibitemShut
	{NoStop}%
	\bibitem [{\citenamefont {Kamilla}\ \emph {et~al.}(1997)\citenamefont
		{Kamilla}, \citenamefont {Jain},\ and\ \citenamefont {Girvin}}]{Kamilla97}%
	\BibitemOpen
	\bibfield  {author} {\bibinfo {author} {\bibfnamefont {R.~K.}\ \bibnamefont
			{Kamilla}}, \bibinfo {author} {\bibfnamefont {J.~K.}\ \bibnamefont {Jain}},\
		and\ \bibinfo {author} {\bibfnamefont {S.~M.}\ \bibnamefont {Girvin}},\
	}\bibfield  {title} {\bibinfo {title} {Fermi-sea-like correlations in a
			partially filled {Landau} level},\ }\href
	{https://doi.org/10.1103/PhysRevB.56.12411} {\bibfield  {journal} {\bibinfo
			{journal} {Phys. Rev. B}\ }\textbf {\bibinfo {volume} {56}},\ \bibinfo
		{pages} {12411} (\bibinfo {year} {1997})}\BibitemShut {NoStop}%
	\bibitem [{\citenamefont {Suen}\ \emph
		{et~al.}(1992{\natexlab{a}})\citenamefont {Suen}, \citenamefont {Engel},
		\citenamefont {Santos}, \citenamefont {Shayegan},\ and\ \citenamefont
		{Tsui}}]{Suen92}%
	\BibitemOpen
	\bibfield  {author} {\bibinfo {author} {\bibfnamefont {Y.~W.}\ \bibnamefont
			{Suen}}, \bibinfo {author} {\bibfnamefont {L.~W.}\ \bibnamefont {Engel}},
		\bibinfo {author} {\bibfnamefont {M.~B.}\ \bibnamefont {Santos}}, \bibinfo
		{author} {\bibfnamefont {M.}~\bibnamefont {Shayegan}},\ and\ \bibinfo
		{author} {\bibfnamefont {D.~C.}\ \bibnamefont {Tsui}},\ }\bibfield  {title}
	{\bibinfo {title} {Observation of a $\nu=1/2$ fractional quantum {Hall} state
			in a double-layer electron system},\ }\href
	{https://doi.org/10.1103/PhysRevLett.68.1379} {\bibfield  {journal} {\bibinfo
			{journal} {Phys. Rev. Lett.}\ }\textbf {\bibinfo {volume} {68}},\ \bibinfo
		{pages} {1379} (\bibinfo {year} {1992}{\natexlab{a}})}\BibitemShut {NoStop}%
	\bibitem [{\citenamefont {Suen}\ \emph
		{et~al.}(1992{\natexlab{b}})\citenamefont {Suen}, \citenamefont {Santos},\
		and\ \citenamefont {Shayegan}}]{Suen92b}%
	\BibitemOpen
	\bibfield  {author} {\bibinfo {author} {\bibfnamefont {Y.~W.}\ \bibnamefont
			{Suen}}, \bibinfo {author} {\bibfnamefont {M.~B.}\ \bibnamefont {Santos}},\
		and\ \bibinfo {author} {\bibfnamefont {M.}~\bibnamefont {Shayegan}},\
	}\bibfield  {title} {\bibinfo {title} {Correlated states of an electron
			system in a wide quantum well},\ }\href
	{https://doi.org/10.1103/PhysRevLett.69.3551} {\bibfield  {journal} {\bibinfo
			{journal} {Phys. Rev. Lett.}\ }\textbf {\bibinfo {volume} {69}},\ \bibinfo
		{pages} {3551} (\bibinfo {year} {1992}{\natexlab{b}})}\BibitemShut {NoStop}%
	\bibitem [{\citenamefont {Liu}\ \emph {et~al.}(2014{\natexlab{a}})\citenamefont
		{Liu}, \citenamefont {Hasdemir}, \citenamefont {Kamburov}, \citenamefont
		{Graninger}, \citenamefont {Shayegan}, \citenamefont {Pfeiffer},
		\citenamefont {West}, \citenamefont {Baldwin},\ and\ \citenamefont
		{Winkler}}]{Liu14b}%
	\BibitemOpen
	\bibfield  {author} {\bibinfo {author} {\bibfnamefont {Y.}~\bibnamefont
			{Liu}}, \bibinfo {author} {\bibfnamefont {S.}~\bibnamefont {Hasdemir}},
		\bibinfo {author} {\bibfnamefont {D.}~\bibnamefont {Kamburov}}, \bibinfo
		{author} {\bibfnamefont {A.~L.}\ \bibnamefont {Graninger}}, \bibinfo {author}
		{\bibfnamefont {M.}~\bibnamefont {Shayegan}}, \bibinfo {author}
		{\bibfnamefont {L.~N.}\ \bibnamefont {Pfeiffer}}, \bibinfo {author}
		{\bibfnamefont {K.~W.}\ \bibnamefont {West}}, \bibinfo {author}
		{\bibfnamefont {K.~W.}\ \bibnamefont {Baldwin}},\ and\ \bibinfo {author}
		{\bibfnamefont {R.}~\bibnamefont {Winkler}},\ }\bibfield  {title} {\bibinfo
		{title} {Even-denominator fractional quantum {Hall} effect at a {Landau}
			level crossing},\ }\href {https://doi.org/10.1103/PhysRevB.89.165313}
	{\bibfield  {journal} {\bibinfo  {journal} {Phys. Rev. B}\ }\textbf {\bibinfo
			{volume} {89}},\ \bibinfo {pages} {165313} (\bibinfo {year}
		{2014}{\natexlab{a}})}\BibitemShut {NoStop}%
	\bibitem [{\citenamefont {Liu}\ \emph {et~al.}(2014{\natexlab{b}})\citenamefont
		{Liu}, \citenamefont {Graninger}, \citenamefont {Hasdemir}, \citenamefont
		{Shayegan}, \citenamefont {Pfeiffer}, \citenamefont {West}, \citenamefont
		{Baldwin},\ and\ \citenamefont {Winkler}}]{Liu14d}%
	\BibitemOpen
	\bibfield  {author} {\bibinfo {author} {\bibfnamefont {Y.}~\bibnamefont
			{Liu}}, \bibinfo {author} {\bibfnamefont {A.~L.}\ \bibnamefont {Graninger}},
		\bibinfo {author} {\bibfnamefont {S.}~\bibnamefont {Hasdemir}}, \bibinfo
		{author} {\bibfnamefont {M.}~\bibnamefont {Shayegan}}, \bibinfo {author}
		{\bibfnamefont {L.~N.}\ \bibnamefont {Pfeiffer}}, \bibinfo {author}
		{\bibfnamefont {K.~W.}\ \bibnamefont {West}}, \bibinfo {author}
		{\bibfnamefont {K.~W.}\ \bibnamefont {Baldwin}},\ and\ \bibinfo {author}
		{\bibfnamefont {R.}~\bibnamefont {Winkler}},\ }\bibfield  {title} {\bibinfo
		{title} {Fractional quantum {Hall} effect at $\ensuremath{\nu}=1/2$ in hole
			systems confined to {GaAs} quantum wells},\ }\href
	{https://doi.org/10.1103/PhysRevLett.112.046804} {\bibfield  {journal}
		{\bibinfo  {journal} {Phys. Rev. Lett.}\ }\textbf {\bibinfo {volume} {112}},\
		\bibinfo {pages} {046804} (\bibinfo {year} {2014}{\natexlab{b}})}\BibitemShut
	{NoStop}%
	\bibitem [{\citenamefont {Mueed}\ \emph {et~al.}(2015)\citenamefont {Mueed},
		\citenamefont {Kamburov}, \citenamefont {Hasdemir}, \citenamefont {Shayegan},
		\citenamefont {Pfeiffer}, \citenamefont {West},\ and\ \citenamefont
		{Baldwin}}]{Mueed15}%
	\BibitemOpen
	\bibfield  {author} {\bibinfo {author} {\bibfnamefont {M.~A.}\ \bibnamefont
			{Mueed}}, \bibinfo {author} {\bibfnamefont {D.}~\bibnamefont {Kamburov}},
		\bibinfo {author} {\bibfnamefont {S.}~\bibnamefont {Hasdemir}}, \bibinfo
		{author} {\bibfnamefont {M.}~\bibnamefont {Shayegan}}, \bibinfo {author}
		{\bibfnamefont {L.~N.}\ \bibnamefont {Pfeiffer}}, \bibinfo {author}
		{\bibfnamefont {K.~W.}\ \bibnamefont {West}},\ and\ \bibinfo {author}
		{\bibfnamefont {K.~W.}\ \bibnamefont {Baldwin}},\ }\bibfield  {title}
	{\bibinfo {title} {Geometric resonance of composite fermions near the
			$\nu=1/2$ fractional quantum {Hall} state},\ }\href
	{https://doi.org/10.1103/PhysRevLett.114.236406} {\bibfield  {journal}
		{\bibinfo  {journal} {Phys. Rev. Lett.}\ }\textbf {\bibinfo {volume} {114}},\
		\bibinfo {pages} {236406} (\bibinfo {year} {2015})}\BibitemShut {NoStop}%
	\bibitem [{\citenamefont {Halperin}(1984)}]{Halperin84}%
	\BibitemOpen
	\bibfield  {author} {\bibinfo {author} {\bibfnamefont {B.~I.}\ \bibnamefont
			{Halperin}},\ }\bibfield  {title} {\bibinfo {title} {Statistics of
			quasiparticles and the hierarchy of fractional quantized {Hall} states},\
	}\href {https://doi.org/10.1103/PhysRevLett.52.1583} {\bibfield  {journal}
		{\bibinfo  {journal} {Phys. Rev. Lett.}\ }\textbf {\bibinfo {volume} {52}},\
		\bibinfo {pages} {1583} (\bibinfo {year} {1984})}\BibitemShut {NoStop}%
	\bibitem [{\citenamefont {He}\ \emph {et~al.}(1993)\citenamefont {He},
		\citenamefont {Das~Sarma},\ and\ \citenamefont {Xie}}]{He93}%
	\BibitemOpen
	\bibfield  {author} {\bibinfo {author} {\bibfnamefont {S.}~\bibnamefont
			{He}}, \bibinfo {author} {\bibfnamefont {S.}~\bibnamefont {Das~Sarma}},\ and\
		\bibinfo {author} {\bibfnamefont {X.~C.}\ \bibnamefont {Xie}},\ }\bibfield
	{title} {\bibinfo {title} {Quantized {Hall} effect and quantum phase
			transitions in coupled two-layer electron systems},\ }\href
	{https://doi.org/10.1103/PhysRevB.47.4394} {\bibfield  {journal} {\bibinfo
			{journal} {Phys. Rev. B}\ }\textbf {\bibinfo {volume} {47}},\ \bibinfo
		{pages} {4394} (\bibinfo {year} {1993})}\BibitemShut {NoStop}%
	\bibitem [{\citenamefont {Peterson}\ \emph {et~al.}(2010)\citenamefont
		{Peterson}, \citenamefont {Papi{\'c}},\ and\ \citenamefont
		{Sarma}}]{Peterson10}%
	\BibitemOpen
	\bibfield  {author} {\bibinfo {author} {\bibfnamefont {M.~R.}\ \bibnamefont
			{Peterson}}, \bibinfo {author} {\bibfnamefont {Z.}~\bibnamefont
			{Papi{\'c}}},\ and\ \bibinfo {author} {\bibfnamefont {S.~D.}\ \bibnamefont
			{Sarma}},\ }\bibfield  {title} {\bibinfo {title} {Fractional quantum {Hall}
			effects in bilayers in the presence of interlayer tunneling and charge
			imbalance},\ }\href@noop {} {\bibfield  {journal} {\bibinfo  {journal}
			{Physical Review B}\ }\textbf {\bibinfo {volume} {82}},\ \bibinfo {pages}
		{235312} (\bibinfo {year} {2010})}\BibitemShut {NoStop}%
	\bibitem [{\citenamefont {Zhu}\ \emph {et~al.}(2016)\citenamefont {Zhu},
		\citenamefont {Liu}, \citenamefont {Haldane},\ and\ \citenamefont
		{Sheng}}]{Zhu16}%
	\BibitemOpen
	\bibfield  {author} {\bibinfo {author} {\bibfnamefont {W.}~\bibnamefont
			{Zhu}}, \bibinfo {author} {\bibfnamefont {Z.}~\bibnamefont {Liu}}, \bibinfo
		{author} {\bibfnamefont {F.~D.~M.}\ \bibnamefont {Haldane}},\ and\ \bibinfo
		{author} {\bibfnamefont {D.~N.}\ \bibnamefont {Sheng}},\ }\bibfield  {title}
	{\bibinfo {title} {Fractional quantum {Hall} bilayers at half filling:
			Tunneling-driven non-abelian phase},\ }\href
	{https://doi.org/10.1103/PhysRevB.94.245147} {\bibfield  {journal} {\bibinfo
			{journal} {Phys. Rev. B}\ }\textbf {\bibinfo {volume} {94}},\ \bibinfo
		{pages} {245147} (\bibinfo {year} {2016})}\BibitemShut {NoStop}%
	\bibitem [{\citenamefont {Zhao}\ \emph {et~al.}(2021)\citenamefont {Zhao},
		\citenamefont {Faugno}, \citenamefont {Pu}, \citenamefont {Balram},\ and\
		\citenamefont {Jain}}]{Zhao21}%
	\BibitemOpen
	\bibfield  {author} {\bibinfo {author} {\bibfnamefont {T.}~\bibnamefont
			{Zhao}}, \bibinfo {author} {\bibfnamefont {W.~N.}\ \bibnamefont {Faugno}},
		\bibinfo {author} {\bibfnamefont {S.}~\bibnamefont {Pu}}, \bibinfo {author}
		{\bibfnamefont {A.~C.}\ \bibnamefont {Balram}},\ and\ \bibinfo {author}
		{\bibfnamefont {J.~K.}\ \bibnamefont {Jain}},\ }\bibfield  {title} {\bibinfo
		{title} {Origin of the $\ensuremath{\nu}=1/2$ fractional quantum {Hall}
			effect in wide quantum wells},\ }\href
	{https://doi.org/10.1103/PhysRevB.103.155306} {\bibfield  {journal} {\bibinfo
			{journal} {Phys. Rev. B}\ }\textbf {\bibinfo {volume} {103}},\ \bibinfo
		{pages} {155306} (\bibinfo {year} {2021})}\BibitemShut {NoStop}%
	\bibitem [{\citenamefont {Dutta}\ \emph {et~al.}(2022)\citenamefont {Dutta},
		\citenamefont {Yang}, \citenamefont {Melcer}, \citenamefont {Kundu},
		\citenamefont {Heiblum}, \citenamefont {Umansky}, \citenamefont {Oreg},
		\citenamefont {Stern},\ and\ \citenamefont {Mross}}]{Dutta22}%
	\BibitemOpen
	\bibfield  {author} {\bibinfo {author} {\bibfnamefont {B.}~\bibnamefont
			{Dutta}}, \bibinfo {author} {\bibfnamefont {W.}~\bibnamefont {Yang}},
		\bibinfo {author} {\bibfnamefont {R.}~\bibnamefont {Melcer}}, \bibinfo
		{author} {\bibfnamefont {H.~K.}\ \bibnamefont {Kundu}}, \bibinfo {author}
		{\bibfnamefont {M.}~\bibnamefont {Heiblum}}, \bibinfo {author} {\bibfnamefont
			{V.}~\bibnamefont {Umansky}}, \bibinfo {author} {\bibfnamefont
			{Y.}~\bibnamefont {Oreg}}, \bibinfo {author} {\bibfnamefont {A.}~\bibnamefont
			{Stern}},\ and\ \bibinfo {author} {\bibfnamefont {D.}~\bibnamefont {Mross}},\
	}\bibfield  {title} {\bibinfo {title} {Distinguishing between non-abelian
			topological orders in a quantum {Hall} system},\ }\href
	{https://doi.org/10.1126/science.abg6116} {\bibfield  {journal} {\bibinfo
			{journal} {Science}\ }\textbf {\bibinfo {volume} {0}},\ \bibinfo {pages}
		{eabg6116} (\bibinfo {year} {2022})},\ \Eprint
	{https://arxiv.org/abs/https://www.science.org/doi/pdf/10.1126/science.abg6116}
	{https://www.science.org/doi/pdf/10.1126/science.abg6116} \BibitemShut
	{NoStop}%
	\bibitem [{\citenamefont {Banerjee}\ \emph {et~al.}(2018)\citenamefont
		{Banerjee}, \citenamefont {Heiblum}, \citenamefont {Umansky}, \citenamefont
		{Feldman}, \citenamefont {Oreg},\ and\ \citenamefont {Stern}}]{Banerjee18}%
	\BibitemOpen
	\bibfield  {author} {\bibinfo {author} {\bibfnamefont {M.}~\bibnamefont
			{Banerjee}}, \bibinfo {author} {\bibfnamefont {M.}~\bibnamefont {Heiblum}},
		\bibinfo {author} {\bibfnamefont {V.}~\bibnamefont {Umansky}}, \bibinfo
		{author} {\bibfnamefont {D.~E.}\ \bibnamefont {Feldman}}, \bibinfo {author}
		{\bibfnamefont {Y.}~\bibnamefont {Oreg}},\ and\ \bibinfo {author}
		{\bibfnamefont {A.}~\bibnamefont {Stern}},\ }\bibfield  {title} {\bibinfo
		{title} {Observation of half-integer thermal hall conductance},\ }\href@noop
	{} {\bibfield  {journal} {\bibinfo  {journal} {Nature}\ }\textbf {\bibinfo
			{volume} {559}},\ \bibinfo {pages} {205} (\bibinfo {year}
		{2018})}\BibitemShut {NoStop}%
	\bibitem [{\citenamefont {Villegas~Rosales}\ \emph
		{et~al.}(2021{\natexlab{b}})\citenamefont {Villegas~Rosales}, \citenamefont
		{Singh}, \citenamefont {Ma}, \citenamefont {Hossain}, \citenamefont {Chung},
		\citenamefont {Pfeiffer}, \citenamefont {West}, \citenamefont {Baldwin},\
		and\ \citenamefont {Shayegan}}]{Rosales21b}%
	\BibitemOpen
	\bibfield  {author} {\bibinfo {author} {\bibfnamefont {K.~A.}\ \bibnamefont
			{Villegas~Rosales}}, \bibinfo {author} {\bibfnamefont {S.~K.}\ \bibnamefont
			{Singh}}, \bibinfo {author} {\bibfnamefont {M.~K.}\ \bibnamefont {Ma}},
		\bibinfo {author} {\bibfnamefont {M.~S.}\ \bibnamefont {Hossain}}, \bibinfo
		{author} {\bibfnamefont {Y.~J.}\ \bibnamefont {Chung}}, \bibinfo {author}
		{\bibfnamefont {L.~N.}\ \bibnamefont {Pfeiffer}}, \bibinfo {author}
		{\bibfnamefont {K.~W.}\ \bibnamefont {West}}, \bibinfo {author}
		{\bibfnamefont {K.~W.}\ \bibnamefont {Baldwin}},\ and\ \bibinfo {author}
		{\bibfnamefont {M.}~\bibnamefont {Shayegan}},\ }\bibfield  {title} {\bibinfo
		{title} {Competition between fractional quantum hall liquid and wigner solid
			at small fillings: Role of layer thickness and landau level mixing},\ }\href
	{https://doi.org/10.1103/PhysRevResearch.3.013181} {\bibfield  {journal}
		{\bibinfo  {journal} {Phys. Rev. Research}\ }\textbf {\bibinfo {volume}
			{3}},\ \bibinfo {pages} {013181} (\bibinfo {year}
		{2021}{\natexlab{b}})}\BibitemShut {NoStop}%
	\bibitem [{\citenamefont {Sajoto}\ \emph {et~al.}(1990)\citenamefont {Sajoto},
		\citenamefont {Suen}, \citenamefont {Engel}, \citenamefont {Santos},\ and\
		\citenamefont {Shayegan}}]{Sajoto90}%
	\BibitemOpen
	\bibfield  {author} {\bibinfo {author} {\bibfnamefont {T.}~\bibnamefont
			{Sajoto}}, \bibinfo {author} {\bibfnamefont {Y.~W.}\ \bibnamefont {Suen}},
		\bibinfo {author} {\bibfnamefont {L.~W.}\ \bibnamefont {Engel}}, \bibinfo
		{author} {\bibfnamefont {M.~B.}\ \bibnamefont {Santos}},\ and\ \bibinfo
		{author} {\bibfnamefont {M.}~\bibnamefont {Shayegan}},\ }\bibfield  {title}
	{\bibinfo {title} {Fractional quantum {Hall} effect in very-low-density
			{Ga}{As}/${\mathrm{al}}_{\mathit{x}}$${\mathrm{ga}}_{1\mathrm{\ensuremath{-}}\mathit{x}}$as
			heterostructures},\ }\href {https://doi.org/10.1103/PhysRevB.41.8449}
	{\bibfield  {journal} {\bibinfo  {journal} {Phys. Rev. B}\ }\textbf {\bibinfo
			{volume} {41}},\ \bibinfo {pages} {8449} (\bibinfo {year}
		{1990})}\BibitemShut {NoStop}%
	\bibitem [{\citenamefont {Chung}\ \emph {et~al.}(2017)\citenamefont {Chung},
		\citenamefont {Baldwin}, \citenamefont {West}, \citenamefont {Kamburov},
		\citenamefont {Shayegan},\ and\ \citenamefont {Pfeiffer}}]{Chung17}%
	\BibitemOpen
	\bibfield  {author} {\bibinfo {author} {\bibfnamefont {Y.~J.}\ \bibnamefont
			{Chung}}, \bibinfo {author} {\bibfnamefont {K.~W.}\ \bibnamefont {Baldwin}},
		\bibinfo {author} {\bibfnamefont {K.~W.}\ \bibnamefont {West}}, \bibinfo
		{author} {\bibfnamefont {D.}~\bibnamefont {Kamburov}}, \bibinfo {author}
		{\bibfnamefont {M.}~\bibnamefont {Shayegan}},\ and\ \bibinfo {author}
		{\bibfnamefont {L.~N.}\ \bibnamefont {Pfeiffer}},\ }\bibfield  {title}
	{\bibinfo {title} {Design rules for modulation-doped alas quantum wells},\
	}\href {https://doi.org/10.1103/PhysRevMaterials.1.021002} {\bibfield
		{journal} {\bibinfo  {journal} {Phys. Rev. Materials}\ }\textbf {\bibinfo
			{volume} {1}},\ \bibinfo {pages} {021002} (\bibinfo {year}
		{2017})}\BibitemShut {NoStop}%
	\bibitem [{\citenamefont {Villegas~Rosales}\ \emph
		{et~al.}(2021{\natexlab{c}})\citenamefont {Villegas~Rosales}, \citenamefont
		{Singh}, \citenamefont {Ma}, \citenamefont {Hossain}, \citenamefont {Chung},
		\citenamefont {Pfeiffer}, \citenamefont {West}, \citenamefont {Baldwin},\
		and\ \citenamefont {Shayegan}}]{Ma21}%
	\BibitemOpen
	\bibfield  {author} {\bibinfo {author} {\bibfnamefont {K.~A.}\ \bibnamefont
			{Villegas~Rosales}}, \bibinfo {author} {\bibfnamefont {S.~K.}\ \bibnamefont
			{Singh}}, \bibinfo {author} {\bibfnamefont {M.~K.}\ \bibnamefont {Ma}},
		\bibinfo {author} {\bibfnamefont {M.~S.}\ \bibnamefont {Hossain}}, \bibinfo
		{author} {\bibfnamefont {Y.~J.}\ \bibnamefont {Chung}}, \bibinfo {author}
		{\bibfnamefont {L.~N.}\ \bibnamefont {Pfeiffer}}, \bibinfo {author}
		{\bibfnamefont {K.~W.}\ \bibnamefont {West}}, \bibinfo {author}
		{\bibfnamefont {K.~W.}\ \bibnamefont {Baldwin}},\ and\ \bibinfo {author}
		{\bibfnamefont {M.}~\bibnamefont {Shayegan}},\ }\bibfield  {title} {\bibinfo
		{title} {Competition between fractional quantum hall liquid and wigner solid
			at small fillings: Role of layer thickness and landau level mixing},\ }\href
	{https://doi.org/10.1103/PhysRevResearch.3.013181} {\bibfield  {journal}
		{\bibinfo  {journal} {Phys. Rev. Research}\ }\textbf {\bibinfo {volume}
			{3}},\ \bibinfo {pages} {013181} (\bibinfo {year}
		{2021}{\natexlab{c}})}\BibitemShut {NoStop}%
	\bibitem [{\citenamefont {d'Ambrumenil}\ and\ \citenamefont
		{Reynolds}(1988)}]{Ambrumenil88}%
	\BibitemOpen
	\bibfield  {author} {\bibinfo {author} {\bibfnamefont {N.}~\bibnamefont
			{d'Ambrumenil}}\ and\ \bibinfo {author} {\bibfnamefont {A.~M.}\ \bibnamefont
			{Reynolds}},\ }\bibfield  {title} {\bibinfo {title} {Fractional quantum
			{Hall} states in higher {Landau} levels},\ }\href
	{http://stacks.iop.org/0022-3719/21/i=1/a=010} {\bibfield  {journal}
		{\bibinfo  {journal} {Journal of Physics C: Solid State Physics}\ }\textbf
		{\bibinfo {volume} {21}},\ \bibinfo {pages} {119} (\bibinfo {year}
		{1988})}\BibitemShut {NoStop}%
	\bibitem [{\citenamefont {Sitko}\ \emph {et~al.}(1996)\citenamefont {Sitko},
		\citenamefont {Yi}, \citenamefont {Yi},\ and\ \citenamefont
		{Quinn}}]{Sitko96}%
	\BibitemOpen
	\bibfield  {author} {\bibinfo {author} {\bibfnamefont {P.}~\bibnamefont
			{Sitko}}, \bibinfo {author} {\bibfnamefont {S.~N.}\ \bibnamefont {Yi}},
		\bibinfo {author} {\bibfnamefont {K.~S.}\ \bibnamefont {Yi}},\ and\ \bibinfo
		{author} {\bibfnamefont {J.~J.}\ \bibnamefont {Quinn}},\ }\bibfield  {title}
	{\bibinfo {title} {"fermi liquid" shell model approach to composite fermion
			excitation spectra in fractional quantum {Hall} states},\ }\href
	{https://doi.org/10.1103/PhysRevLett.76.3396} {\bibfield  {journal} {\bibinfo
			{journal} {Phys. Rev. Lett.}\ }\textbf {\bibinfo {volume} {76}},\ \bibinfo
		{pages} {3396} (\bibinfo {year} {1996})}\BibitemShut {NoStop}%
	\bibitem [{\citenamefont {Lee}\ \emph {et~al.}(2001)\citenamefont {Lee},
		\citenamefont {Scarola},\ and\ \citenamefont {Jain}}]{Lee01}%
	\BibitemOpen
	\bibfield  {author} {\bibinfo {author} {\bibfnamefont {S.-Y.}\ \bibnamefont
			{Lee}}, \bibinfo {author} {\bibfnamefont {V.~W.}\ \bibnamefont {Scarola}},\
		and\ \bibinfo {author} {\bibfnamefont {J.~K.}\ \bibnamefont {Jain}},\
	}\bibfield  {title} {\bibinfo {title} {Stripe formation in the fractional
			quantum {Hall} regime},\ }\href
	{https://doi.org/10.1103/PhysRevLett.87.256803} {\bibfield  {journal}
		{\bibinfo  {journal} {Phys. Rev. Lett.}\ }\textbf {\bibinfo {volume} {87}},\
		\bibinfo {pages} {256803} (\bibinfo {year} {2001})}\BibitemShut {NoStop}%
	\bibitem [{\citenamefont {Lee}\ \emph {et~al.}(2002)\citenamefont {Lee},
		\citenamefont {Scarola},\ and\ \citenamefont {Jain}}]{Lee02}%
	\BibitemOpen
	\bibfield  {author} {\bibinfo {author} {\bibfnamefont {S.-Y.}\ \bibnamefont
			{Lee}}, \bibinfo {author} {\bibfnamefont {V.~W.}\ \bibnamefont {Scarola}},\
		and\ \bibinfo {author} {\bibfnamefont {J.~K.}\ \bibnamefont {Jain}},\
	}\bibfield  {title} {\bibinfo {title} {Structures for interacting composite
			fermions: Stripes, bubbles, and fractional quantum {Hall} effect},\ }\href
	{https://doi.org/10.1103/PhysRevB.66.085336} {\bibfield  {journal} {\bibinfo
			{journal} {Phys. Rev. B}\ }\textbf {\bibinfo {volume} {66}},\ \bibinfo
		{pages} {085336} (\bibinfo {year} {2002})}\BibitemShut {NoStop}%
	\bibitem [{\citenamefont {Balram}(2016)}]{Balram16c}%
	\BibitemOpen
	\bibfield  {author} {\bibinfo {author} {\bibfnamefont {A.~C.}\ \bibnamefont
			{Balram}},\ }\bibfield  {title} {\bibinfo {title} {Interacting composite
			fermions: Nature of the 4/5, 5/7, 6/7, and 6/17 fractional quantum {Hall}
			states},\ }\href {https://doi.org/10.1103/PhysRevB.94.165303} {\bibfield
		{journal} {\bibinfo  {journal} {Phys. Rev. B}\ }\textbf {\bibinfo {volume}
			{94}},\ \bibinfo {pages} {165303} (\bibinfo {year} {2016})}\BibitemShut
	{NoStop}%
	\bibitem [{\citenamefont {Balram}\ and\ \citenamefont
		{Jain}(2017{\natexlab{a}})}]{Balram17}%
	\BibitemOpen
	\bibfield  {author} {\bibinfo {author} {\bibfnamefont {A.~C.}\ \bibnamefont
			{Balram}}\ and\ \bibinfo {author} {\bibfnamefont {J.~K.}\ \bibnamefont
			{Jain}},\ }\bibfield  {title} {\bibinfo {title} {Fermi wave vector for the
			partially spin-polarized composite-fermion {Fermi} sea},\ }\href
	{https://doi.org/10.1103/PhysRevB.96.235102} {\bibfield  {journal} {\bibinfo
			{journal} {Phys. Rev. B}\ }\textbf {\bibinfo {volume} {96}},\ \bibinfo
		{pages} {235102} (\bibinfo {year} {2017}{\natexlab{a}})}\BibitemShut
	{NoStop}%
	\bibitem [{\citenamefont {Balram}\ and\ \citenamefont
		{Jain}(2017{\natexlab{b}})}]{Balram17b}%
	\BibitemOpen
	\bibfield  {author} {\bibinfo {author} {\bibfnamefont {A.~C.}\ \bibnamefont
			{Balram}}\ and\ \bibinfo {author} {\bibfnamefont {J.~K.}\ \bibnamefont
			{Jain}},\ }\bibfield  {title} {\bibinfo {title} {Particle-hole symmetry for
			composite fermions: An emergent symmetry in the fractional quantum {Hall}
			effect},\ }\href {https://doi.org/10.1103/PhysRevB.96.245142} {\bibfield
		{journal} {\bibinfo  {journal} {Phys. Rev. B}\ }\textbf {\bibinfo {volume}
			{96}},\ \bibinfo {pages} {245142} (\bibinfo {year}
		{2017}{\natexlab{b}})}\BibitemShut {NoStop}%
	\bibitem [{\citenamefont {Balram}\ and\ \citenamefont
		{W\'ojs}(2020)}]{Balram20b}%
	\BibitemOpen
	\bibfield  {author} {\bibinfo {author} {\bibfnamefont {A.~C.}\ \bibnamefont
			{Balram}}\ and\ \bibinfo {author} {\bibfnamefont {A.}~\bibnamefont
			{W\'ojs}},\ }\bibfield  {title} {\bibinfo {title} {Fractional quantum {Hall}
			effect at $\ensuremath{\nu}=2+4/9$},\ }\href
	{https://doi.org/10.1103/PhysRevResearch.2.032035} {\bibfield  {journal}
		{\bibinfo  {journal} {Phys. Rev. Research}\ }\textbf {\bibinfo {volume}
			{2}},\ \bibinfo {pages} {032035} (\bibinfo {year} {2020})}\BibitemShut
	{NoStop}%
	\bibitem [{\citenamefont {Zhao}(2021)}]{Zhao21b}%
	\BibitemOpen
	\bibfield  {author} {\bibinfo {author} {\bibfnamefont {T.}~\bibnamefont
			{Zhao}},\ }\bibfield  {title} {\bibinfo {title} {Fixed-phase diffusion monte
			carlo study of activation gap and skyrmion excitations of a
			$\ensuremath{\nu}=1$ system in the presence of charged impurities},\ }\href
	{https://doi.org/10.1103/PhysRevB.104.115303} {\bibfield  {journal} {\bibinfo
			{journal} {Phys. Rev. B}\ }\textbf {\bibinfo {volume} {104}},\ \bibinfo
		{pages} {115303} (\bibinfo {year} {2021})}\BibitemShut {NoStop}%
	\bibitem [{\citenamefont {Hossain}\ \emph {et~al.}(2021)\citenamefont
		{Hossain}, \citenamefont {Zhao}, \citenamefont {Pu}, \citenamefont {Mueed},
		\citenamefont {Ma}, \citenamefont {Rosales}, \citenamefont {Chung},
		\citenamefont {Pfeiffer}, \citenamefont {West}, \citenamefont {Baldwin} \emph
		{et~al.}}]{Hossain21}%
	\BibitemOpen
	\bibfield  {author} {\bibinfo {author} {\bibfnamefont {M.~S.}\ \bibnamefont
			{Hossain}}, \bibinfo {author} {\bibfnamefont {T.}~\bibnamefont {Zhao}},
		\bibinfo {author} {\bibfnamefont {S.}~\bibnamefont {Pu}}, \bibinfo {author}
		{\bibfnamefont {M.}~\bibnamefont {Mueed}}, \bibinfo {author} {\bibfnamefont
			{M.}~\bibnamefont {Ma}}, \bibinfo {author} {\bibfnamefont {K.~V.}\
			\bibnamefont {Rosales}}, \bibinfo {author} {\bibfnamefont {Y.}~\bibnamefont
			{Chung}}, \bibinfo {author} {\bibfnamefont {L.}~\bibnamefont {Pfeiffer}},
		\bibinfo {author} {\bibfnamefont {K.}~\bibnamefont {West}}, \bibinfo {author}
		{\bibfnamefont {K.}~\bibnamefont {Baldwin}}, \emph {et~al.},\ }\bibfield
	{title} {\bibinfo {title} {Bloch ferromagnetism of composite fermions},\
	}\href@noop {} {\bibfield  {journal} {\bibinfo  {journal} {Nature Physics}\
		}\textbf {\bibinfo {volume} {17}},\ \bibinfo {pages} {48} (\bibinfo {year}
		{2021})}\BibitemShut {NoStop}%
	\bibitem [{\citenamefont {Zhao}\ \emph {et~al.}(2022)\citenamefont {Zhao},
		\citenamefont {Kudo}, \citenamefont {Faugno}, \citenamefont {Balram},\ and\
		\citenamefont {Jain}}]{Zhao22}%
	\BibitemOpen
	\bibfield  {author} {\bibinfo {author} {\bibfnamefont {T.}~\bibnamefont
			{Zhao}}, \bibinfo {author} {\bibfnamefont {K.}~\bibnamefont {Kudo}}, \bibinfo
		{author} {\bibfnamefont {W.~N.}\ \bibnamefont {Faugno}}, \bibinfo {author}
		{\bibfnamefont {A.~C.}\ \bibnamefont {Balram}},\ and\ \bibinfo {author}
		{\bibfnamefont {J.~K.}\ \bibnamefont {Jain}},\ }\bibfield  {title} {\bibinfo
		{title} {Revisiting excitation gaps in the fractional quantum {Hall}
			effect},\ }\href {https://doi.org/10.1103/PhysRevB.105.205147} {\bibfield
		{journal} {\bibinfo  {journal} {Phys. Rev. B}\ }\textbf {\bibinfo {volume}
			{105}},\ \bibinfo {pages} {205147} (\bibinfo {year} {2022})}\BibitemShut
	{NoStop}%
	\bibitem [{\citenamefont {Melik-Alaverdian}\ \emph {et~al.}(1999)\citenamefont
		{Melik-Alaverdian}, \citenamefont {Bonesteel},\ and\ \citenamefont
		{Ortiz}}]{Melik-Alaverdian99}%
	\BibitemOpen
	\bibfield  {author} {\bibinfo {author} {\bibfnamefont {V.}~\bibnamefont
			{Melik-Alaverdian}}, \bibinfo {author} {\bibfnamefont {N.~E.}\ \bibnamefont
			{Bonesteel}},\ and\ \bibinfo {author} {\bibfnamefont {G.}~\bibnamefont
			{Ortiz}},\ }\bibfield  {title} {\bibinfo {title} {Skyrmion physics beyond the
			lowest {Landau}-level approximation},\ }\href
	{https://doi.org/10.1103/PhysRevB.60.R8501} {\bibfield  {journal} {\bibinfo
			{journal} {Phys. Rev. B}\ }\textbf {\bibinfo {volume} {60}},\ \bibinfo
		{pages} {R8501} (\bibinfo {year} {1999})}\BibitemShut {NoStop}%
	\bibitem [{\citenamefont {G\"u\ifmmode~\mbox{\c{c}}\else \c{c}\fi{}l\"u}\ \emph
		{et~al.}(2005)\citenamefont {G\"u\ifmmode~\mbox{\c{c}}\else \c{c}\fi{}l\"u},
		\citenamefont {Jeon}, \citenamefont {Umrigar},\ and\ \citenamefont
		{Jain}}]{Guclu05a}%
	\BibitemOpen
	\bibfield  {author} {\bibinfo {author} {\bibfnamefont {A.~D.}\ \bibnamefont
			{G\"u\ifmmode~\mbox{\c{c}}\else \c{c}\fi{}l\"u}}, \bibinfo {author}
		{\bibfnamefont {G.~S.}\ \bibnamefont {Jeon}}, \bibinfo {author}
		{\bibfnamefont {C.~J.}\ \bibnamefont {Umrigar}},\ and\ \bibinfo {author}
		{\bibfnamefont {J.~K.}\ \bibnamefont {Jain}},\ }\bibfield  {title} {\bibinfo
		{title} {Quantum monte carlo study of composite fermions in quantum dots: The
			effect of landau-level mixing},\ }\href
	{https://doi.org/10.1103/PhysRevB.72.205327} {\bibfield  {journal} {\bibinfo
			{journal} {Phys. Rev. B}\ }\textbf {\bibinfo {volume} {72}},\ \bibinfo
		{pages} {205327} (\bibinfo {year} {2005})}\BibitemShut {NoStop}%
	\bibitem [{\citenamefont {Wu}\ and\ \citenamefont {Yang}(1976)}]{Wu76}%
	\BibitemOpen
	\bibfield  {author} {\bibinfo {author} {\bibfnamefont {T.~T.}\ \bibnamefont
			{Wu}}\ and\ \bibinfo {author} {\bibfnamefont {C.~N.}\ \bibnamefont {Yang}},\
	}\bibfield  {title} {\bibinfo {title} {Dirac monopole without strings:
			Monopole harmonics},\ }\href {https://doi.org/10.1016/0550-3213(76)90143-7}
	{\bibfield  {journal} {\bibinfo  {journal} {Nucl. Phys. B}\ }\textbf
		{\bibinfo {volume} {107}},\ \bibinfo {pages} {365} (\bibinfo {year}
		{1976})}\BibitemShut {NoStop}%
	\bibitem [{\citenamefont {Laughlin}(1983)}]{Laughlin83}%
	\BibitemOpen
	\bibfield  {author} {\bibinfo {author} {\bibfnamefont {R.~B.}\ \bibnamefont
			{Laughlin}},\ }\bibfield  {title} {\bibinfo {title} {Anomalous quantum {Hall}
			effect: An incompressible quantum fluid with fractionally charged
			excitations},\ }\href {https://doi.org/10.1103/PhysRevLett.50.1395}
	{\bibfield  {journal} {\bibinfo  {journal} {Phys. Rev. Lett.}\ }\textbf
		{\bibinfo {volume} {50}},\ \bibinfo {pages} {1395} (\bibinfo {year}
		{1983})}\BibitemShut {NoStop}%
	\bibitem [{\citenamefont {Grimm}\ and\ \citenamefont {Storer}(1971)}]{Grimm71}%
	\BibitemOpen
	\bibfield  {author} {\bibinfo {author} {\bibfnamefont {R.}~\bibnamefont
			{Grimm}}\ and\ \bibinfo {author} {\bibfnamefont {R.}~\bibnamefont {Storer}},\
	}\bibfield  {title} {\bibinfo {title} {Monte-carlo solution of
			schr{\"o}dinger's equation},\ }\href@noop {} {\bibfield  {journal} {\bibinfo
			{journal} {Journal of Computational Physics}\ }\textbf {\bibinfo {volume}
			{7}},\ \bibinfo {pages} {134} (\bibinfo {year} {1971})}\BibitemShut {NoStop}%
	\bibitem [{\citenamefont {Ceperley}\ and\ \citenamefont
		{Kalos}(1979)}]{Ceperley79}%
	\BibitemOpen
	\bibfield  {author} {\bibinfo {author} {\bibfnamefont {D.~M.}\ \bibnamefont
			{Ceperley}}\ and\ \bibinfo {author} {\bibfnamefont {M.}~\bibnamefont
			{Kalos}},\ }\bibfield  {title} {\bibinfo {title} {Monte carlo methods in
			statistical physics},\ }\href@noop {} {\bibfield  {journal} {\bibinfo
			{journal} {Topics in current physics}\ }\textbf {\bibinfo {volume} {7}},\
		\bibinfo {pages} {145} (\bibinfo {year} {1979})}\BibitemShut {NoStop}%
\end{thebibliography}

%

\end{document}


\title{Supplemental Material for ``Composite fermion pairing induced by Landau level mixing"}
\author{Tongzhou Zhao$^1$,  Ajit C. Balram$^{2,3}$, and J. K. Jain$^4$}
\affiliation{$^1$Institute of Physics, Chinese Academy of Sciences, Beijing 100190, China}
\affiliation{$^{2}$Institute of Mathematical Sciences, CIT Campus, Chennai 600113, India}
\affiliation{$^{3}$Homi Bhabha National Institute, Training School Complex, Anushaktinagar, Mumbai 400094, India}   
 \affiliation{$^4$Department of Physics, 104 Davey Lab, Pennsylvania State University, University Park, Pennsylvania 16802, USA}
\maketitle

\renewcommand{\thesection}{S\arabic{section}}  
\renewcommand{\thetable}{S\arabic{table}}  
\renewcommand{\thefigure}{S\arabic{figure}}
\renewcommand{\theequation}{S\arabic{equation}}

The supplemental material contains the following.  In Sec.~\ref{sec: review_FPDMC}, we provide a brief review, for completeness, of the fixed-phase diffusion Monte Carlo method. In Secs. ~\ref{sec: energy_extrapolations} and~\ref{sec: pair_correlations}, we show the extrapolations to the thermodynamic limit of the energies and the pair-correlation functions of various states as a function of the Landau level mixing (LLM). Sec.~\ref{sec:intuitive} has a brief discussion on the nature of pairing.

\section{Fixed-phase diffusion Monte Carlo method}\label{sec: review_FPDMC}
In this section, we outline the fixed-phase diffusion Monte Carlo (FPDMC) method for completeness. The FPDMC is a variant of the diffusion Monte Carlo (DMC) method. We refer readers to Refs.~\cite{Foulkes01,Reynolds82} for the review of the standard DMC method and its most widely used variant for fermionic systems, namely the fixed-node DMC, of which the FPDMC is a generalization. Both the DMC and the fixed-node DMC rely on interpreting the wave function as the density distribution function of diffusing particles. Neither is useful for the fractional quantum Hall effect (FQHE), where the wave function is necessarily complex due to the breaking of the time-reversal symmetry by the magnetic field. The FPDMC was first introduced to overcome this difficulty by Ortiz, Ceperley and Martin~\cite{Ortiz93}, first implemented on in the periodic (torus) geometry. Later the method was generalized to the Haldane sphere~\cite{Melik-Alaverdian97}, which is the geometry we employ in this article. Since then, many results have been obtained based on this method~\cite{Melik-Alaverdian99,Guclu05a,Zhang16,Zhao18,Zhao21,Zhao21b,Zhao22}. Some useful formulas for calculating the drift velocity and other related quantities (see below) can be found in the supplemental material of Ref.~\cite{Zhang16}.

Our approach is based on Ref.~\cite{Melik-Alaverdian97}. In the following paragraphs, the length is measured in units of magnetic length $\ell$ and the energy in units of  $e^2/\varepsilon \ell$. We will write every physical quantity in a dimensionless form, which is slightly different from the notation in the main text.  On the sphere, it is convenient to work with the stereographic coordinates $\vec{r}=\left(x, y\right)=\left( \cos \phi \tan\left(\theta/2\right), \sin \phi \tan\left(\theta/2\right)\right)$ in addition to the spinor coordinates $(u, v)$ defined in the main text (we also use $\mathcal{R}=\left(\vec{r}_1,\vec{r}_2,...,\vec{r}_N\right)$ to label the collective coordinates of the system in the paragraphs below). We also define the complex stereographic coordinate as $z=x+i y$. We note that in practice, one often needs to switch between $(\theta, \phi)$, $(z, \bar{z})$, $(x, y)$ and $(u, v)$, depending on which form gives the most efficient way of calculating physical quantities. The Hamiltonian on the sphere is written as:
\begin{align}
H = -\frac{N}{2\kappa}+\frac{1}{2\kappa}\sum_i D(\vec{r}_i)[-i\vec{\nabla}_i+\vec{A}(\vec{r}_i)]^2+U\left(\mathcal{R}\right),
\end{align}
where $D(\vec{r}_i) = \frac{(1+r_i^2)^2}{4Q}, \vec{A} =\frac{2Q}{1+r_i^2}(y_i, -x_i)$ (we have chosen the Wu-Yang gauge~\cite{Wu76} in our study), $U\left(\mathcal{R}\right)=\sum_{i<j}\frac{1}{d_{ij}}+E_\text{bg}$ is the potential energy of the system,  where $\frac{1}{d_{ij}}$ is the Coulomb potential energy betwen the $i$'th and the $j$'th electrons, where $d_{ij}=\frac{2\sqrt{Q}|\vec{r}_i-\vec{r}_j|}{(1+r_i^2)(1+r_j^2)}=2\sqrt{Q}(u_i v_j-u_j v_i)$ is the chord distance; $E_\text{bg} = -\frac{N^2}{2\sqrt{Q}}$ is the potential energy, which consists of the background-background repulsion and the electron-background attraction; see Appendix I in Ref.~\cite{Jain07} for details. The appearance of $\kappa$ on the denominator of the kinetic energy is due to the fact that the kinetic energy is measured in units of $e^2/\varepsilon \ell$, and the LLL kinetic energy $\frac{N}{2\kappa}$ is subtracted conventionally.

The general idea of the DMC is that, by setting the time to be an imaginary number $\tau=i t$, the Schr\"{o}dinger equation becomes
\begin{equation}
\label{eqn: im_time_schrodinger}
H \Psi\left(\mathcal{R}, \tau \right)=-\frac{\partial}{\partial \tau}\Psi\left(\mathcal{R},\tau\right),
\end{equation}
and the ground state $|\Psi\rangle$ can be obtained from any initial state $|\Psi_\text{T}\rangle$ after a long-time evolution $|\Psi\rangle=\lim_{\tau\to\infty}e^{-\tau H} |\Psi_\text{T}\rangle$, given the condition $\langle\Psi|\Psi_\text{T}\rangle\neq0$.

The FPDMC method proceeds by writing the wave function as $\Psi\left(\mathcal{R}\right)=\left|\Psi\left(\mathcal{R}\right)\right|e^{i \varphi_\text{T}\left(\mathcal{R}\right)}$, where the phase  $\varphi_\text{T}$ is taken to be a fixed function of the particle coordinates. The expectation value of energy is given by $\langle \Psi | H | \Psi \rangle=\langle |\Psi| | H_R | |\Psi| \rangle$ where 
\begin{equation}
H_\text{R}=-\frac{N}{2\kappa}-\frac{1}{2\kappa}\sum_i D(\vec{r}_i)\left[\vec{\nabla}_i^2-\tilde{ \vec{A}}^2(\vec{r}_i)\right]+U\left(\mathcal{R}\right)
\end{equation}
with $\tilde{ \vec{A}}(\vec{r}_i) = \vec{A}(\vec{r}_i)+\vec{\nabla}_i \varphi_\text{T}$.
The energy minimization in the fixed phase sector is equivalent to solving the Schr\"odinger equation:
\begin{equation}
H_\text{R}\left|\Psi\left(\mathcal{R}, \tau \right)\right|=-\frac{\partial}{\partial \tau} |\Psi\left(\mathcal{R}, \tau \right)|.
\label{eqn: coupled-schrodinger}
\end{equation}
The phase $ \varphi_\text{T}$ is generally chosen as the phase of a trial wavefunciton $\Psi_\text{T}=\left|\Psi_\text{T}\right|e^{i \varphi_\text{T}}$, which in turn is selected based on a variational approach. For example, in Ref.~\cite{Melik-Alaverdian97}, the trial wave function for the FQHE at $\nu=1/3$ was chosen to be the Laughlin state~\cite{Laughlin83}, which is an excellent representation of the exact ground state in the LLL~\cite{Ambrumenil88, Balram20b}, and therefore its phase may remain a good approximation at least for small LLM. While one may expect that the phases of ``nice" trial wave functions ought to provide low energies, there is no simple rule that tells us how good a given $\varphi_\text{T}$ is, which is why we consider many different candidate states in our study. Nonetheless, we stress that all of our results and conclusions are subject to our phase choice.

Once the phase is fixed, the standard DMC technique is employed to solve for the real and non-negative wave function $\left|\Psi\right|$ in Eq.~\eqref{eqn: im_time_schrodinger} through the so-called importance-sampling method~\cite{Grimm71,Ceperley79,Reynolds82}. On the Haldane sphere, one defines the importance-sampled density distribution function as:
\begin{equation}
P\left(\mathcal{R}, \tau \right) = \left|\Psi(\mathcal{R}, \tau)\right| \left|\Psi_\text{T}(\mathcal{R}, \tau)\right|\prod_{i=1}^N \frac{1}{D\left(\vec{r}_i\right)}.
\end{equation}
With the help of the mixed estimator $\langle |\Psi|\left|H_\text{R}\right||\Psi_\text{T}|\rangle$~\cite{Foulkes01}, the expectation value of the ground state energy can be written in terms of $P$ as follows:
\begin{equation}\label{eqn: mixed_energy}
\begin{aligned}
\langle \Psi\left|H\right|\Psi\rangle =& \langle |\Psi|\left|H_\text{R}\right||\Psi|\rangle\\
= & \lim_{\tau\to \infty}\frac{\langle  e^{-\tau H_\text{R}/2}|\Psi_\text{T}|\left|H_\text{R}\right|e^{-\tau H_\text{R}/2} |\Psi_\text{T}|\rangle}{\langle  e^{-\tau H_\text{R}/2}|\Psi_\text{T}||e^{-\tau H_\text{R}/2} |\Psi_\text{T}|\rangle}\\
= &  \lim_{\tau\to \infty}\frac{\langle  e^{-\tau H_\text{R}}|\Psi_\text{T}|\left|H_\text{R}\right| |\Psi_\text{T}|\rangle}{\langle  e^{-\tau H_\text{R}}\left|\Psi_\text{T}\right||\left|\Psi_\text{T}\right|\rangle}\\
= & \frac{\langle |\Psi|\left|H_\text{R}\right||\Psi_\text{T}|\rangle}{\langle |\Psi|||\Psi_\text{T}|\rangle}\\
= & \lim_{\tau\to \infty}\frac{\int P\left(\mathcal{R},\tau\to \infty\right)E_L\left(\mathcal{R}\right)d\tilde{\mathcal{R}}}{\int P\left(\mathcal{R},\tau\to \infty\right) d\tilde{\mathcal{R}}}\\
\approx & \frac{1}{M}\sum_{m}E_L\left(\mathcal{R}_m\right),
\end{aligned}
\end{equation}
where $d\tilde{\mathcal{R}}$ is shorthand for $d\mathcal{R} \prod_{i=1}^N \frac{1}{D\left(\vec{r}_i\right)}$. $E_L\left(\mathcal{R}\right)$ is the local energy of the system:
\begin{equation}\label{eqn: local_energy}
\begin{aligned}
E_L = \frac{H_\text{R} \left|\Psi_\text{T}\right|}{\left|\Psi_\text{T}\right|}  =  \Re{\frac{H \Psi_\text{T}}{\Psi_\text{T}}},
\end{aligned}
\end{equation}
The last line of Eq.~\eqref{eqn: mixed_energy} represents the average over the Monte Carlo generated collection of $\mathcal{R}$'s, with $M$ being the total number of Monte Carlo steps. Eq.~\eqref{eqn: local_energy} indicates that, for the LLL-projected states, one does not need to explicitly calculate the kinetic energy. This is very useful since usually these states cannot be written in terms of simple closed-form analytic expressions and the computation of their derivatives (which is required in evaluating $H_\text{R} \left|\Psi_\text{T}\right|$) which would be extremely tedious, can be avoided based on Eq.~\eqref{eqn: local_energy}. Another reason to write $\langle H\rangle$ in the above form is that, by including the trial wave function $\left|\Psi_\text{T}\right|$, the Monte Carlo moves are guided to the regions where $\Psi$ is significant (this is why the algorithm is called the importance-sampling). This is more efficient than directly sampling $\left|\Psi\right|$ based on Eq.~\eqref{eqn: im_time_schrodinger}~\cite{Foulkes01}.

The time-evolution equation satisfied by $P$ is
\begin{equation}\label{eqn: evolution_P}
\begin{aligned}
-\frac{\partial}{\partial \tau} P\left(\mathcal{R},\tau\right) &= \sum_{i=1}^{N} \left[-\frac{1}{2}\vec{\nabla}_i^2 \left[D\left(\vec{r}_i\right)P\left(\mathcal{R},\tau\right)\right]\right.\\
&\left.+\vec{\nabla}_i \cdot \left[ D\left(\vec{r}_i\right)\vec{F}_i\left(\mathcal{R}\right)P\left(\mathcal{R},\tau\right) \right]  \right]\\
&+\left[E_L\left(\mathcal{R}\right)-E_T\right]P\left(\mathcal{R},\tau\right),
\end{aligned}
\end{equation}
where $\vec{F}_i =\vec{\nabla}_i\ln\left|\Psi_T\right|$ is called the drift velocity, whose meaning will be explained in the following paragraphs. In Eq.~\eqref{eqn: evolution_P}, $E_T$ is a constant within each iteration of the DMC simulation. It is updated in the algorithm to be equal the system's average energy. Eq.~\eqref{eqn: evolution_P} represents nothing but the diffusion process of many randomly moving particles (called random walkers). For the convenience of the stochastical simulation, the above equation can be written in the form of an integral equation
\begin{equation}
P\left(\mathcal{R},\tau+\Delta\tau\right) = \int G\left(\mathcal{R}\to\mathcal{R}',\Delta\tau\right) P\left(\mathcal{R},\tau\right) d\mathcal{R}.
\end{equation}
For the short-time evolution $\Delta \tau\ll1 $, $G$ is approximated as~\cite{Reynolds82}
\begin{equation}
\begin{aligned}
&G\left(\mathcal{R} \to \mathcal{R}', \Delta \tau \right)\\ &=A\left(\Delta \tau \right)G_\text{d}\left(\mathcal{R} \to \mathcal{R}', \Delta \tau \right) G_\text{b}\left(\mathcal{R} \to \mathcal{R}', \Delta \tau \right),
\end{aligned}
\end{equation}
where $A\left(\Delta \tau\right)$ is a normalization factor the wave function, which can be ignored. The next two terms $G_d$ and $G_b$ controls the drifting motion and the branching rate of the random walkers, respectively. They are:
\begin{equation}
\begin{aligned}
G_\text{d}\left(\mathcal{R} \to \mathcal{R}', \Delta \tau \right) &=\prod_{i=1}^{N} \left\{ \frac{1}{2\pi D\left(\vec{r}_i\right) \Delta \tau}\right.\\
&\left.\times \exp\left[\frac{-\left[\vec{r}_i'-\vec{r}_i-D\left(\vec{r}_i\right)\Delta \tau \vec{F}_i\left(\mathcal{R}\right) \right]^2 }{2D\left(\vec{r}_i\right)\Delta\tau}\right]\right\}
\end{aligned}
\end{equation}
and
\begin{equation}
\begin{aligned}
&G_\text{b}\left(\mathcal{R} \to \mathcal{R}', \Delta \tau \right)\\
&=\exp\left[ -\kappa \Delta \tau \left( \frac{E_L\left(\mathcal{R}\right)+E_L\left(\mathcal{R'}\right)}{2}-E_T\right)\right].
\end{aligned}
\end{equation}
Essentially, the DMC calculation is the simulation of the diffusion process of many random walkers. The program generates a collection of random walkers, each of which represents a copy of the system in the ensemble. $G_\text{d}$ controls the drifting motion of the random walkers. The $i$'th particle of the system drifts at velocity $\vec{F}_i\left(\mathcal{R}\right)$ while also making a random motion simultaneously. On the other hand, these walkers are created or annihilated constantly (this is the so-called {\it branching process}, see Ref.~\cite{Foulkes01} for additional information), and the survival probability of the random walkers over the time interval $\Delta \tau$ is $G_\text{b}$. Both $G_\text{d}$ and $G_\text{b}$ affect the final distribution of the random walkers and hence the wave function $\left|\Psi\right|$. We refer readers to Ref.~\cite{Foulkes01} for details of the implementation of the DMC simulation, which are by now standard. Here we mention certain features of the FPDMC algorithm and its implementation which have enabled us to accurately evaluate the quantities of our interest:\\

1) As done for the energy calculation, we also use the mixed estimator to evaluate the pair-correlation function in our study. The definition of the pair-correlation function is~\cite{Ortiz93}
\begin{equation}
\begin{aligned}
g\left(r\right) = [2/\rho(N-1)]\times\langle\sum_{i\neq j}\delta\left(r-r_{ij}\right)\rangle_{\tilde{P}\left(\mathcal{R},\tau\to\infty\right)},
\end{aligned}
\end{equation}
where we have assumed that the density $\rho$ is unitary for the FQHE state. We note here that generically, the mixed estimator introduces some numerical error in $g\left(r\right)$, and there are more accurate but more complicated ways to calculate $g\left(r\right)$ (e.g., see Ref.~\cite{Foulkes01} for the definition of the extrapolated estimator). However, since we are only concerned with the qualitative behavior of $g\left(r\right)$, this form is sufficient for our purpose.

2) When $\kappa\to0$, we have $G_b\to1$. In this limit, the branching process is suppressed, and the DMC algorithm resembles the usual variational Monte Carlo (VMC) simulation, in which the final state is identical to the initial trial state. This is because the branching process eliminates the high-energy components in a wave function. Without branching, the system stays at the initial status.

3) The accuracy of the FPDMC depends on the choice of the trial phase. Each trial wave function specifies a fixed phase $\varphi_\text{T}$, and the FPDMC gives the lowest energy state within each phase sector. If $\varphi_\text{T}$ were exactly known (which is not the case for our calculations), the FPDMC would give the energy of the true ground state.

4) While the final state is different from the trial wave function, it has non-zero and sometimes large 
overlap with the initial state. This is guaranteed because the initial and the final states are related by the time-evolution. Furthermore, the phase, assumed to be invariant, encodes the topological content of the wave function, which is why it is reasonable to label the final state by the initial trial wave function, as we have done in the main text.

5) The FPDMC method automatically includes the LLM, because the algorithm minimizes the total energy including both the kinetic and the potential energies by optimizing the wave function in the Hilbert space that is not restricted to the LLL. The FPDMC is also non-perturbative in $\kappa$.\\

\section{Energy extrapolations}\label{sec: energy_extrapolations}
\begin{figure*}
	\includegraphics[width = 0.32 \textwidth]{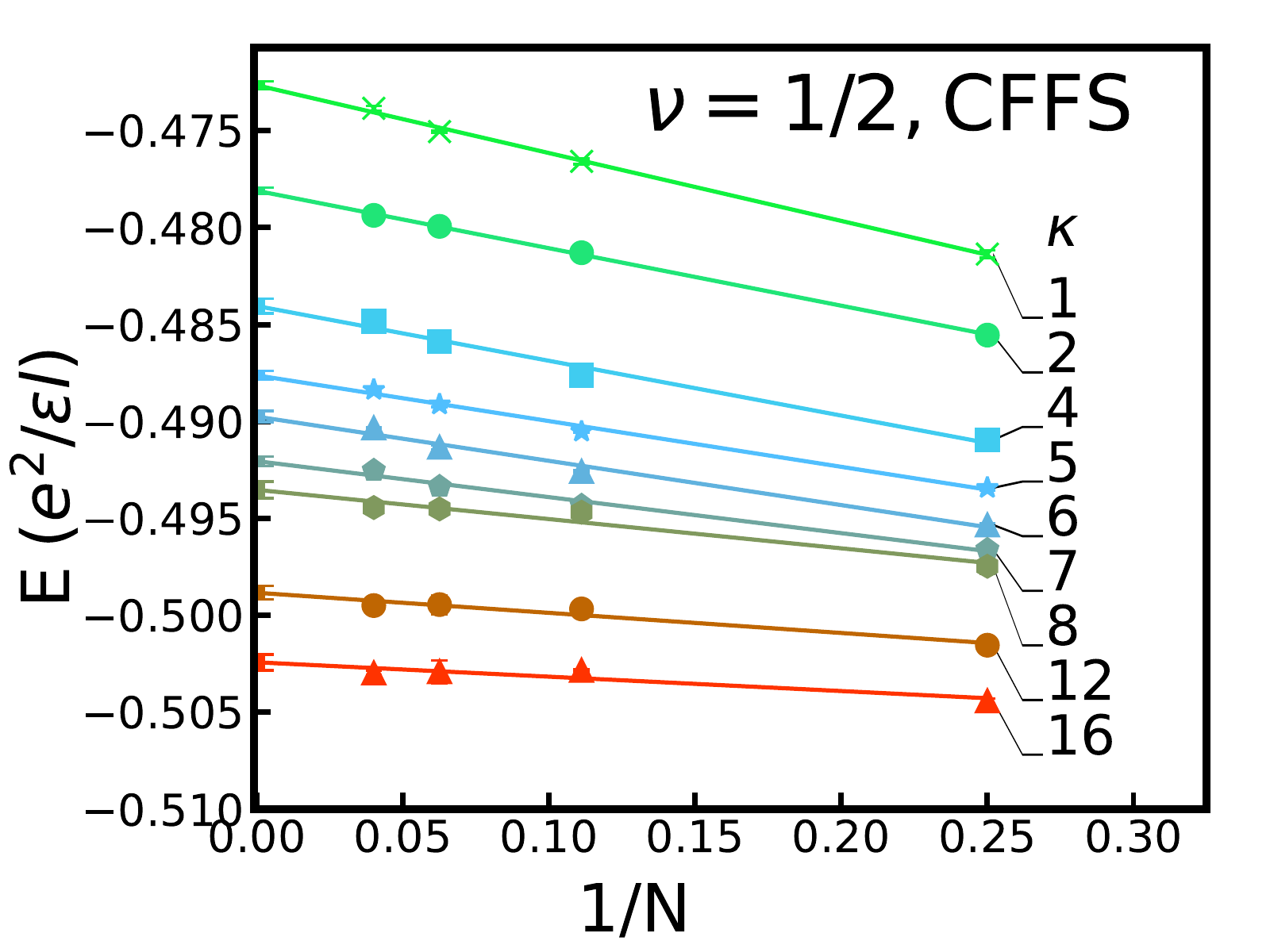}
	\includegraphics[width = 0.32 \textwidth]{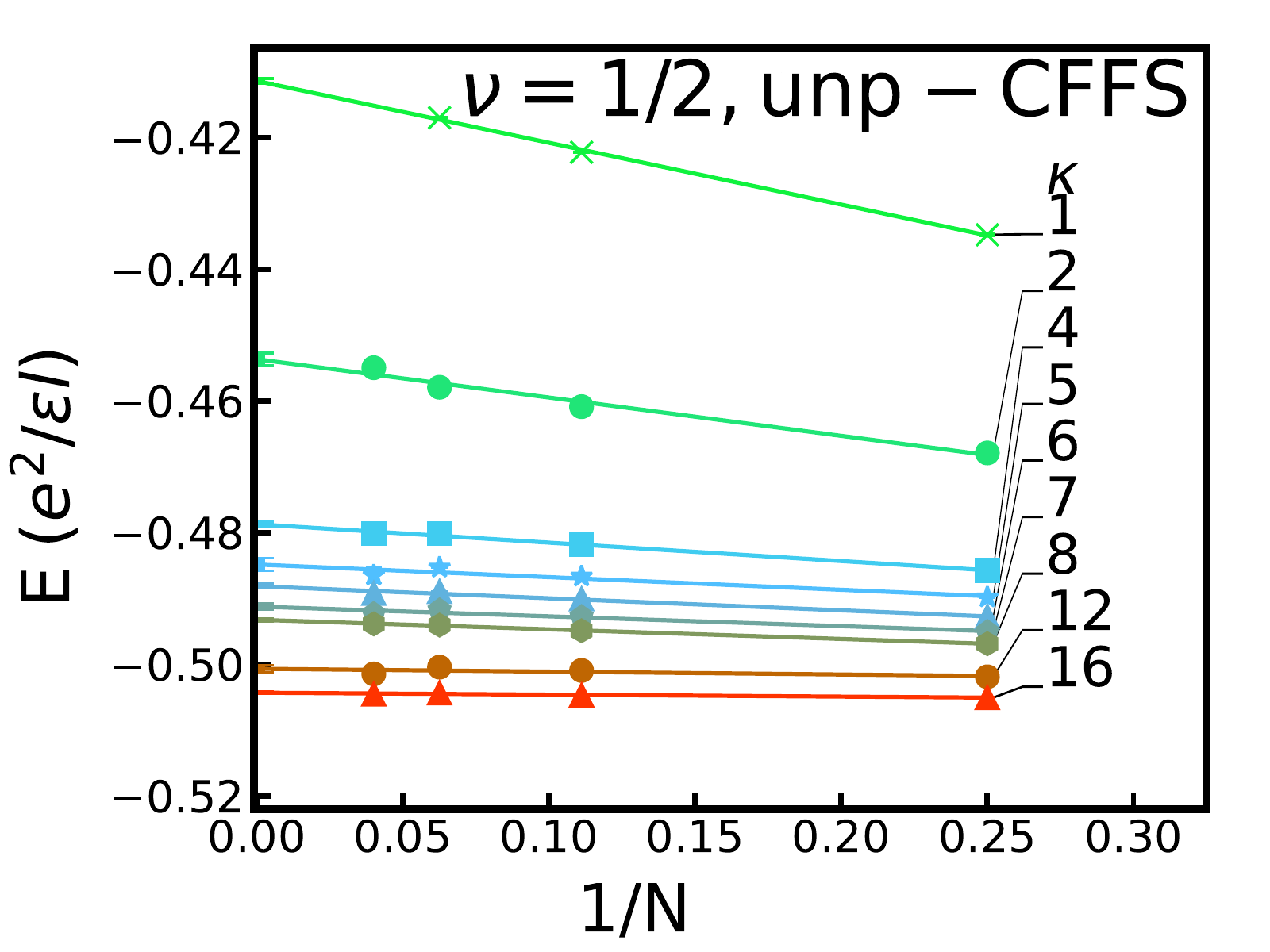}
	\includegraphics[width = 0.32 \textwidth]{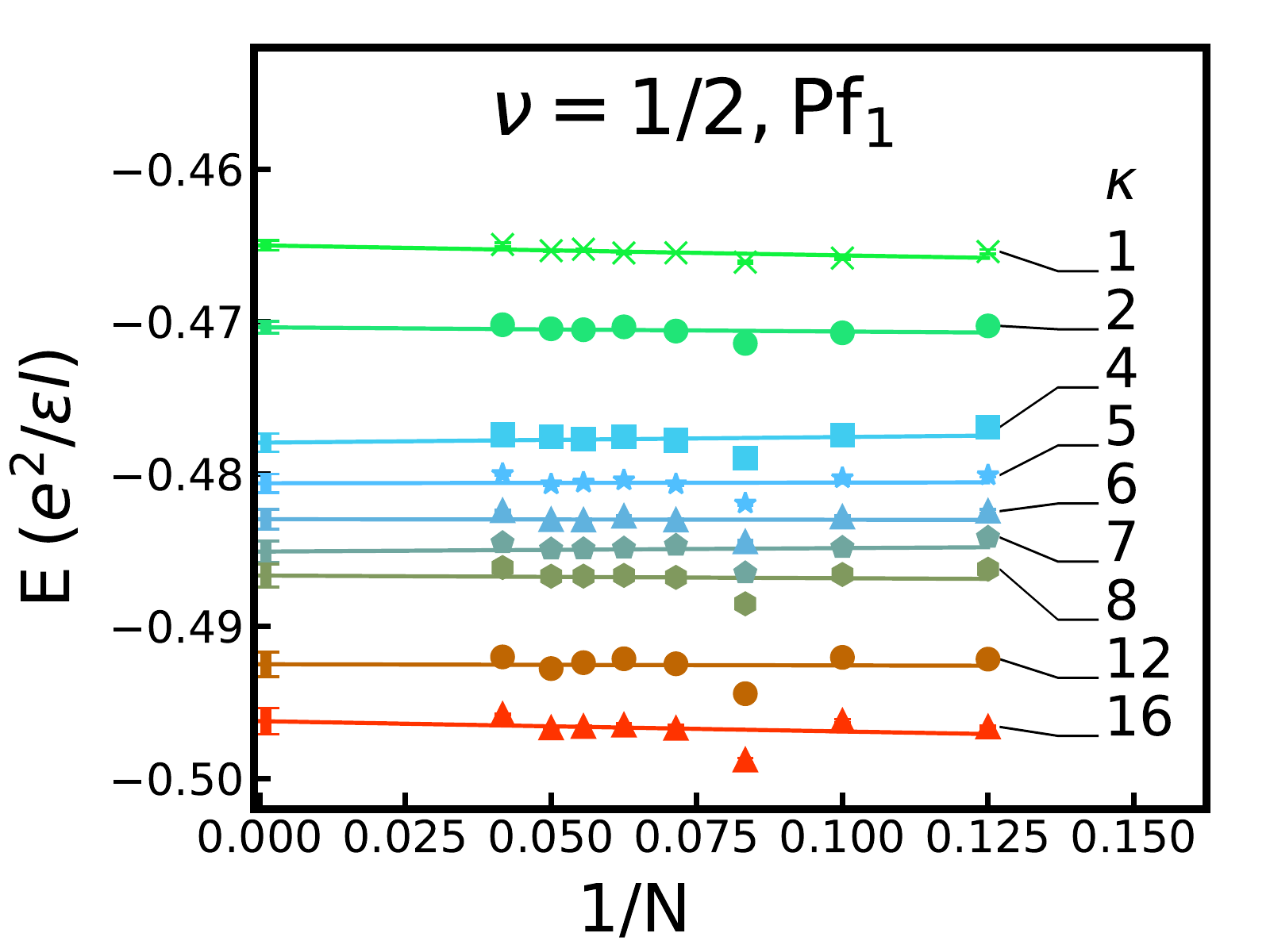}

	\includegraphics[width = 0.32 \textwidth]{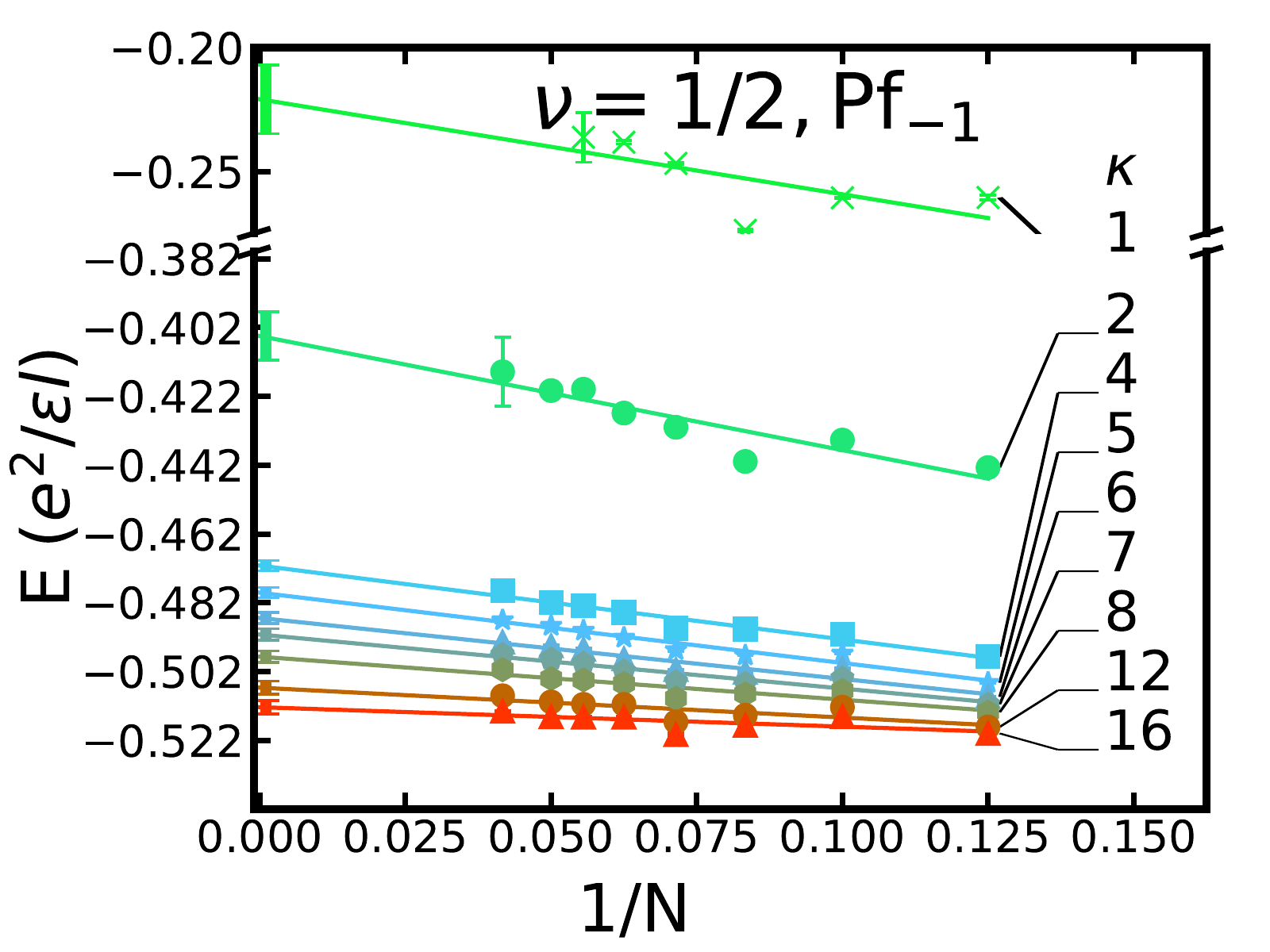}	
	\includegraphics[width = 0.32 \textwidth]{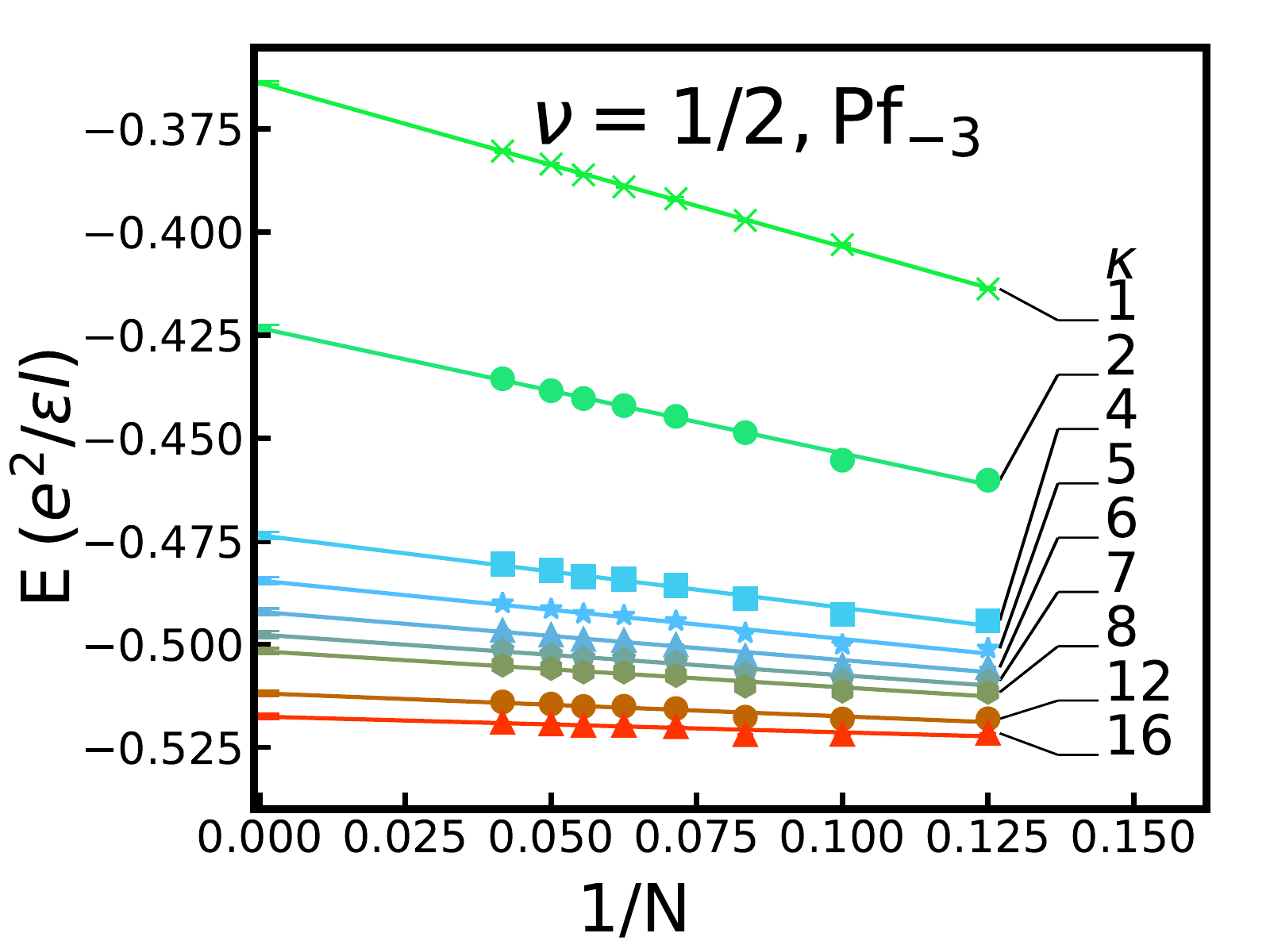}
	\includegraphics[width = 0.32 \textwidth]{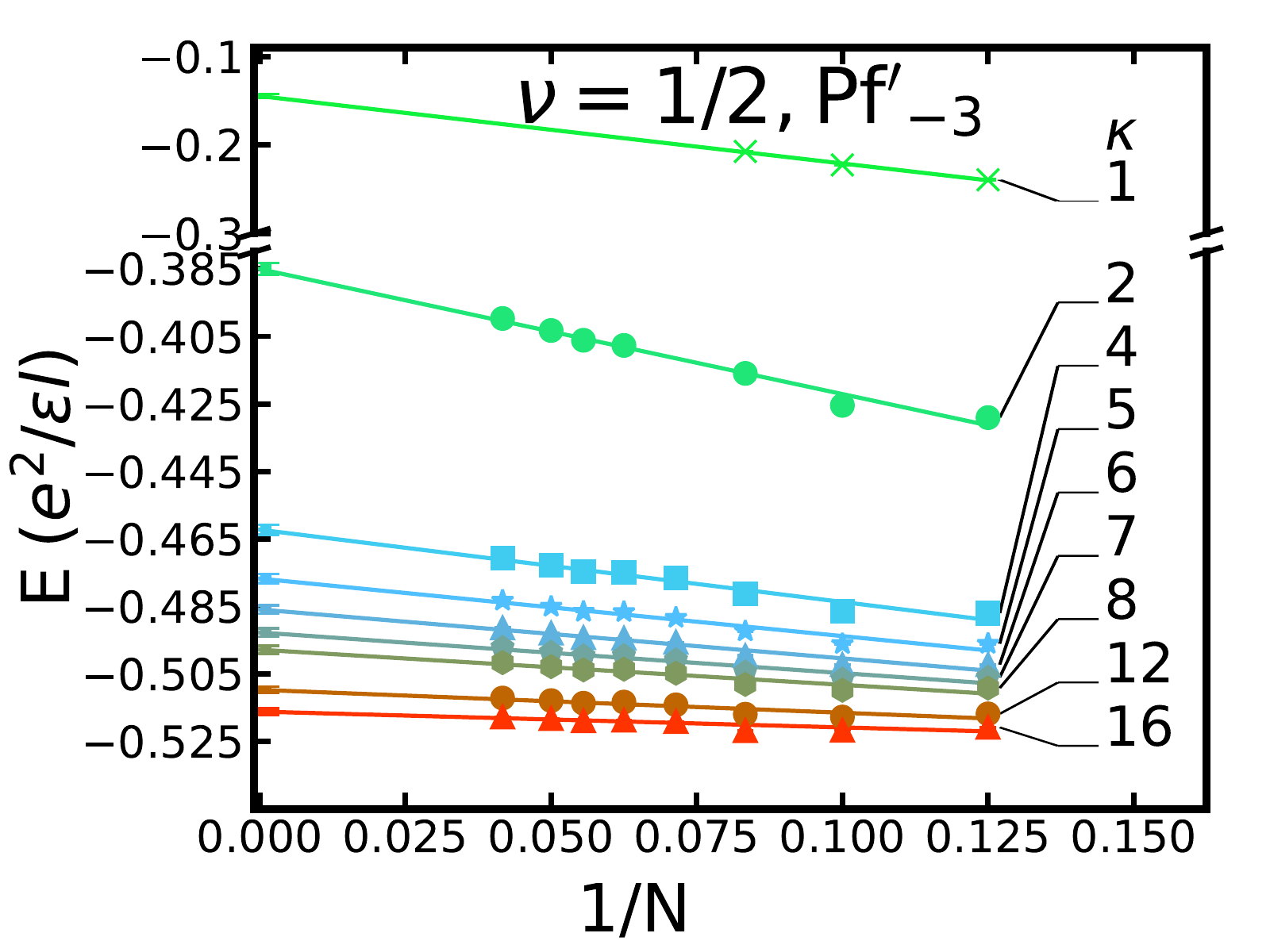}
	\includegraphics[width = 0.32 \textwidth]{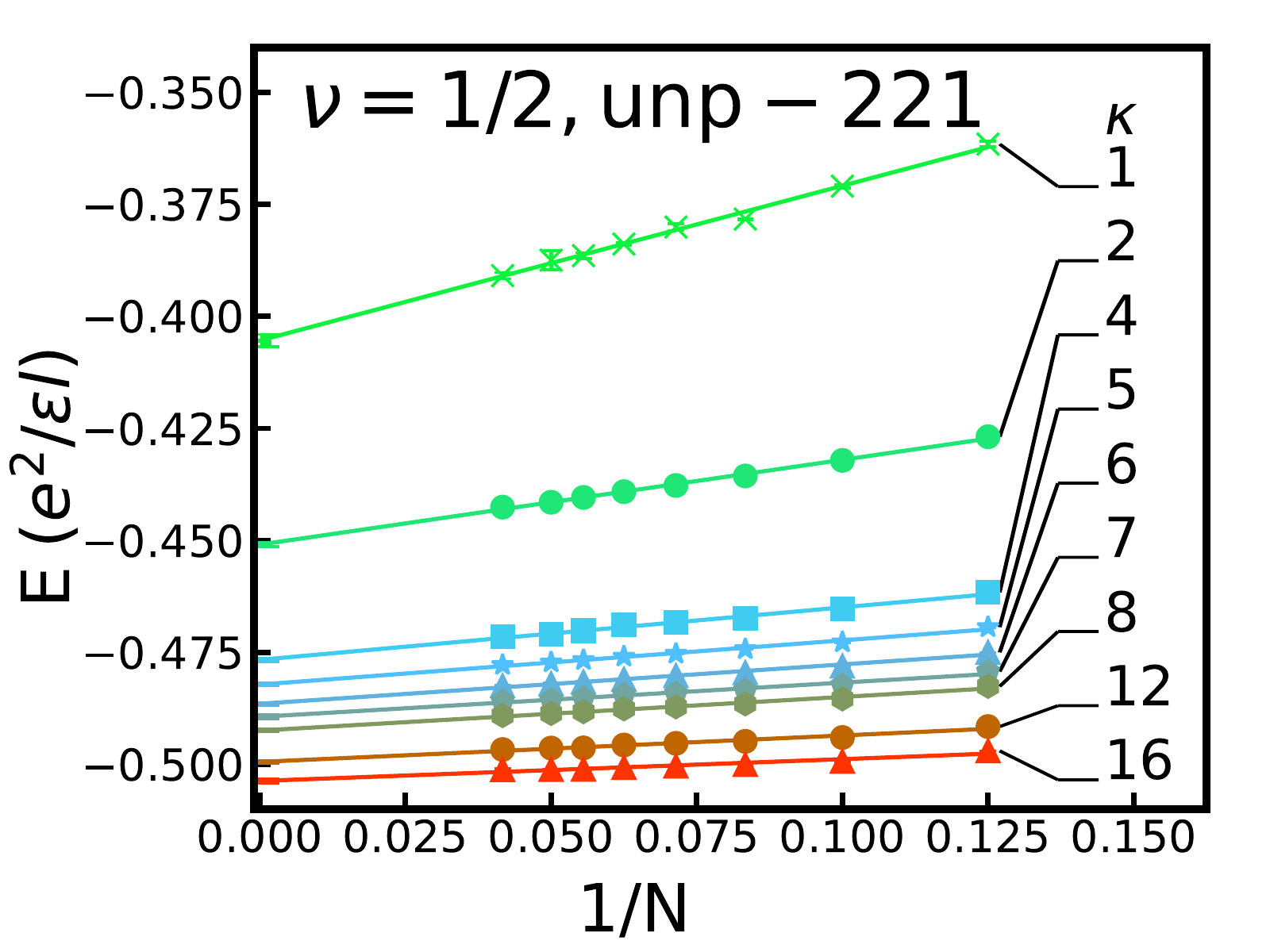}
	\includegraphics[width = 0.32 \textwidth]{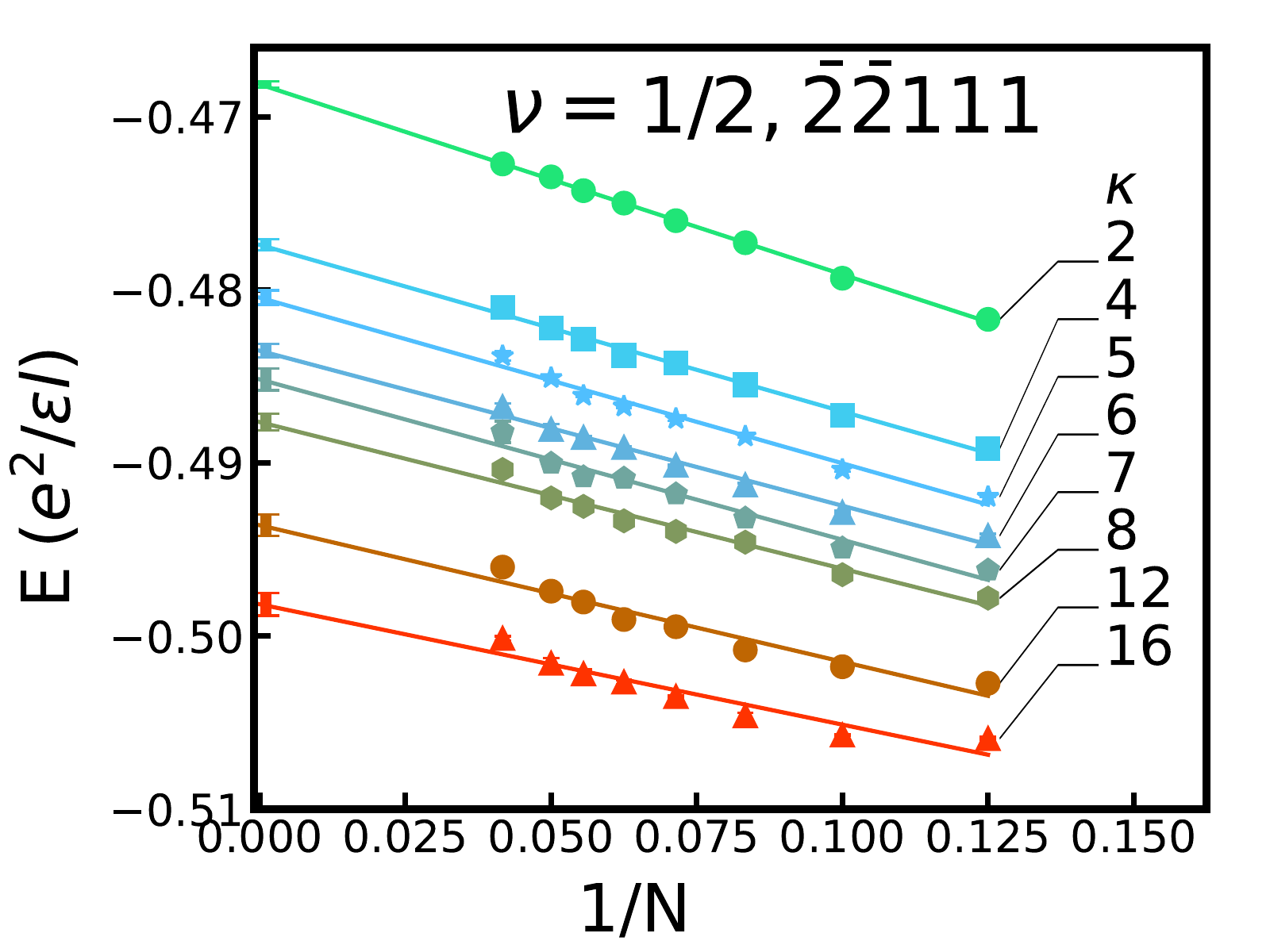}
	\includegraphics[width = 0.32 \textwidth]{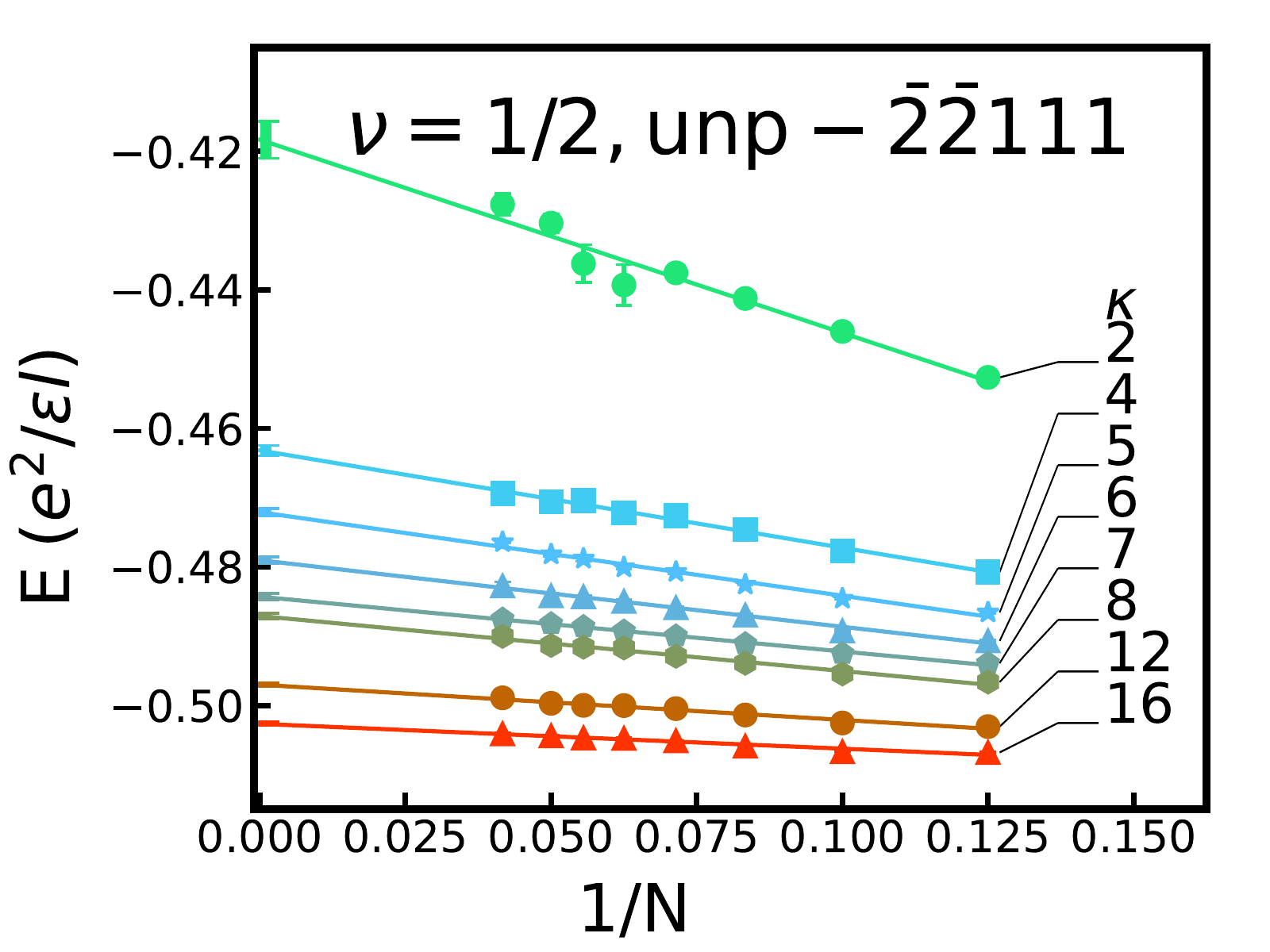}
	\caption{\label{fig: other_extrapolation_12}The energies of all candidate states considered for $\nu=1/2$ as a function of $1/N$ for different values of the LL mixing parameter $\kappa$ (labeled on plots). The thermodynamic values of energies obtained by linear regression are shown near the vertical axes.}
\end{figure*}
\begin{figure*}
	\includegraphics[width = 0.32 \textwidth]{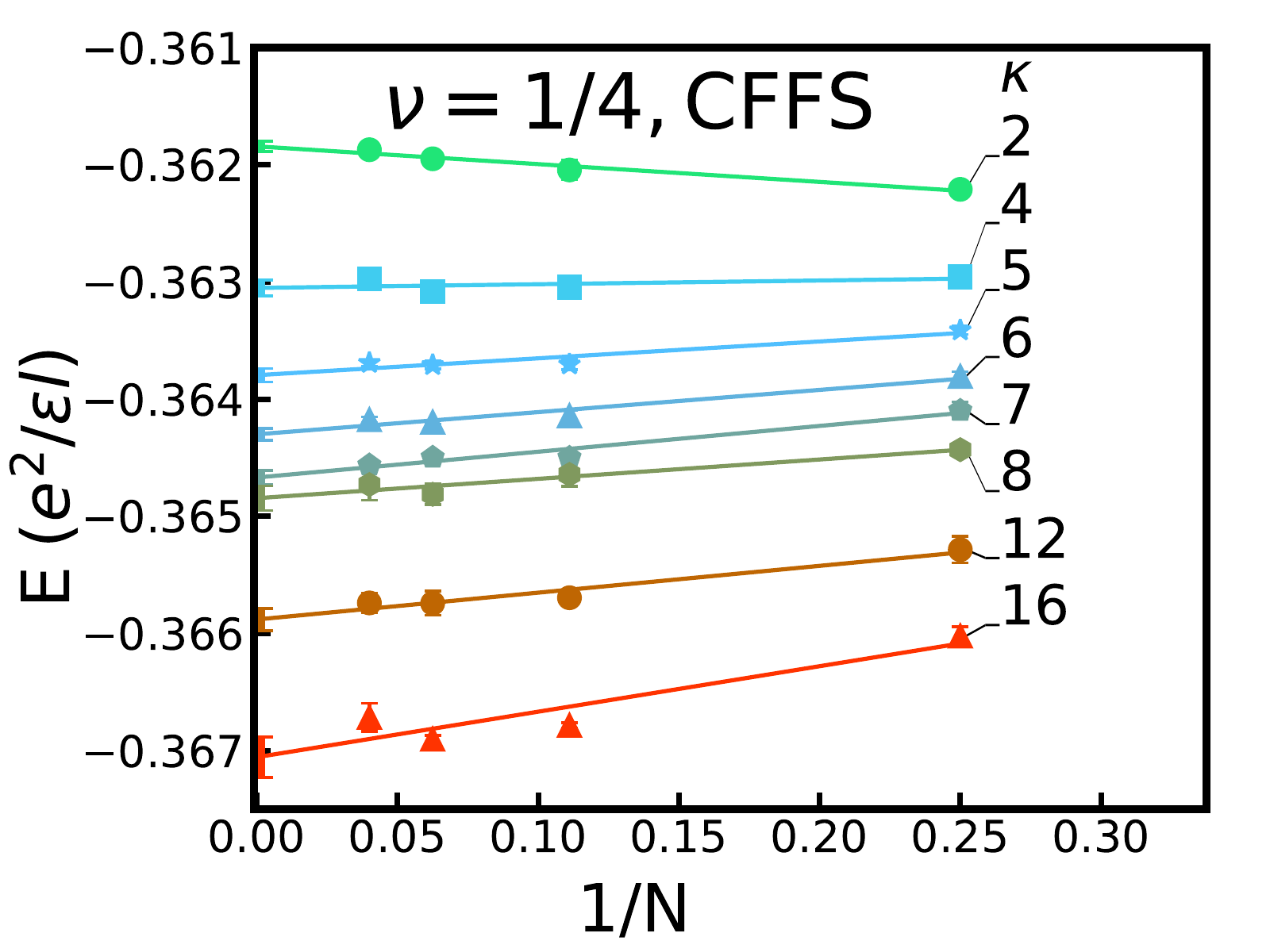}
	\includegraphics[width = 0.32 \textwidth]{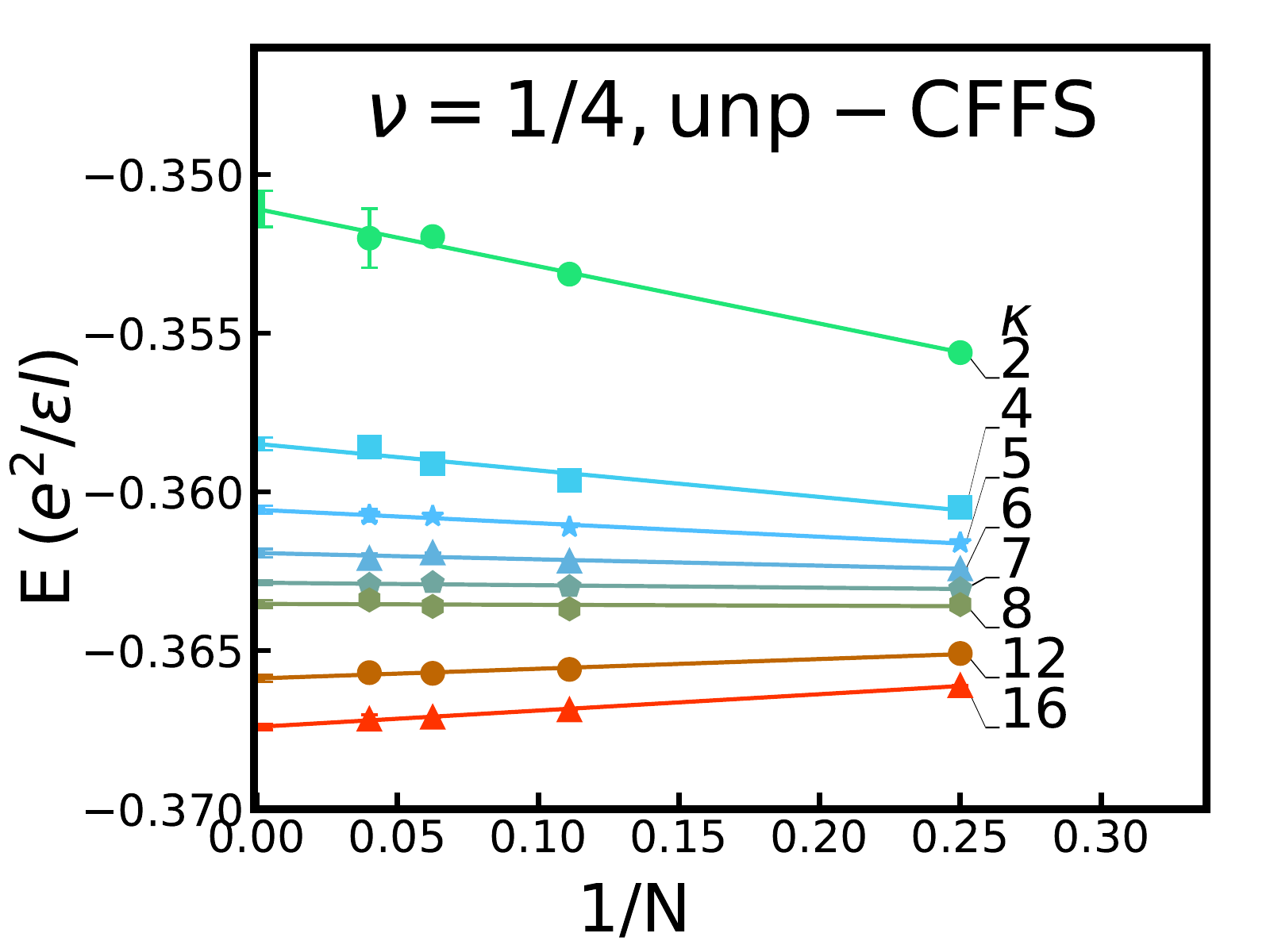}
	\includegraphics[width = 0.32 \textwidth]{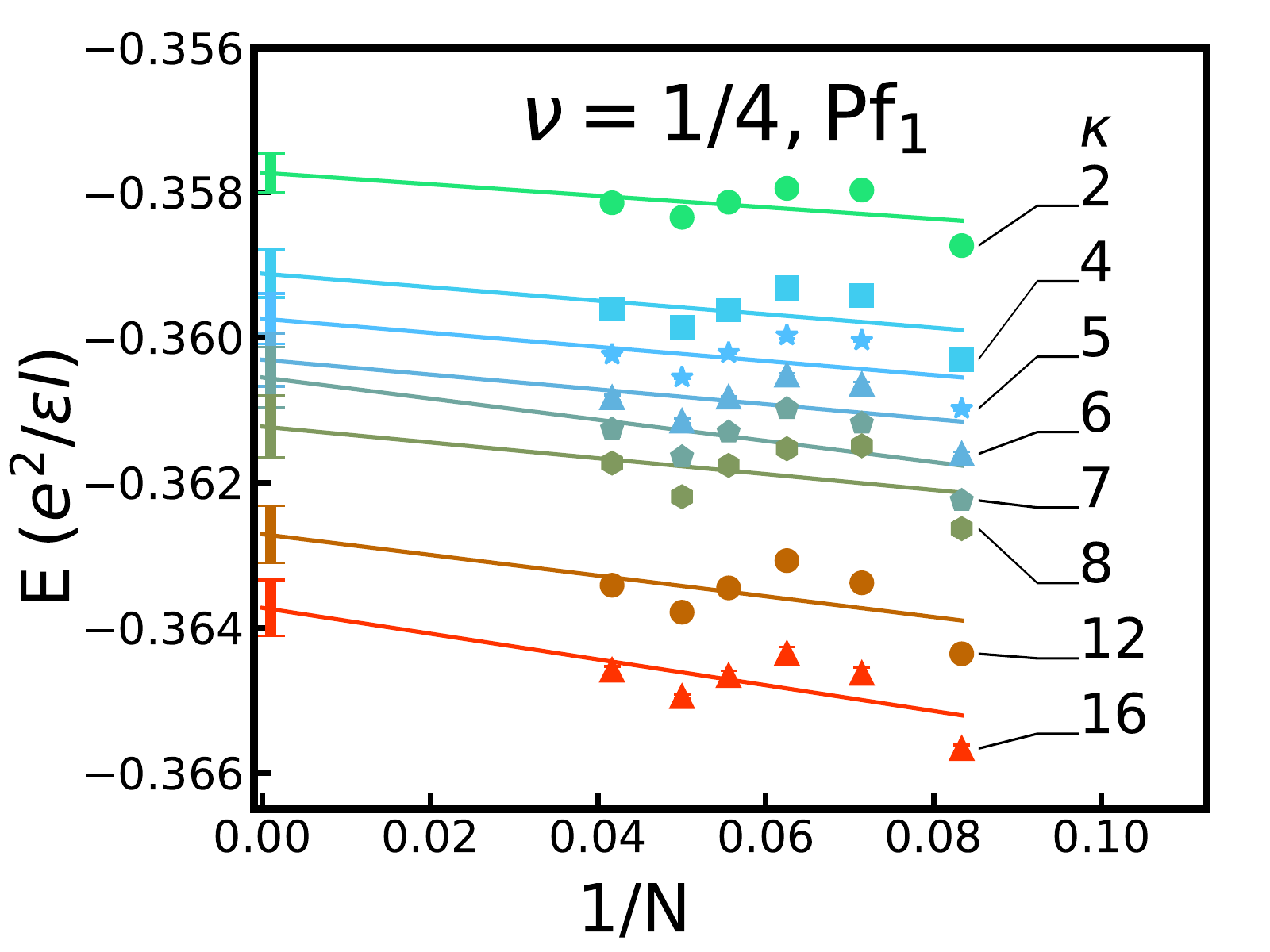}
	
	\includegraphics[width = 0.32 \textwidth]{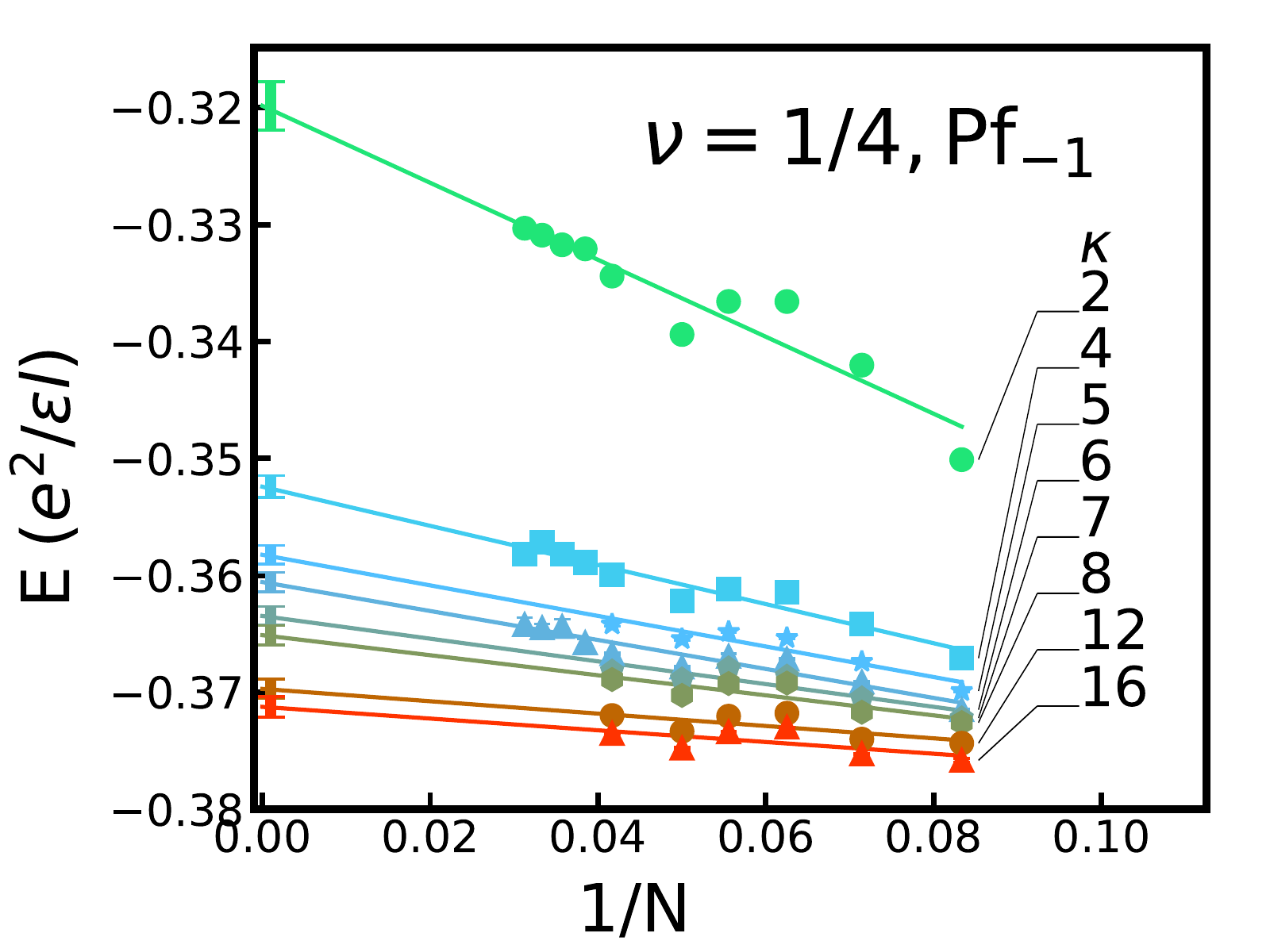}
	\includegraphics[width = 0.32 \textwidth]{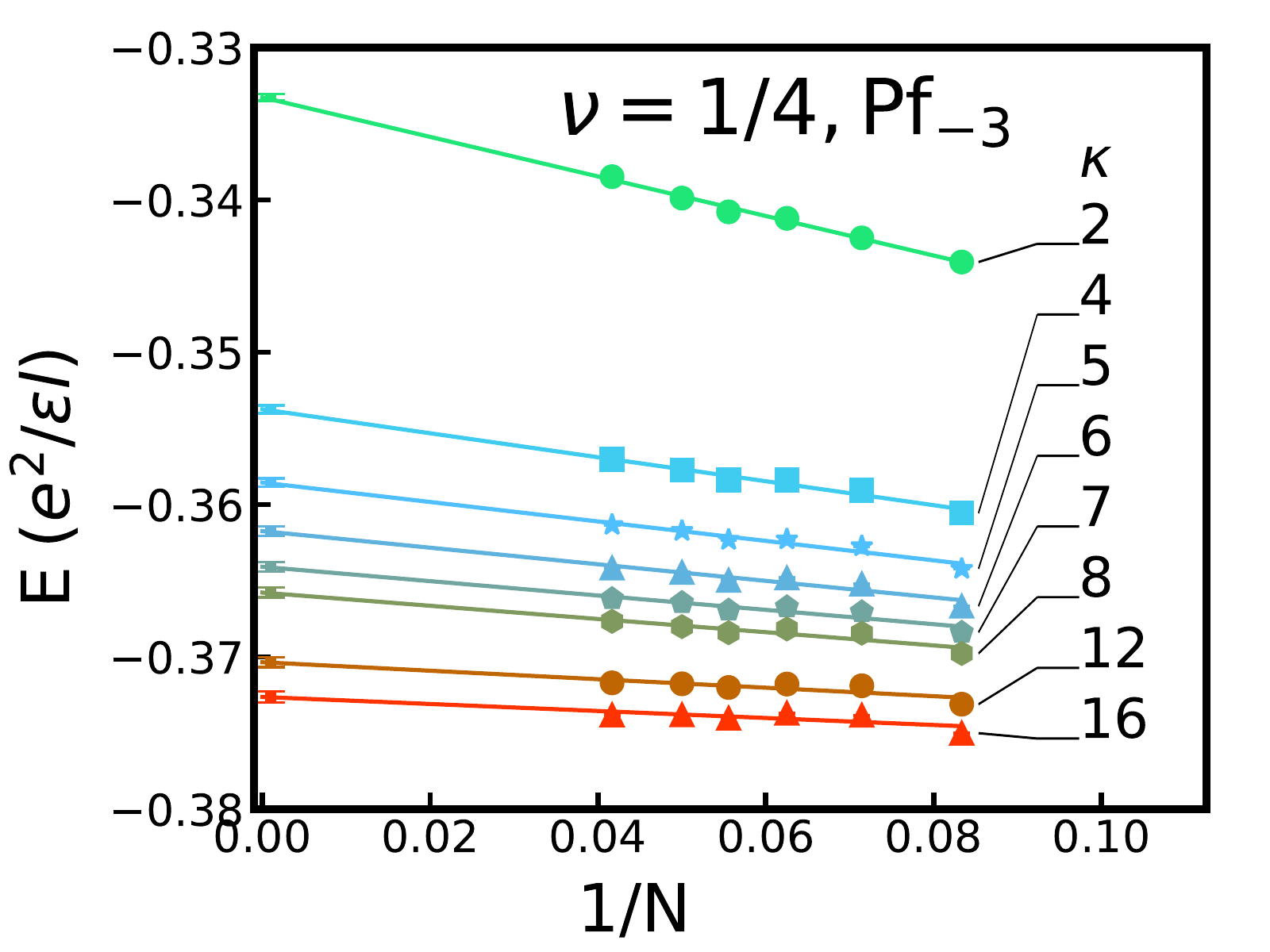}
	\includegraphics[width = 0.32 \textwidth]{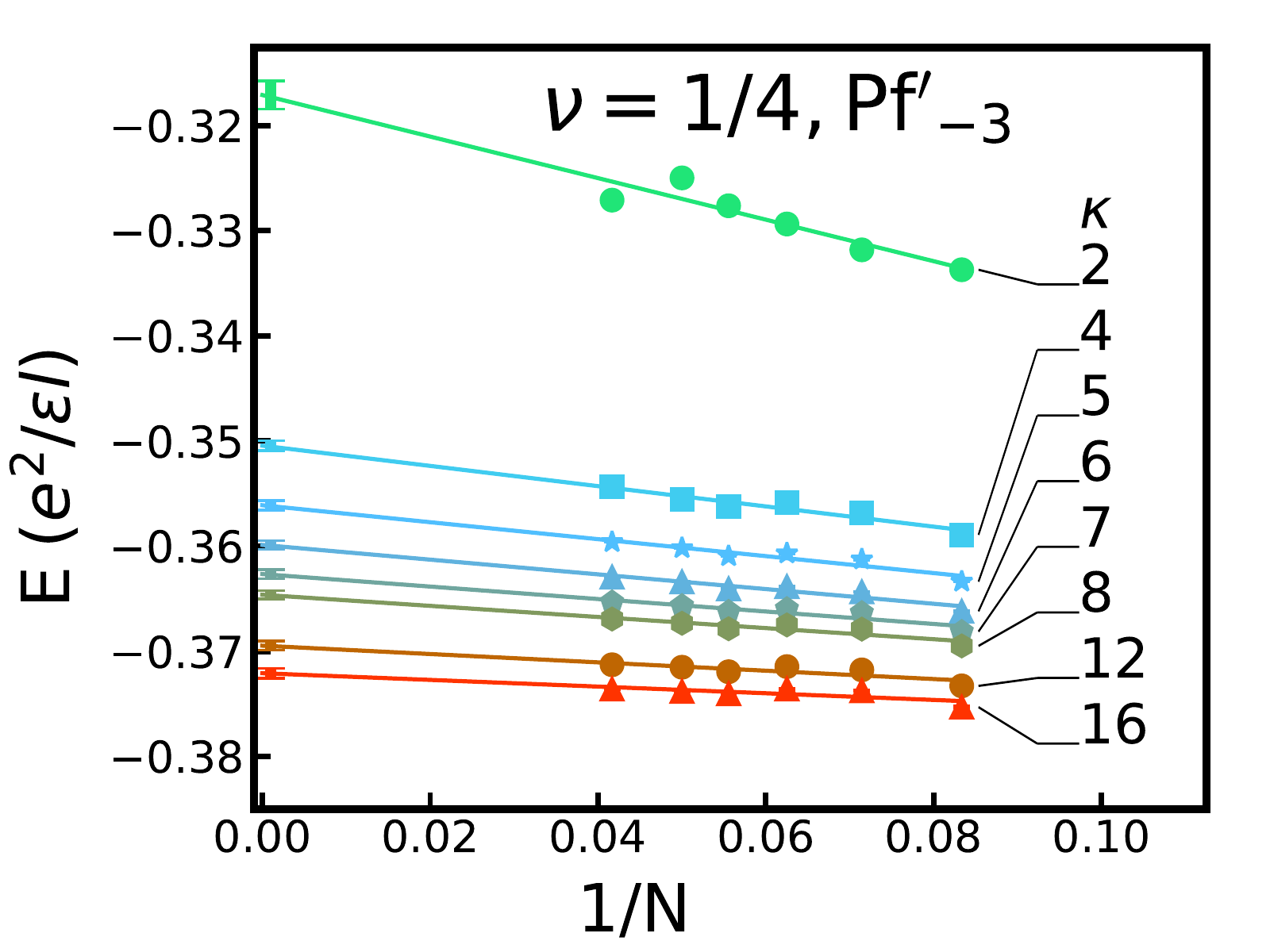}
	
	\includegraphics[width = 0.32 \textwidth]{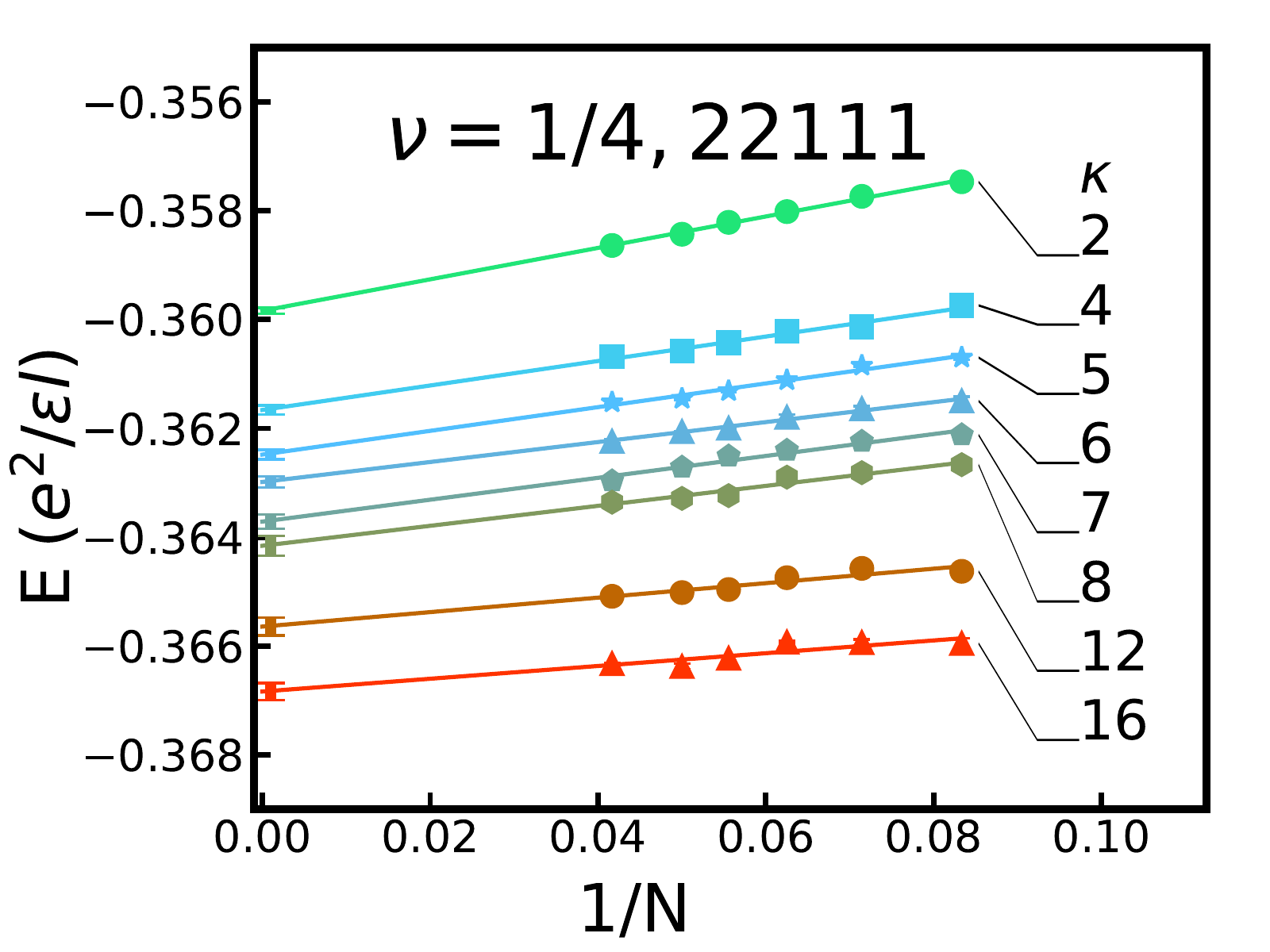}
	\includegraphics[width = 0.32 \textwidth]{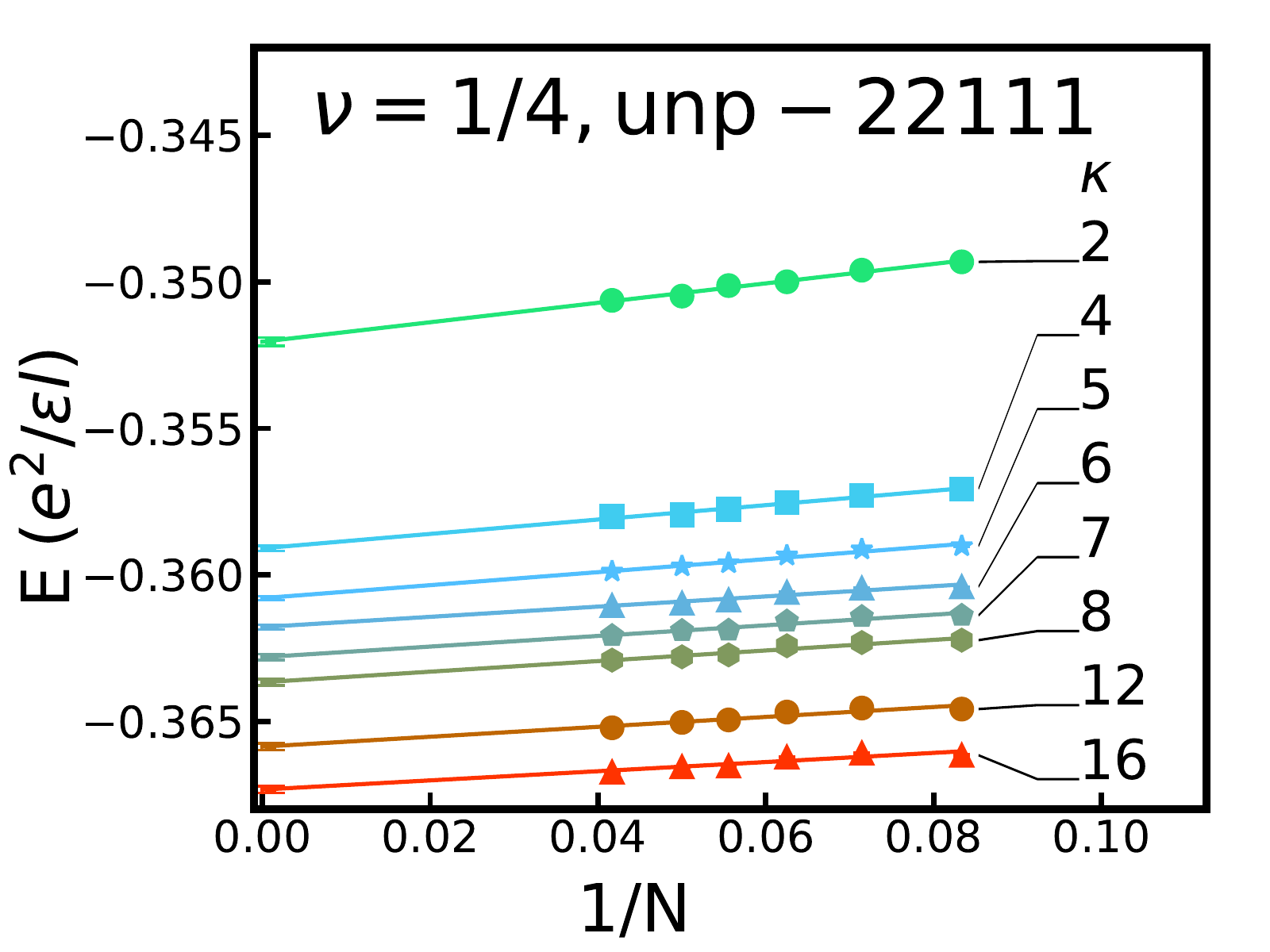}
	\includegraphics[width = 0.32 \textwidth]{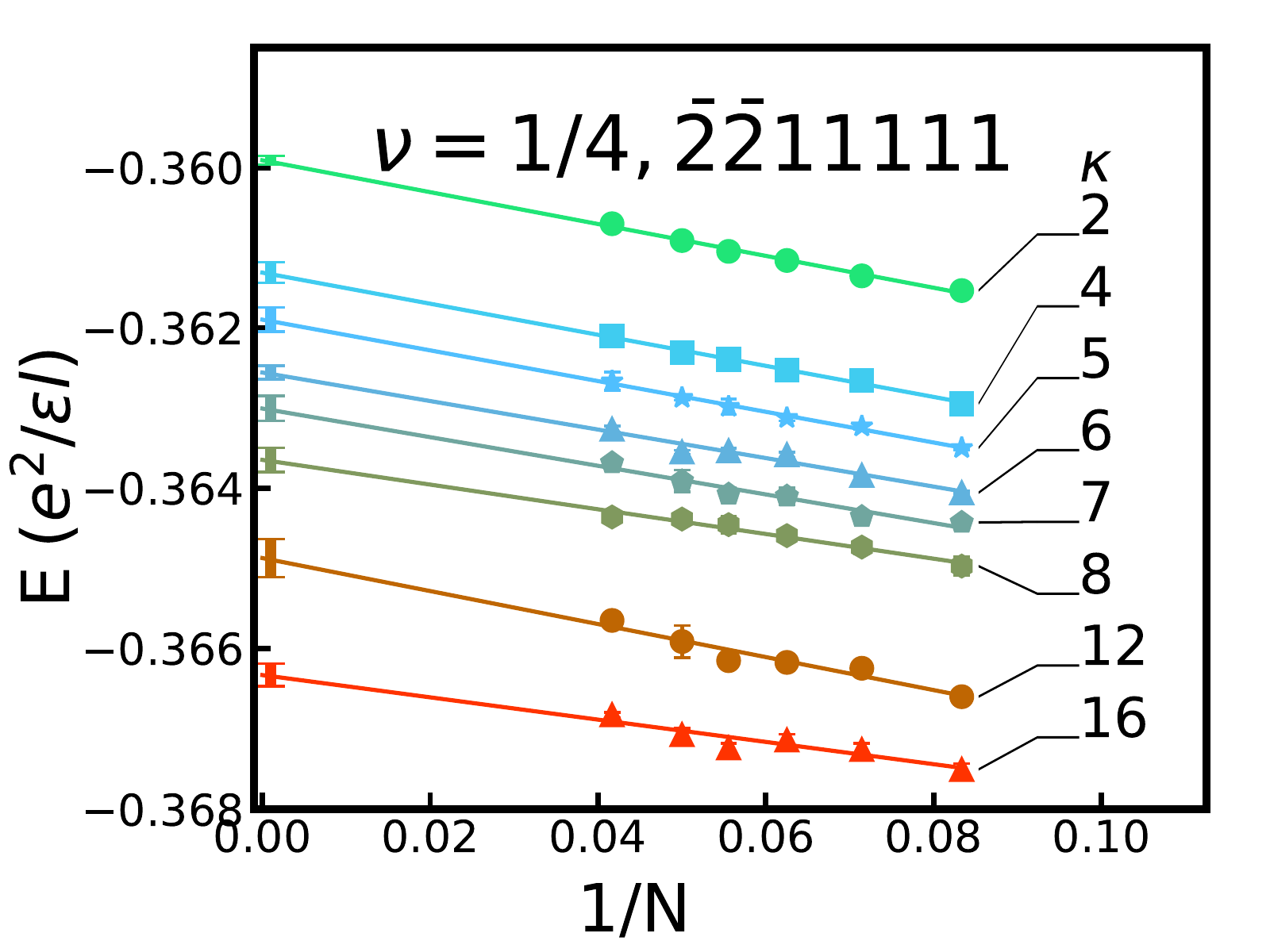}
	\includegraphics[width = 0.32 \textwidth]{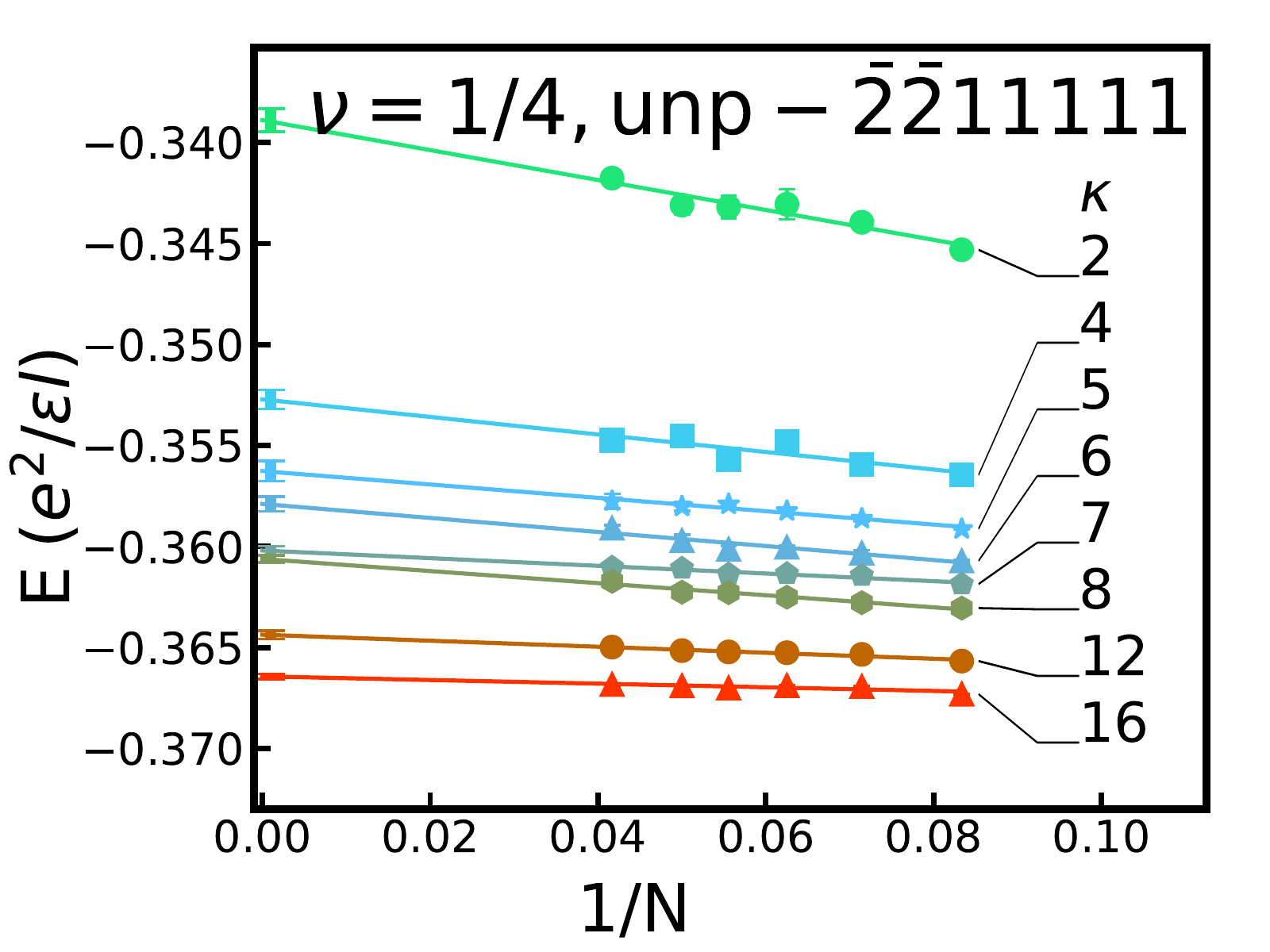}
	\caption{\label{fig: other_extrapolation_14}The energies of all candidate states considered for $\nu=1/4$ as a function of $1/N$ for different values of the LL mixing parameter $\kappa$ (labeled on plots). The thermodynamic values of energies obtained by linear regressions are shown near the vertical axes.}
\end{figure*}

We show the energy extrapolations of all the states considered in this article, for both $\nu=1/2$ (Fig.~\ref{fig: other_extrapolation_12}) and $\nu=1/4$ (Fig.~\ref{fig: other_extrapolation_14}). Due to the enormous time cost of the FPDMC calculation, the largest systems are limited to around $N=25$ particles. We have not included the data points for $N\leq10$ for the $\nu=1/4$ $\mathrm{Pf}_{-1}$ state since the energies for these systems do not fit to a linear regression in $1/N$, indicating strong finite size correction; for this particular state, we have performed additional calculations at $\kappa=2, 4, 6$ for system sizes $N=26, 28, 30, 32$ to confirm that our extrapolations with $N$ ranging from $12$ to $24$ are sufficiently accurate. We note that at $\nu=1/2$, the energies for $N$ ranging from $8$ to $24$ fit well to straight lines in $1/N$ for all states, so one does not have the same issue that we had for the $\nu=1/4$ $\mathrm{Pf}_{-1}$ state. For $\nu=1/2$ unprojected states, when $\kappa=1$, the kinetic energy is large and also very non-uniform, causing the FPDMC program to frequently create or annihilate random walkers, which prevents convergence in a reasonable number of steps.
As a result, some data points for unprojected states at $\kappa=1$ for large system sizes are missing in Fig.~\ref{fig: other_extrapolation_12}. Fortunately this does not affect our conclusion, since we are primarily interested in larger values of $\kappa$. For the same reason, we have not carried out any calculations at $\kappa=1$ at $\nu=1/4$.

\section{Pair-correlation functions}\label{sec: pair_correlations}
\begin{figure*}
	\includegraphics[width = 0.32 \textwidth]{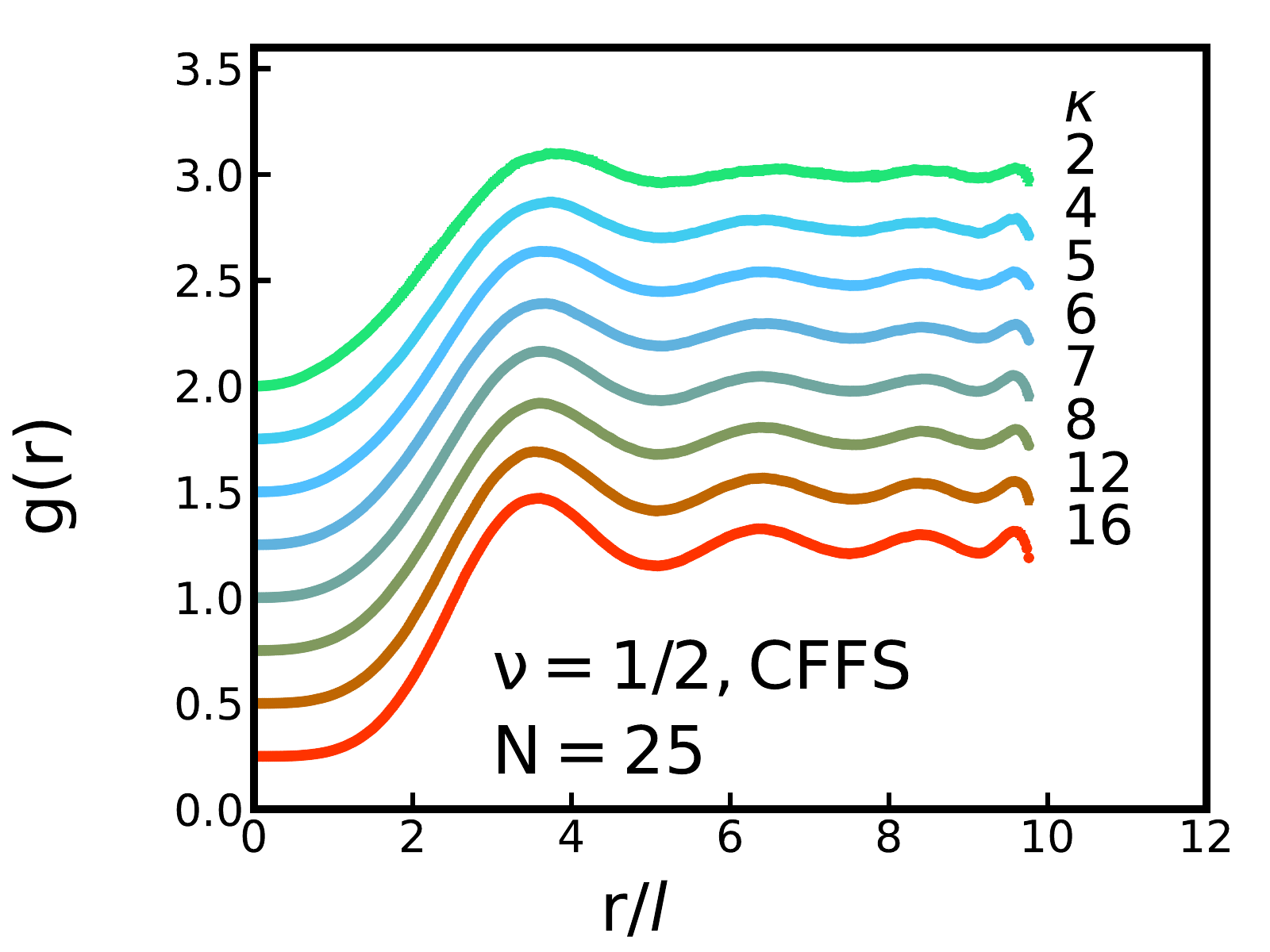}
	\includegraphics[width = 0.32 \textwidth]{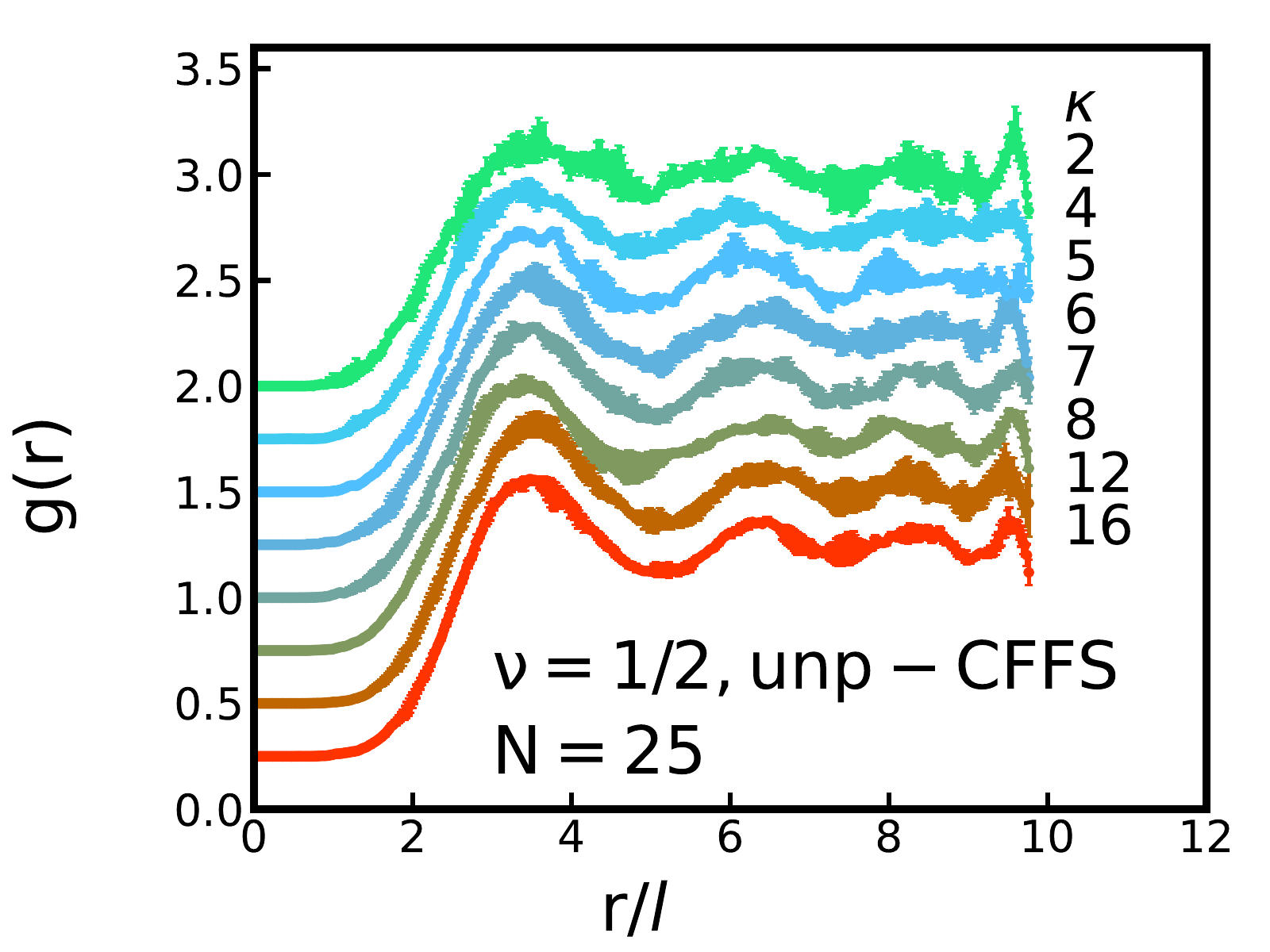}
	\includegraphics[width = 0.32 \textwidth]{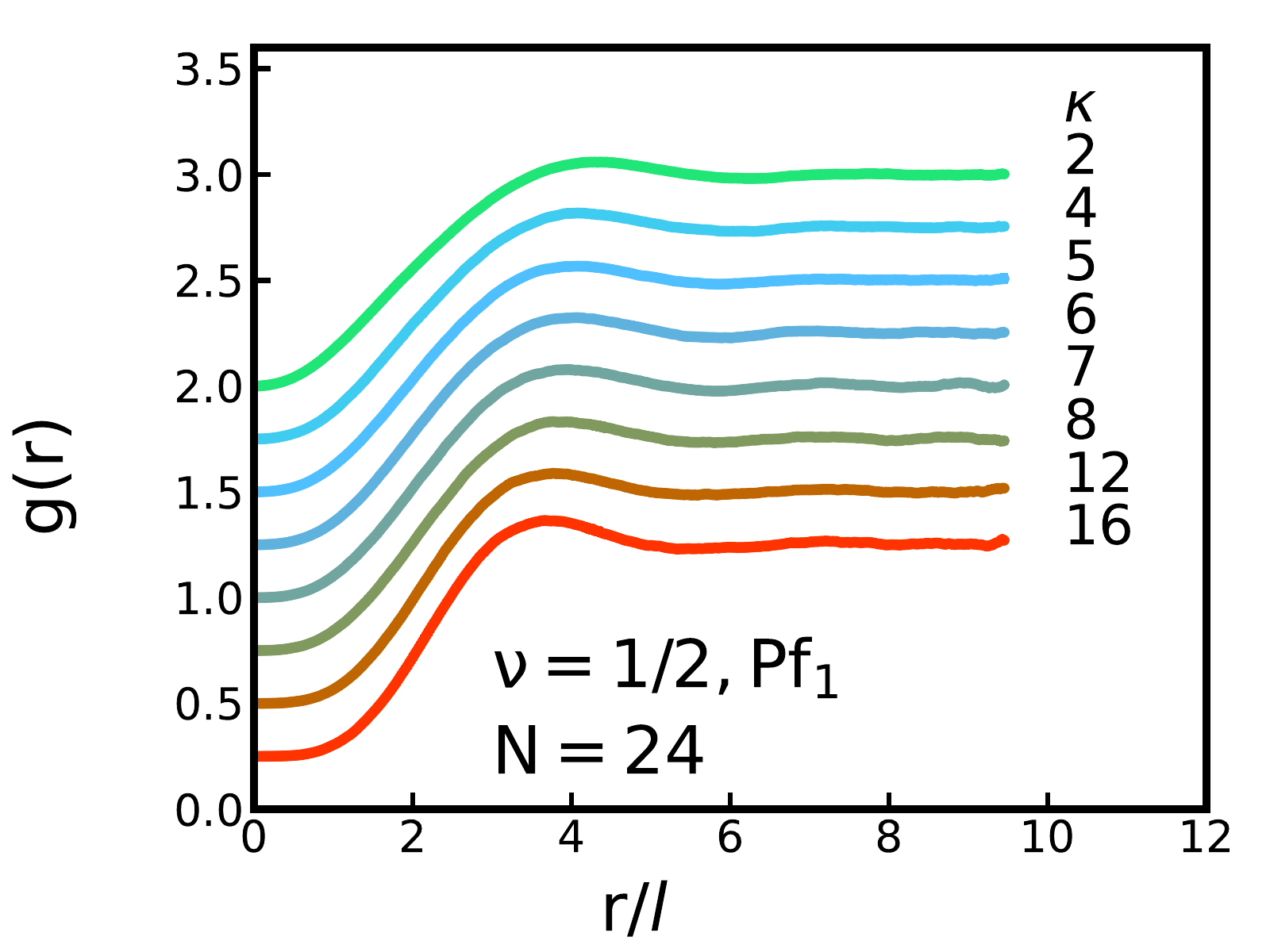}
	
	\includegraphics[width = 0.32 \textwidth]{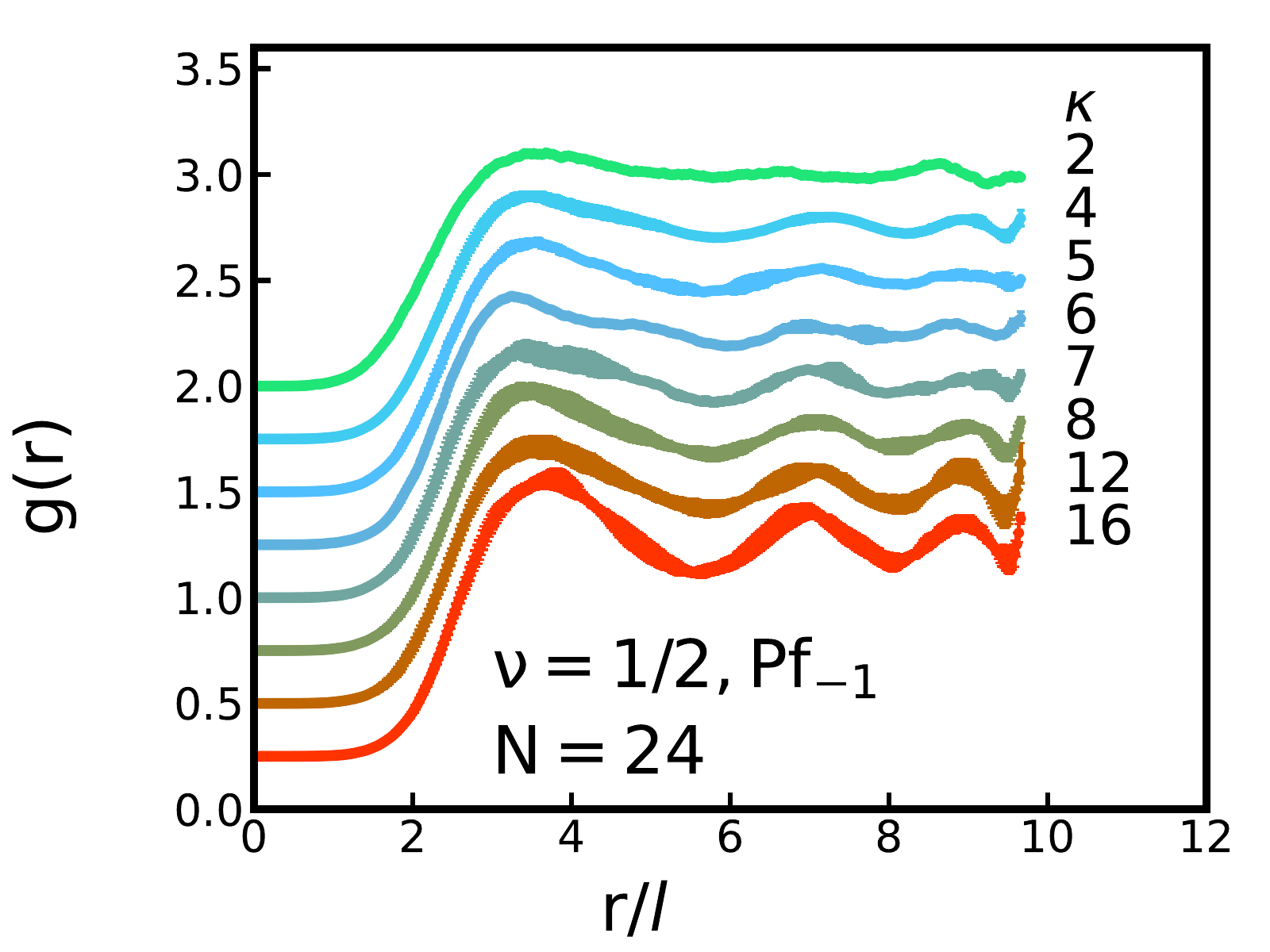}
	\includegraphics[width = 0.32 \textwidth]{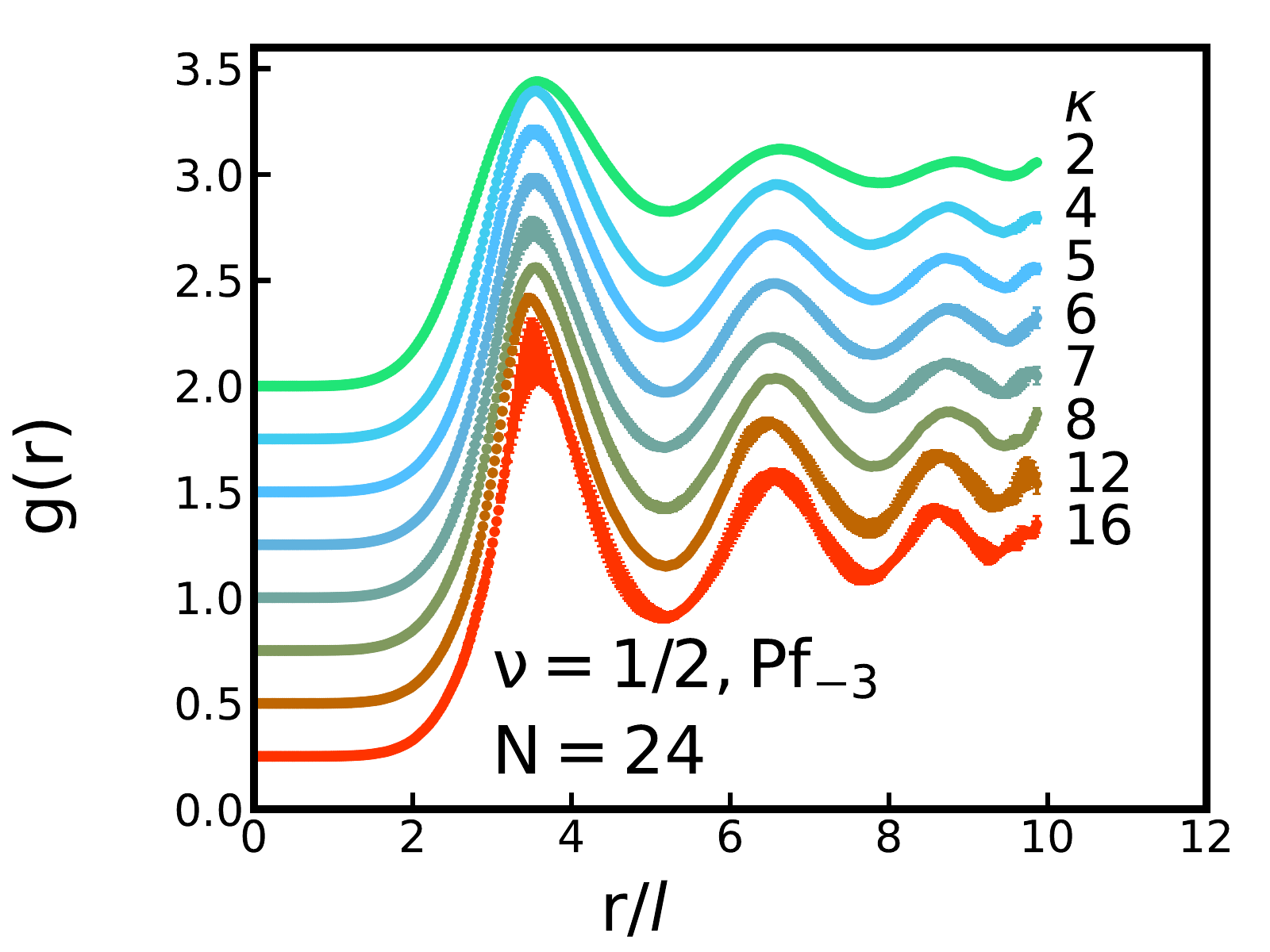}
	\includegraphics[width = 0.32 \textwidth]{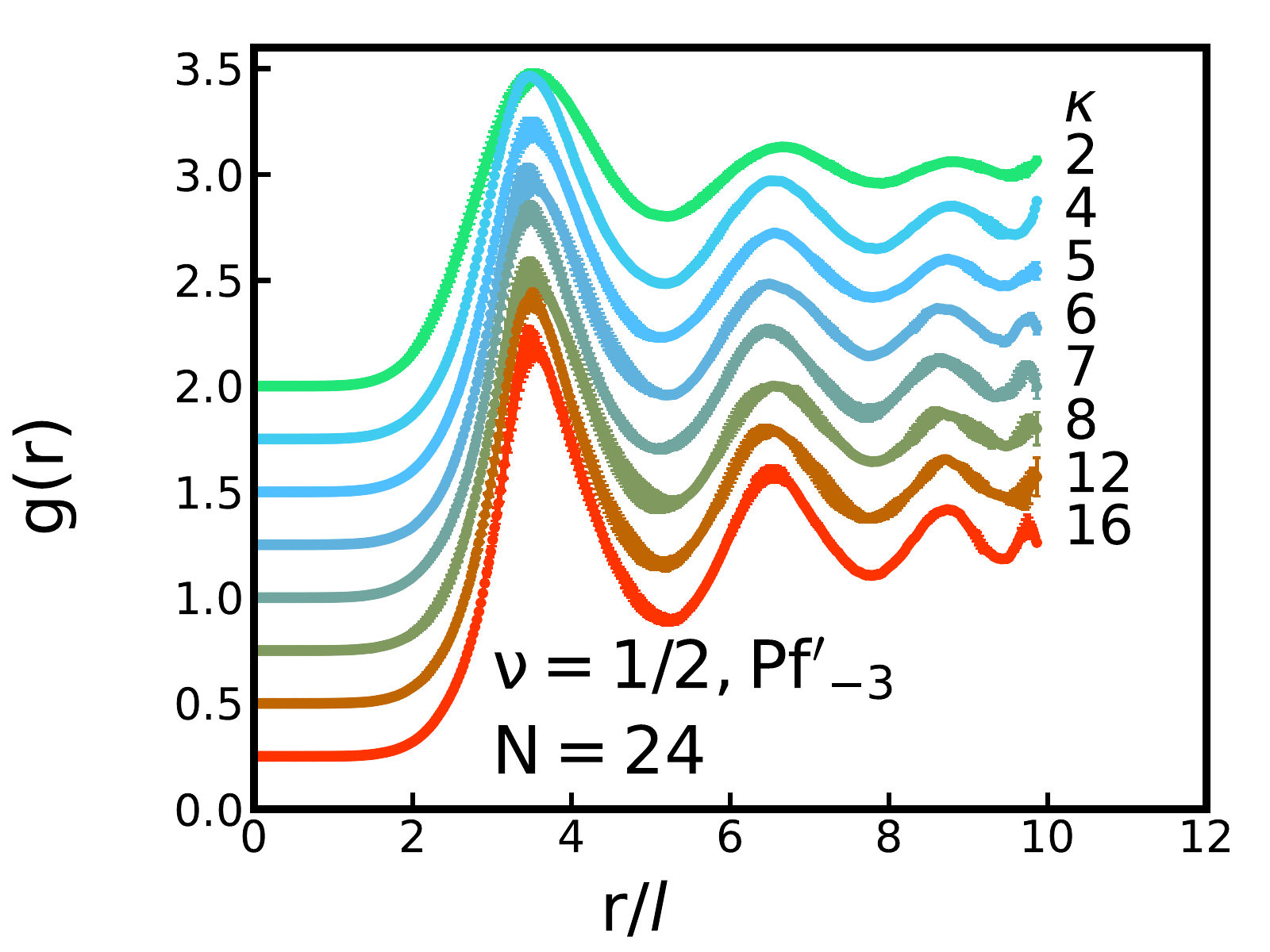}
	
	\includegraphics[width = 0.32 \textwidth]{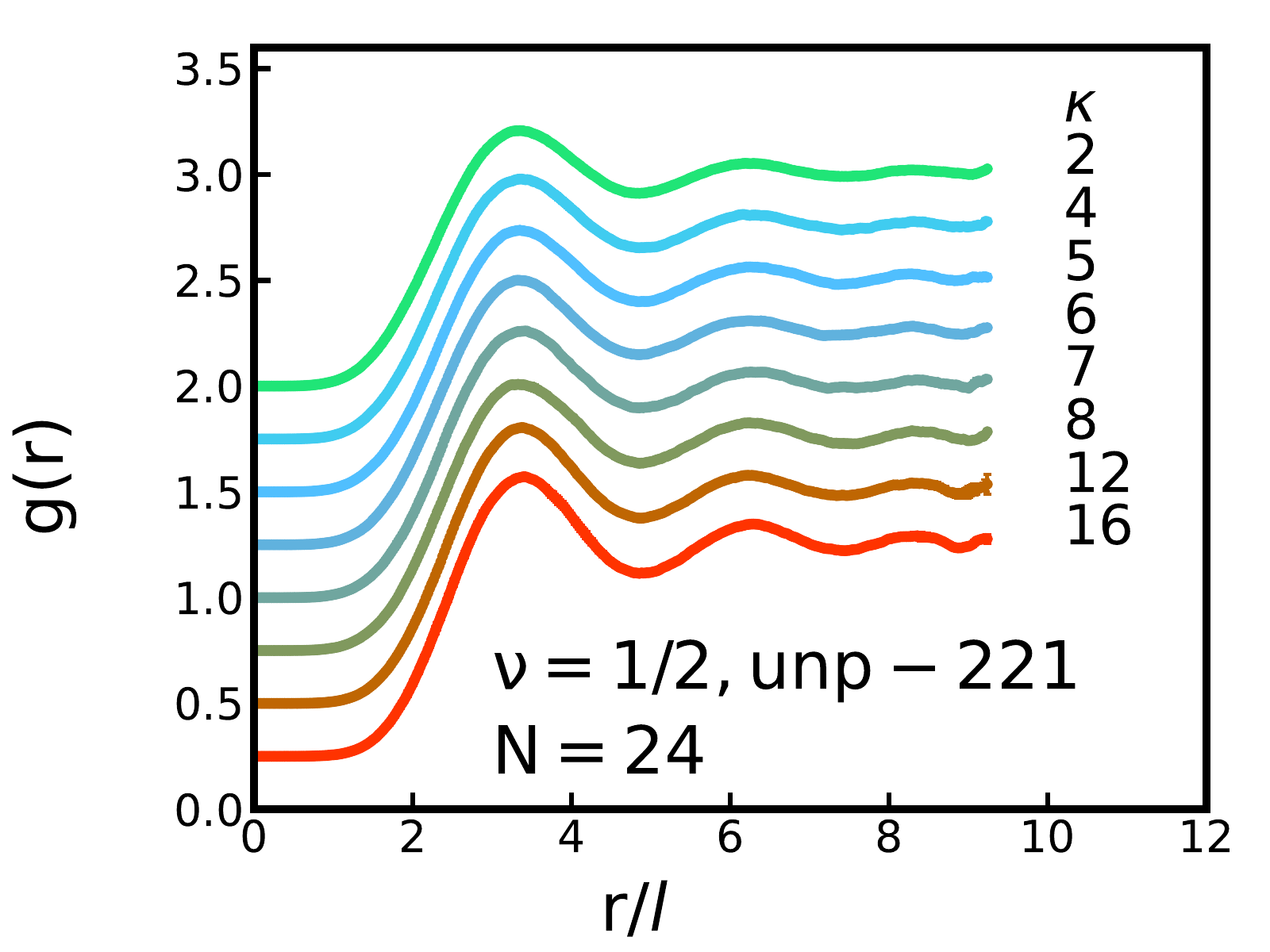}
	\includegraphics[width = 0.32 \textwidth]{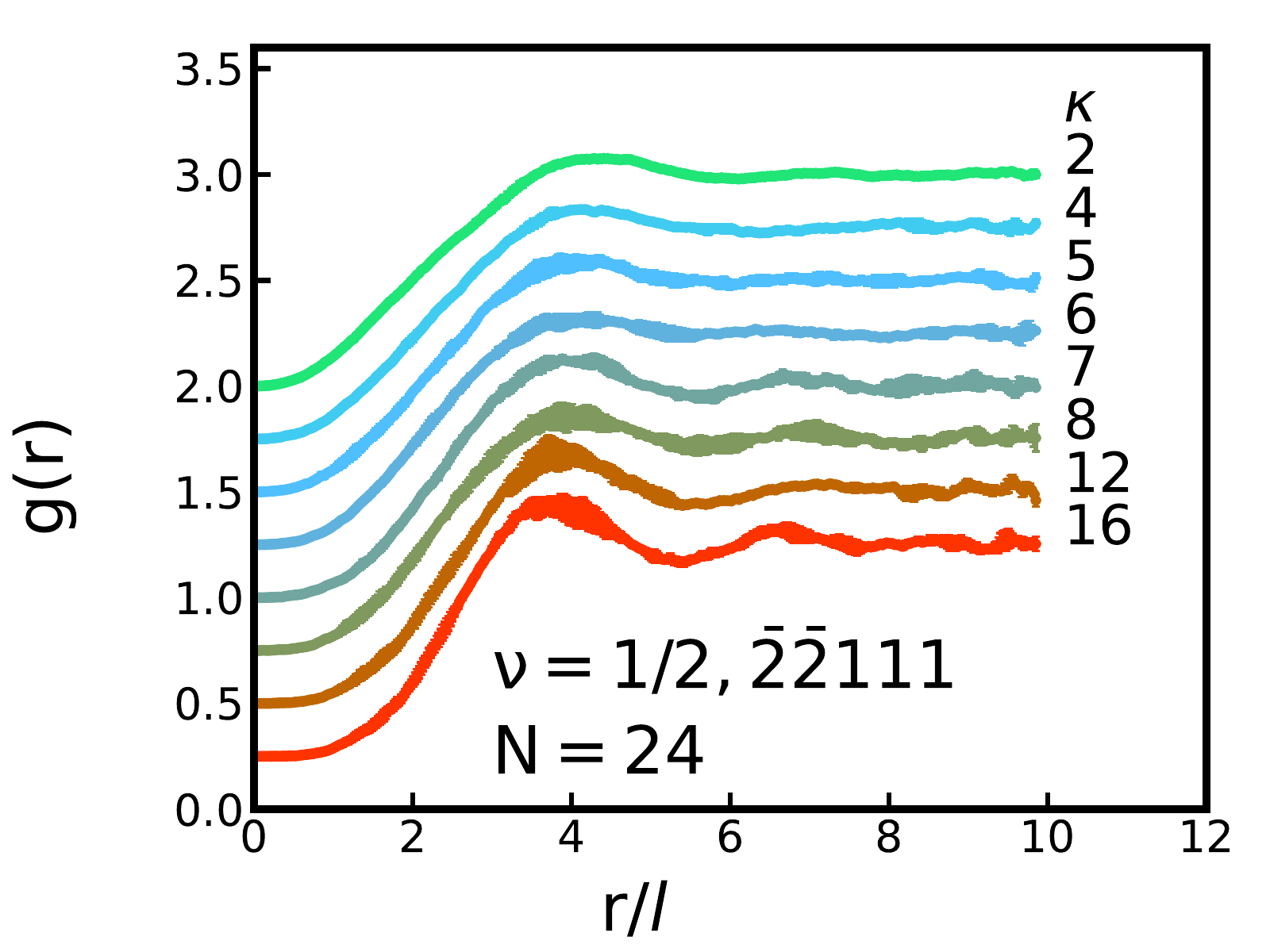}
	\includegraphics[width = 0.32 \textwidth]{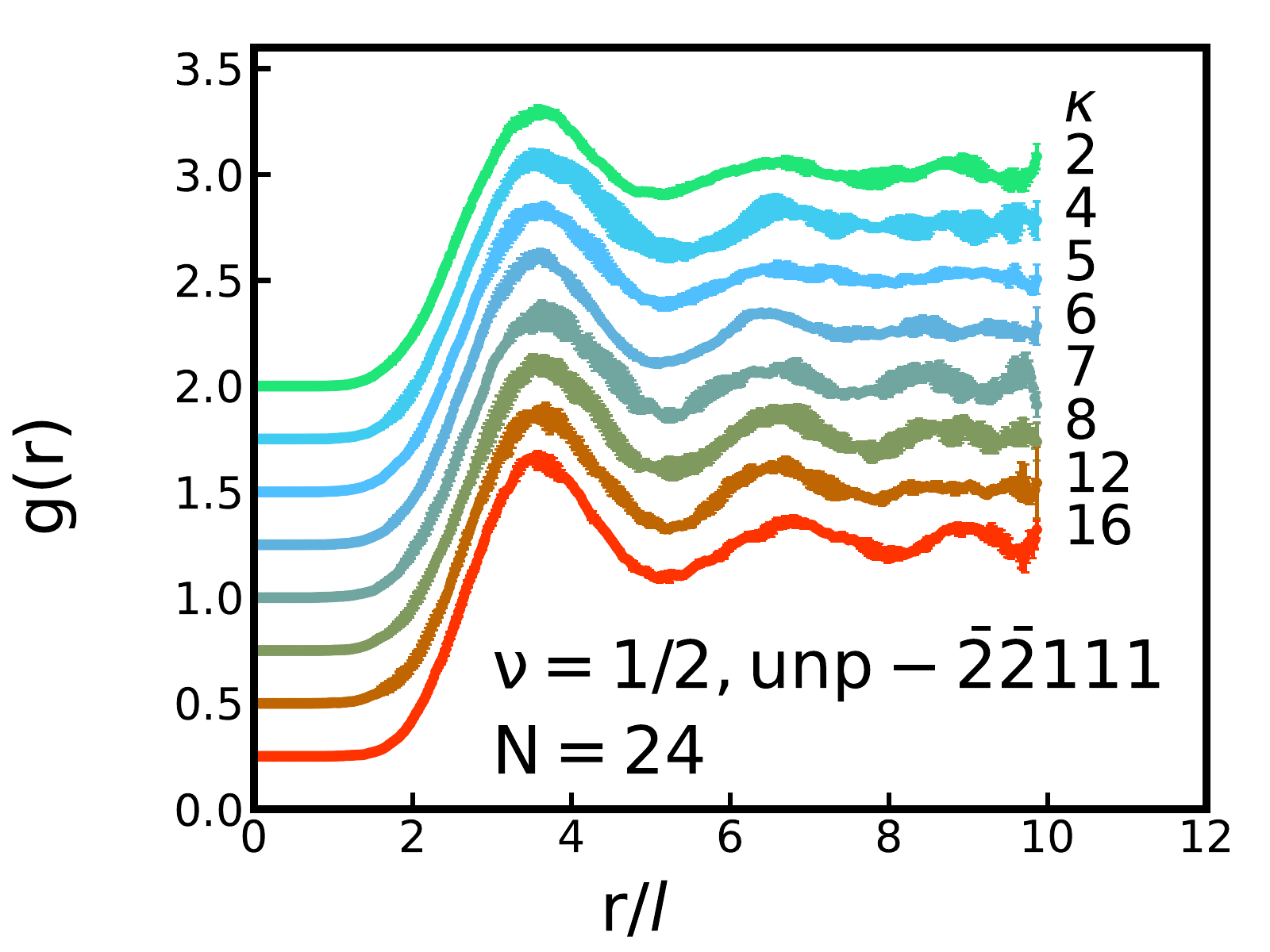}
		
	\caption{\label{fig: pair_corre_12}Pair-correlation functions for various candidate states at the filling factor $\nu=1/2$. The curves corresponding to different LLM are marked by different colors, with the $\kappa$ value for each curve shown on the right. The curves are shifted vertically for clarity.}
\end{figure*}

\begin{figure*}
	\includegraphics[width = 0.32 \textwidth]{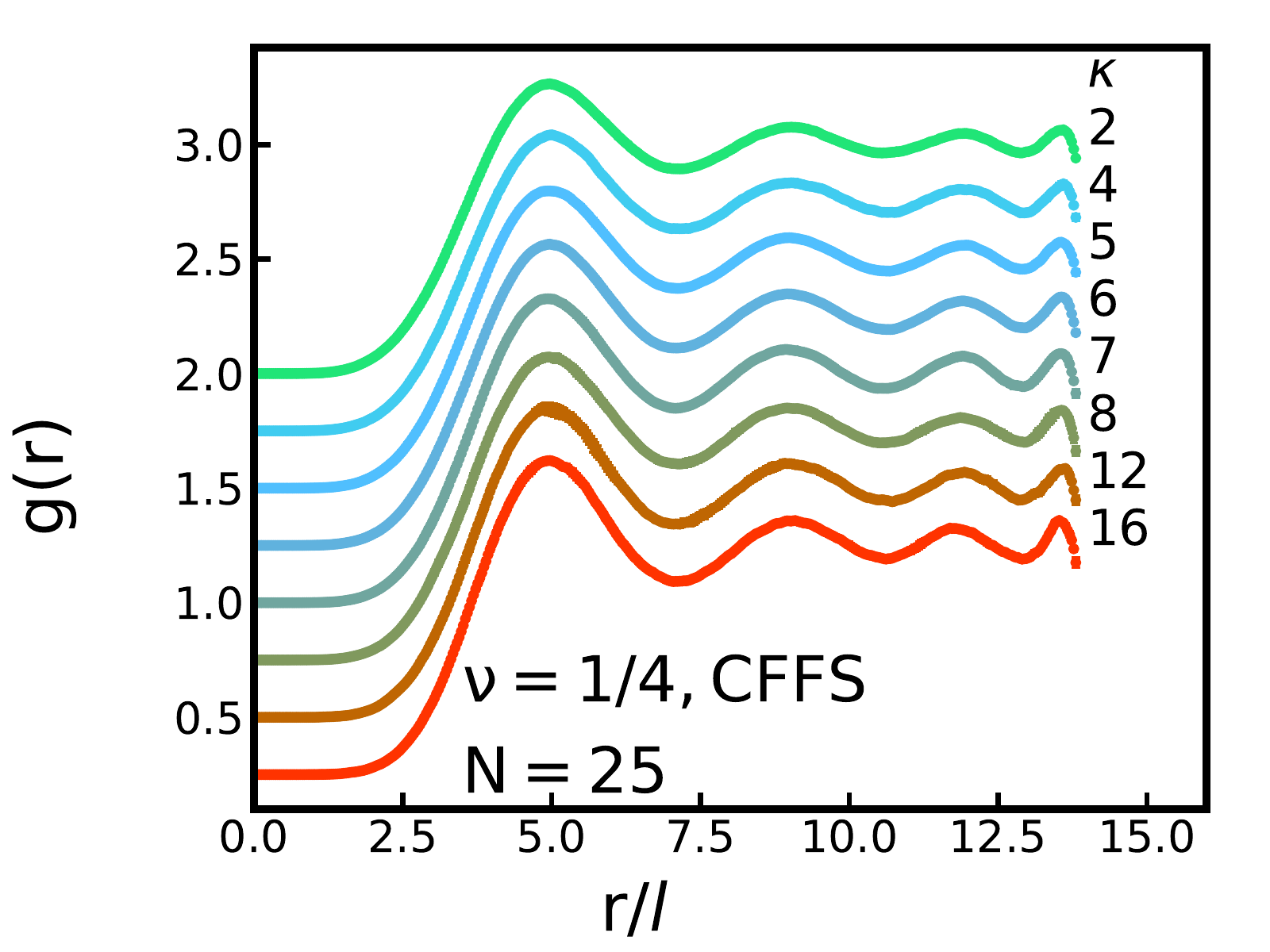}
	\includegraphics[width = 0.32 \textwidth]{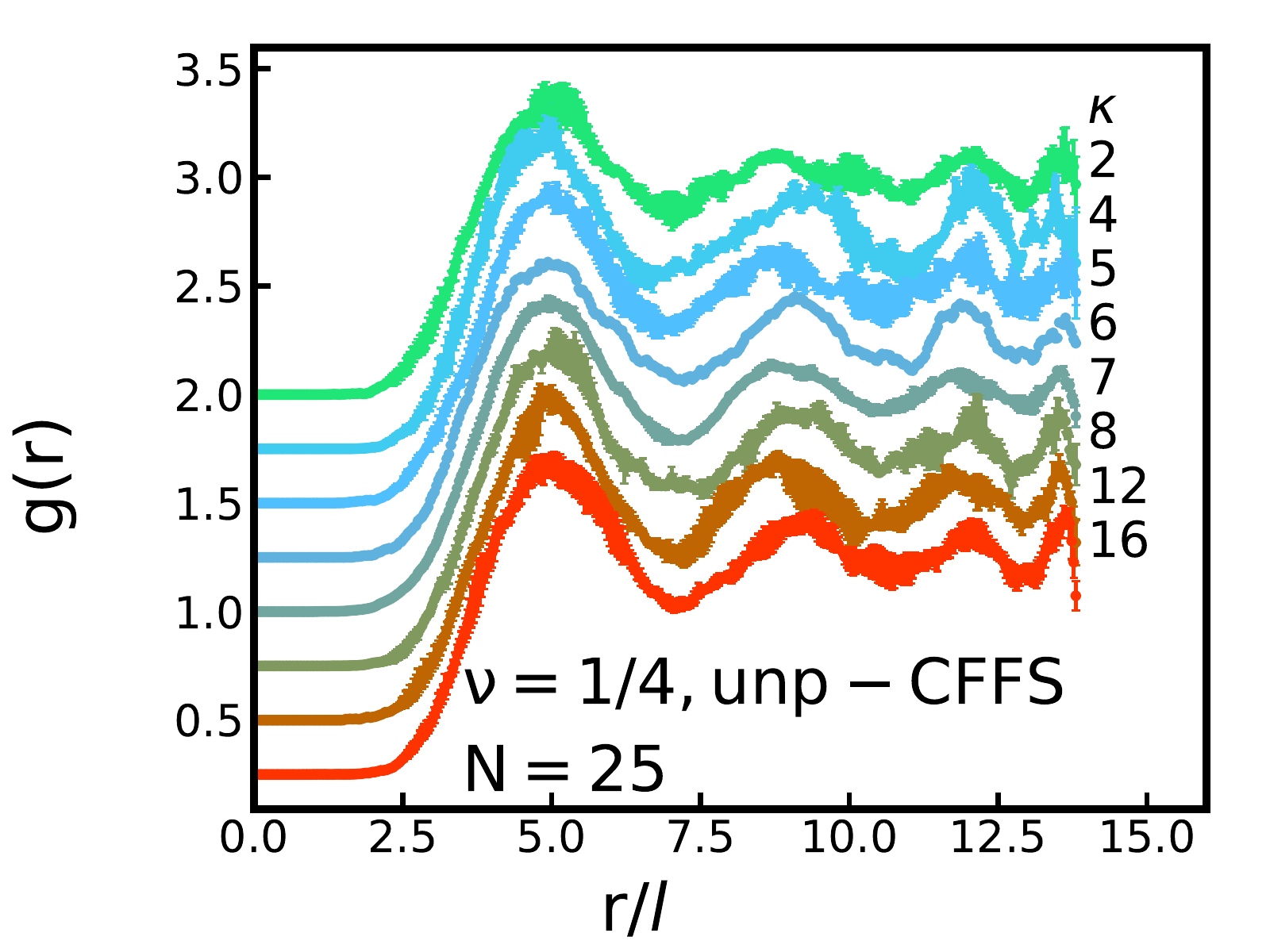}
	\includegraphics[width = 0.32 \textwidth]{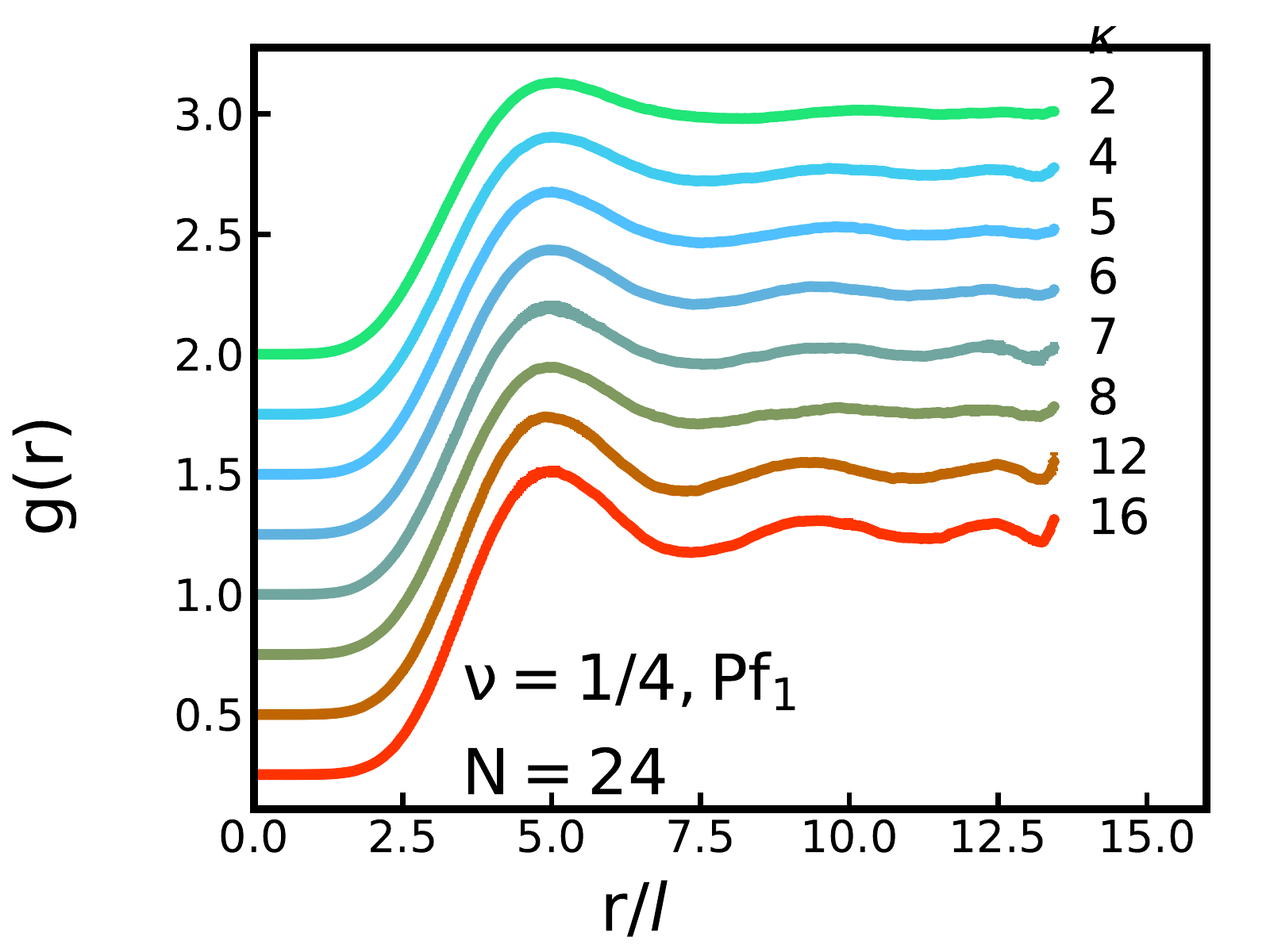}
	
	\includegraphics[width = 0.32 \textwidth]{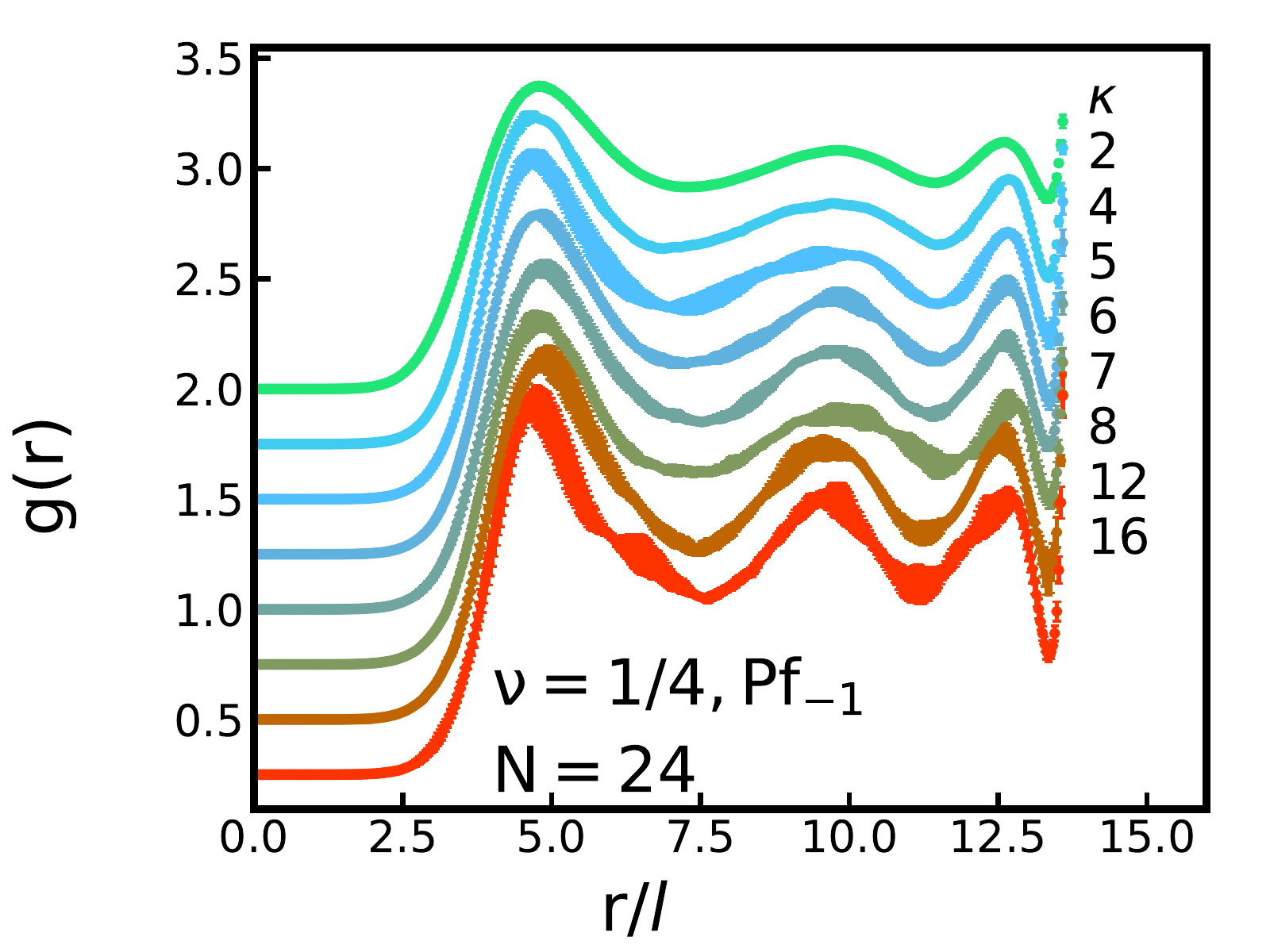}
	\includegraphics[width = 0.32 \textwidth]{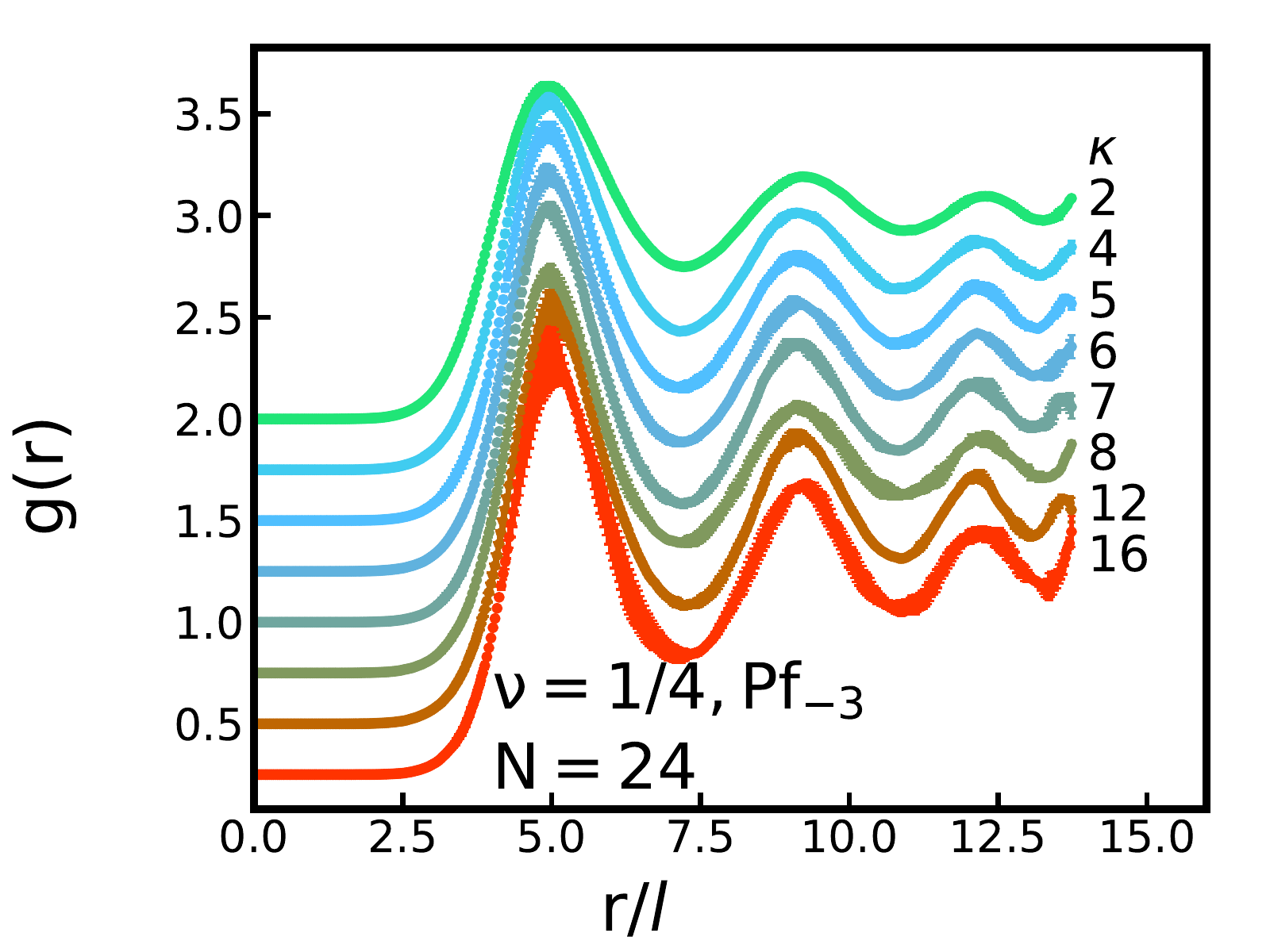}
	\includegraphics[width = 0.32 \textwidth]{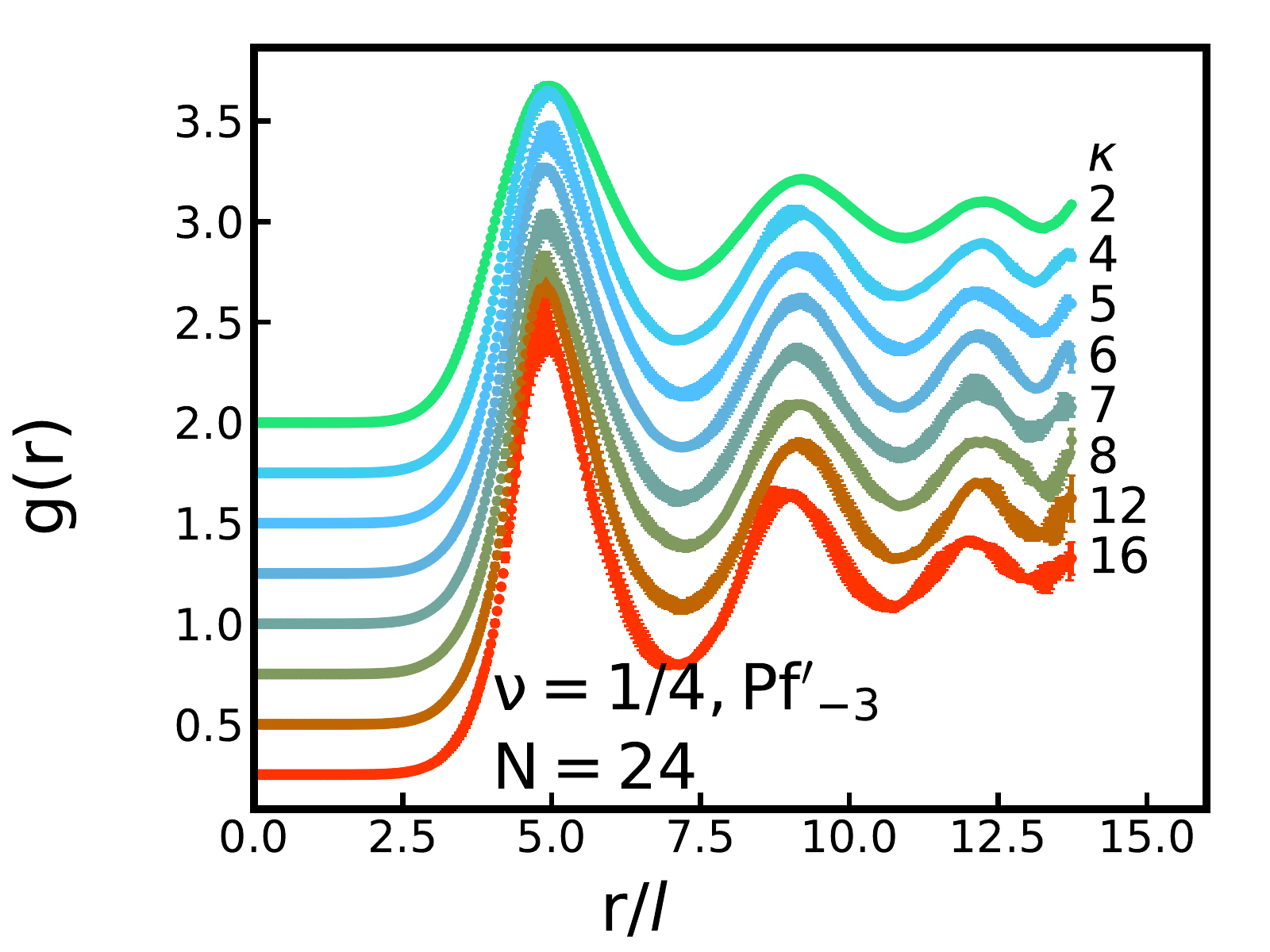}
	
	\includegraphics[width = 0.32 \textwidth]{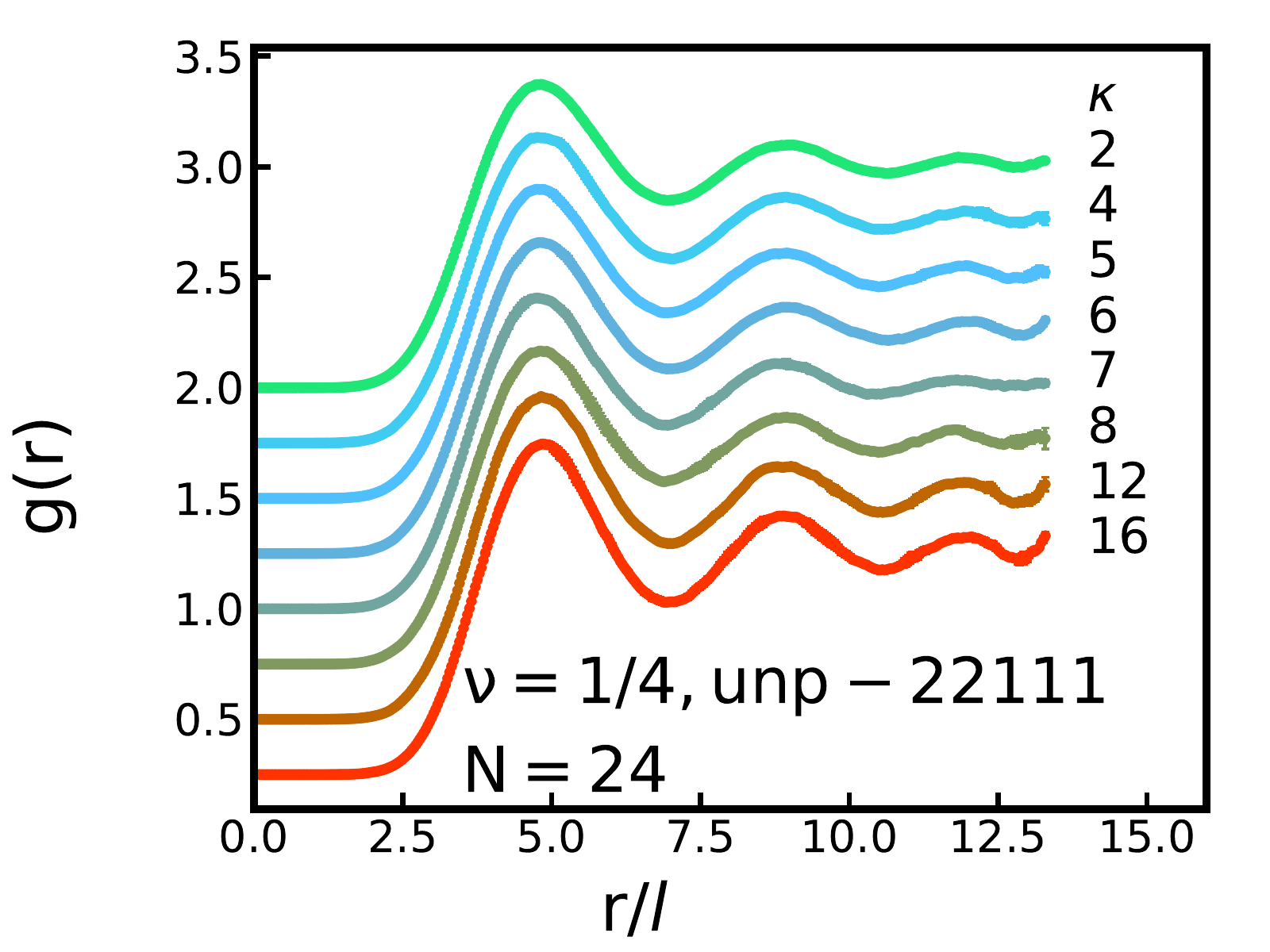}
	\includegraphics[width = 0.32 \textwidth]{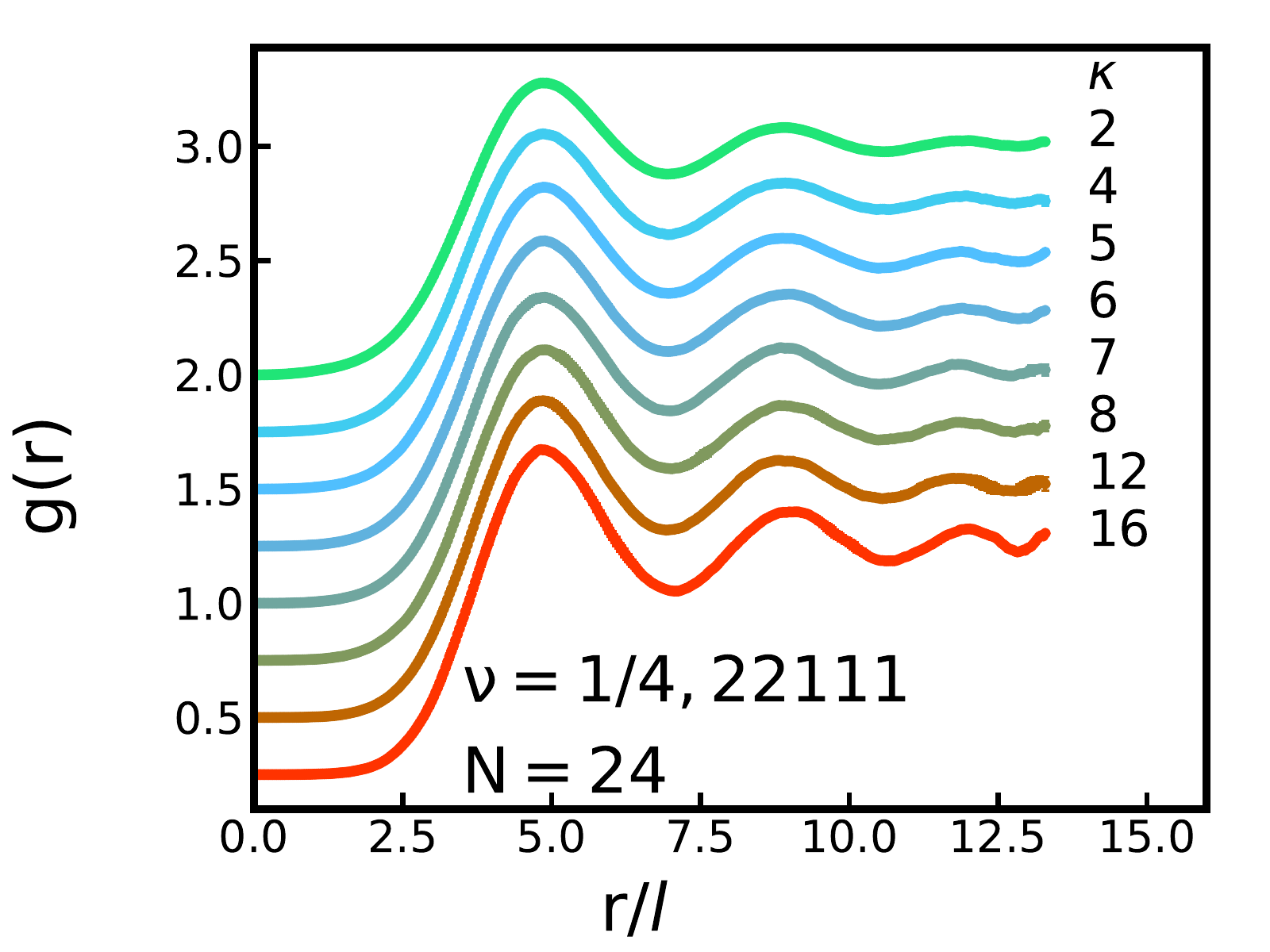}
	\includegraphics[width = 0.32 \textwidth]{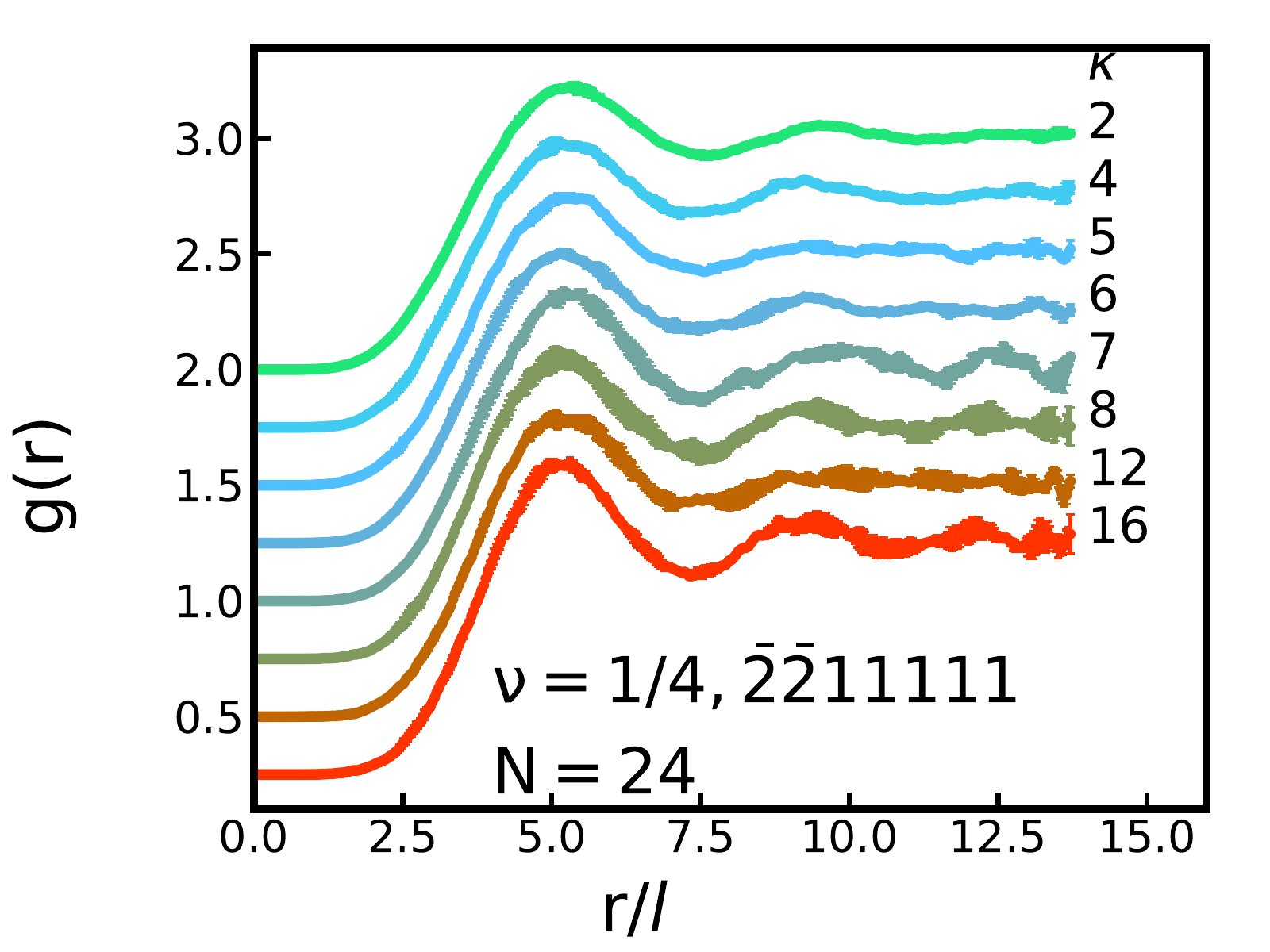}
	
	\includegraphics[width = 0.32 \textwidth]{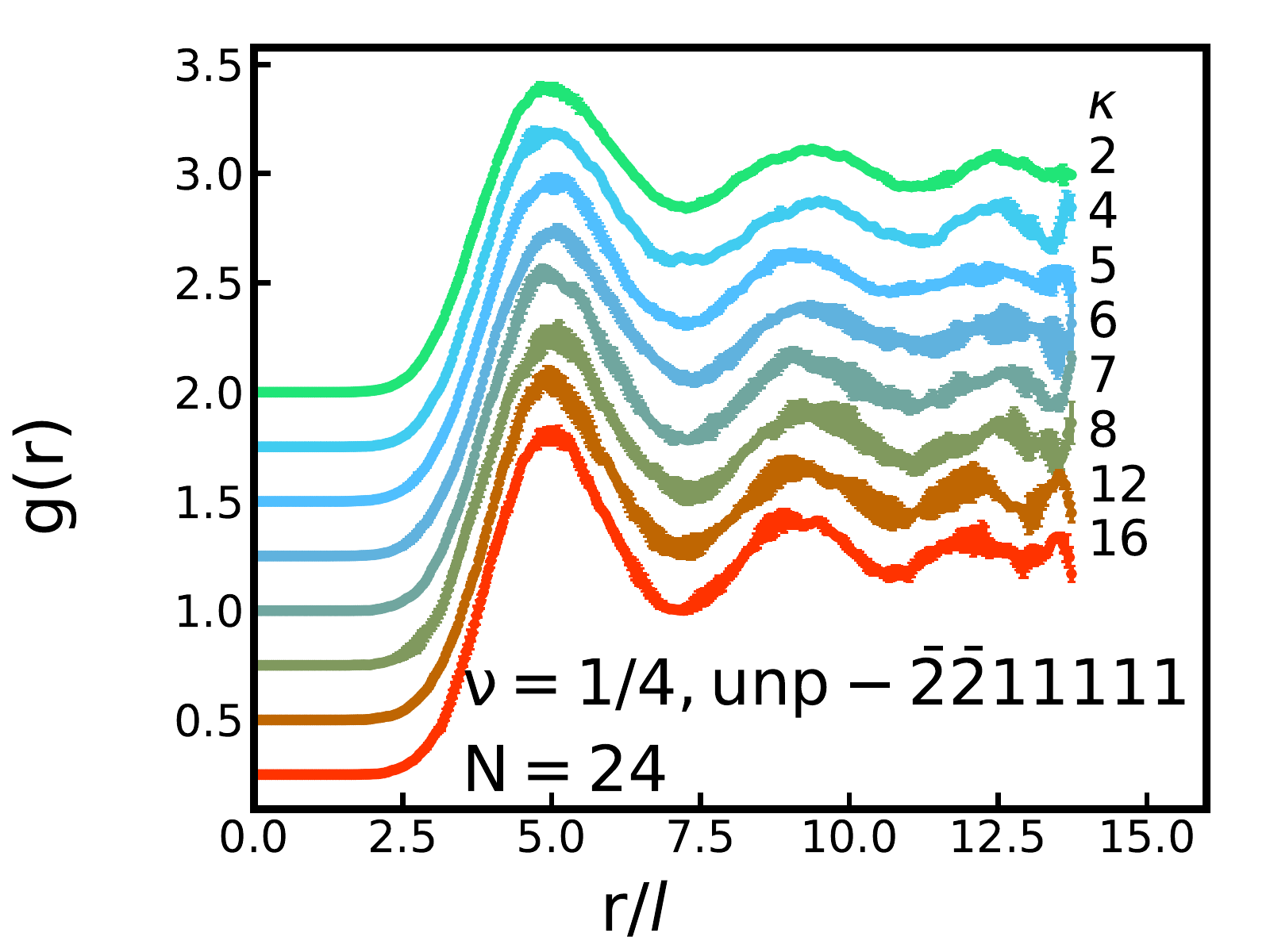}
	\caption{\label{fig: pair_corre_14}Pair-correlation functions for various candidate states at the filling factor $\nu=1/4$. The curves corresponding to different LLM are marked by different colors, with the $\kappa$ value for each curve shown on the right. The curves are shifted vertically for clarity.}
\end{figure*}

In this section, we show the pair-correlation functions of different states at $\nu=1/2$ (Fig.~\ref{fig: pair_corre_12}) and $\nu=1/4$ (Fig.~\ref{fig: pair_corre_14}), calculated with the largest system sizes that we have studied. For both filling factors, the pair-correlation functions for $\mathrm{\Psi_{1/2, 1/4}^{Pf_{-3}}}$ and $\mathrm{\Psi_{1/2, 1/4}^{Pf'_{-3}}}$ show oscillations that decay with distance and converge to unity in the limit $r\to \infty$, as expected for an incompressible state~\cite{Kamilla97, Balram15b, Balram17}. We note that the oscillations in $g\left(r\right)$ for $\mathrm{\Psi_{1/2, 1/4}^{Pf_{-3}}}$ and $\mathrm{\Psi_{1/2, 1/4}^{Pf'_{-3}}}$ are strong, which suggests that the FQH states predicted by our calculations might be fragile with small gaps.

At $\kappa=0$ the unprojected Pf$_{-1}$ state starts off with a nice pair-correlation function that is identical to that of the Moore-Read Pfaffian state~\cite{Moore91}, but with LLM it shows noise like fluctuations. 
At $\nu=1/4$, as we have mentioned earlier, finite-size effects may have a stronger influence on Pf$_{-1}$ than on the other states we considered. However, we have not pursued this issue further because Pf$_{-1}$ is energetically less favourable than Pf$_{-3}$.


\section{Mechanism and Nature of Pairing}
\label{sec:intuitive}

The interaction between composite fermions is a complex function of the interaction between electrons and LL mixing~\cite{Sitko96, Lee01, Lee02, Balram16c, Balram17b}. For the underlying physical mechanism of pairing, we have given the following qualitative picture. Weakly interacting CFs are produced when the interaction between electrons is strongly repulsive at short range.  LL mixing screens the short-distance interaction between electrons~\cite{Bishara09, Sodemann13, Peterson13}, which, as shown previously, can lead to an attractive interaction between composite fermions and hence CF pairing~\cite{Scarola00b}. Such a reduction of the short range repulsion is also the reason why a paired CF state occurs in the second LL~\cite{Willett87} even though we have a CF Fermi sea in the lowest LL~\cite{Halperin93, Halperin20, Shayegan20} . 

We do not have a simple argument for why the $l=-3$ pairing is favored over other pairing channels. It is worth noting that this is a complicated issue given the rather non-trivial nature of the state. To give an example, a simpler question would be whether the 5/2 state is a Pfaffian~\cite{Moore91} ($l=1$) or an anti-Pfaffian~\cite{Lee07, Levin07} ($l=-3$). These two states are equally plausible in the absence of LL mixing, and LL mixing breaks the tie between them. The modified effective interaction between electrons can be determined, in principle, by treating LL mixing in a perturbative fashion~\cite{MacDonald84,Melik-Alaverdian95,Murthy02,Bishara09,Wojs10, Sodemann13, Simon13,Peterson13,Peterson14}.  However, no simple argument can tell us which of the two states wins. Neither mean field theory nor Chern Simons field theory~\cite{Halperin93} provides any insight. It has taken many years of detailed, exact-diagonalization calculations by many groups~\cite{Bishara09, Wojs10, Rezayi11, Peterson13, Pakrouski15, Rezayi17, Sreejith17, Herviou22} to conclude that the anti-Pfaffian wins, and actually 3-body pseudopotential in the relative angular momentum 9 channel breaks the tie~\cite{Rezayi17}.  For the present purposes, the determination of the effective electron-electron is further complicated by the necessity of going beyond a perturbative approach. However, even if we were to find the electron-electron interaction pseudopotentials, that would not give deeper insight into the nature of pairing than what our calculation has given, because one cannot read the nature of pairing simply by looking at the electron-electron interaction pseudopotentials; for this purpose one would need to perform exact diagonalization with the effective interaction.